\PassOptionsToPackage{main=english}{babel}
\documentclass{iopjournal}

\usepackage{amsmath}
\usepackage{amssymb}
\usepackage{graphicx}          
\usepackage{caption}           
\usepackage{subcaption}
\usepackage{hyperref}          
\usepackage{bm}
\usepackage{bbold}
\usepackage{stfloats}
\usepackage{comment}
\usepackage{booktabs}
\usepackage{tabularx}
\usepackage{multirow}
\usepackage{enumitem}
\usepackage{array}

\usepackage[T1]{fontenc}   
\usepackage[autostyle]{csquotes}

\usepackage{tikz}
\usetikzlibrary{positioning, arrows.meta, calc, fit,decorations.pathmorphing, matrix}

\begin{document}

\pagestyle{plain}

\title{Neural posterior estimation of the neutrino direction in IceCube using transformer-encoded normalizing flows on the sphere}

\author{\textbf{The IceCube Collaboration}\\R. Abbasi$^{1}$, M. Ackermann$^{2}$, J. Adams$^{3}$, J. A. Aguilar$^{4}$, M. Ahlers$^{5}$, J.M. Alameddine$^{6}$, S. Ali$^{7}$, N. M. Amin$^{8}$, K. Andeen$^{9}$, C. Arg{\"u}elles$^{10}$
% ...
}

\affil{$^{1}$Department of Physics, Loyola University Chicago, Chicago, IL 60660, USA}
\affil{$^{2}$Deutsches Elektronen-Synchrotron DESY, Platanenallee 6, D-15738 Zeuthen, Germany}
\affil{$^{3}$Dept. of Physics and Astronomy, University of Canterbury, Private Bag 4800, Christchurch, New Zealand}
\affil{$^{4}$Universit{\'e} Libre de Bruxelles, Science Faculty CP230, B-1050 Brussels, Belgium}
\affil{$^{5}$Niels Bohr Institute, University of Copenhagen, DK-2100 Copenhagen, Denmark}
\affil{$^{6}$Dept. of Physics, TU Dortmund University, D-44221 Dortmund, Germany}
\affil{$^{7}$Dept. of Physics and Astronomy, University of Kansas, Lawrence, KS 66045, USA}
\affil{$^{8}$Bartol Research Institute and Dept. of Physics and Astronomy, University of Delaware, Newark, DE 19716, USA}
\affil{$^{9}$Department of Physics, Marquette University, Milwaukee, WI 53201, USA}
\affil{$^{10}$Department of Physics and Laboratory for Particle Physics and Cosmology, Harvard University, Cambridge, MA 02138, USA}
% ...

\email{analysis@icecube.wisc.edu}

\begin{abstract}
IceCube is a cubic-kilometer-scale neutrino detector located at the geographic South Pole. A precise directional reconstruction of IceCube neutrinos is vital for associations with astronomical objects. 
In this context, we discuss neural posterior estimation of the neutrino direction via a transformer encoder that maps to a normalizing flow on the 2-sphere. It achieves a new state-of-the-art angular resolution for the two main event morphologies in IceCube - tracks and showers - while being significantly faster than traditional B-spline-based likelihood reconstructions. All-sky scans can be performed within seconds rather than hours, and take constant computation time, regardless of whether the posterior extent is arc-minutes or spans the whole sky. We utilize a combination of $C^2$-smooth rational-quadratic splines, scale transformations and rotations to define a novel spherical normalizing-flow distribution whose parameters are predicted as a whole as the output of the transformer encoder. We test several structural choices diverting from the vanilla transformer architecture. In particular, we find dual residual streams, nonlinear QKV projection and a separate class token with its own cross-attention processing to boost test-time performance.

The angular resolution for both showers and tracks improves substantially over the whole trained energy range from 100 GeV to 100 PeV. At 100 TeV deposited energy, for example, the median angular resolution improves by a factor of $1.3$ for throughgoing tracks, by a factor of $1.7$ for showers and by a factor of $2.5$ for starting tracks compared to state-of-the art likelihood reconstructions based on B-splines. While previous machine-learning (ML) efforts have managed to obtain competitive shower resolutions, this is the first time an ML-based method outperforms likelihood-based muon reconstructions above 100 GeV. 
\end{abstract}


\section{Introduction}

The IceCube neutrino detector \cite{icecube_instrumentation} is the world's largest high-energy neutrino observatory and has made several breakthrough discoveries over the last decade. Among them, we find the association of high-energy neutrinos with specific astrophysical sources including the active galactic nucleus NGC 1068 \cite{ngc_1086} and the Galactic Plane \cite{gal_plane_paper}. These results were achieved through statistical analyses that rely fundamentally on two observables reconstructed for each individual neutrino event: the deposited energy and the arrival direction. The arrival direction is particularly important for point-source searches, where the flux sensitivity scales approximately with the angular resolution through improved background suppression. We focus on the direction reconstruction in this paper.

\subsection{Motivation}
IceCube detects neutrinos via Cherenkov light \cite{cherenkov_radiation} which is emitted from charged relativistic particles that originate in the neutrino interaction.
The traditional approach to inferring the direction is a maximum-likelihood estimation based on the observed Cherenkov light distributions across the detector's optical modules. Over the years, dedicated reconstructions have been developed for the distinct event morphologies that arise from different neutrino interaction channels. Three major morphology classes are useful to be differentiated in this context: shower-like events, throughgoing tracks, and starting tracks.

Shower-like events, produced by charged-current $\nu_e$ interactions, neutral-current interactions of all flavors, and sub-PeV charged-current $\nu_\tau$ interactions, exhibit approximately spherical Cherenkov light patterns. The angular resolution of showers is typically on the order of 5 to 10 degrees and the associated uncertainty contours are often non-Gaussian. This class of events was instrumental in the discovery of the astrophysical diffuse neutrino flux \cite{hese_paper} and neutrino emission from the Galactic Plane \cite{gal_plane_paper}. The Galactic Plane measurement relied on a neural-network-based likelihood approach for shower reconstruction \cite{event_generator_paper}, combined with CNN-based regression networks \cite{icecube_cnn_paper} in the event selection stage. Recently, a B-spline-based likelihood reconstruction achieved state-of-the-art angular resolution for showers \cite{monopod_update} by modeling the shower extension and improving the treatment of systematic uncertainties. We refer to this method as \textit{Taupede2024} for the remainder of the paper.

Throughgoing track events originate from charged-current $\nu_\mu$ interactions that occur outside the instrumented volume, producing muons that traverse the full detector. The resulting elongated light patterns typically yield angular resolutions below one degree, making this the most important event morphology for point-source studies. The primary reconstruction algorithm, referred to as \textit{SplineMPEMax} in the following, models the muon track as an effective averaged energy-loss pattern \cite{schatto_thesis} and has served as the backbone of all point-source analyses to date, including the identification of TXS 0506+056 \cite{txs_paper} and NGC 1068 \cite{ngc_1086} as neutrino sources. At energies above a few TeV, and especially beyond 100 TeV, stochastic energy losses begin to dominate and the quasi-continuous modeling assumption becomes less accurate. An extension that explicitly models stochastic losses along the track was developed in \cite{segmented_spline_paper}. However, its substantially longer runtime, numerical instabilities, and only marginal resolution improvements have so far precluded its adoption in practice.

Starting track events, in which the $\nu_\mu$ interaction vertex lies within the instrumented volume, present a distinct reconstruction challenge. The Cherenkov light from the initial hadronic cascade overlaps with that of the typically short outgoing muon track, complicating directional inference. Nevertheless, this event class holds substantial promise for southern-sky neutrino astronomy in IceCube. Atmospheric muon veto techniques \cite{hese_paper} can yield highly pure starting-track samples with a direct line of sight toward the Galactic Center, and the angular resolution is considerably better than that of pure showers. A recent southern-sky diffuse flux measurement incorporating a dedicated starting-track selection \cite{estes_paper} has demonstrated their potential for southern-sky measurements. However, the full power of starting tracks to resolve Galactic structure has yet to be utilized.

A unique challenge for reconstruction in IceCube, compared to artificially constructed detectors, is that the detection medium is a naturally formed glacier. While the ice at several kilometers depth is exceptionally transparent on average, inhomogeneous dust deposits introduce position-dependent variations in scattering and absorption \cite{ftp_paper} that must be accounted for in the reconstruction. Most of the aforementioned likelihood-based approaches rely on B-spline parameterizations of the expected Cherenkov photon arrival time distributions  \cite{bspline_paper}. Incorporating the high-dimensional ice optical properties into these models is challenging. Beyond systematic uncertainties, complex event signatures such as high-energy tracks possess intrinsic degrees of freedom, notably the stochastic energy-loss profile, that are difficult to model in their own right \cite{millipede_paper}\cite{segmented_spline_paper}. A directional likelihood estimation must jointly profile over all these combined nuisance parameters, which is both numerically challenging and computationally expensive. An alternative to B-spline parameterizations that has emerged in recent years is to model the likelihood using neural networks \cite{event_generator_paper}. This approach, combined with small feed-forward networks for the event selection stage, enabled the first measurement of neutrinos from the Galactic Plane \cite{gal_plane_paper}. Neural-network-based likelihoods typically offer an easier incorporation of systematic uncertainties into the modeling process. However, the challenge of parameterizing complex intrinsic degrees of freedom, such as the muon energy-loss profile, persists, and the computational cost of likelihood evaluation remains comparable to or greater than that of B-spline methods.

In general, a low processing time is desirable for many applications. It is particularly important for real-time neutrino alerts \cite{realtime_paper} that are sent out to the astronomical community. Here, the goal is to obtain precise localization contours as fast as possible. Currently, expensive profile-likelihood scans can take up to several hours per event, which delays localization for astronomical follow-up observations by other experiments. One way to get a fast prediction is to switch from an explicit likelihood model to a neural-network regression model that operates in the inverse direction, i.e. predicts the direction from the data. However, such inverse models typically either perform point predictions or assume Gaussian uncertainties \cite{icecube_cnn_paper}, which can lead to biases or a lack of precision. 

In this paper, we describe a method that also operates in the inverse direction but goes beyond point estimation and Gaussian posterior assumptions to predict the full directional posterior distribution. Rather than explicitly parameterizing nuisance degrees of freedom in the forward direction, the method implicitly learns to marginalize over them during training. At inference time, the complete posterior is obtained in a single neural-network forward pass, with the computational cost effectively offloaded to the training stage. The result is a reconstruction framework that combines the speed of direct inference with precise angular resolution, robust non-Gaussian uncertainty quantification, and the flexibility to incorporate complex systematic effects. Because it is both fast and precise, it is applicable to event selections, final-level analysis reconstructions, and real-time astronomical alerts.

\subsection{Outline}
The proposed inference procedure is based on amortized neural posterior estimation (NPE) with conditional normalizing flows. In standard neural posterior estimation \cite{npe_definition}, one is typically interested in obtaining the posterior $p(\theta|x_o)$ for one particular event with observed data $x_o$. The procedure involves iterative re-simulations of both a proposal prior and a posterior approximation, and is situated in the broader framework of simulation-based inference (SBI) \cite{sbi_review}. Here, we are interested in all posteriors at the same time, and just apply one single approximate posterior training. Hence it is a form of \textit{amortized} neural posterior estimation since the cost of estimation of a single posterior is amortized by a neural network. 

Conditional normalizing flows amortized with neural networks have previously shown promise for reconstructing the posterior distribution of the energy and direction of neutrinos with high fidelity \cite{icrc_proceeding_23}, without being bound to Gaussian uncertainty assumptions. In that prior work we used a compression of the per-module photon data into a summary statistic that is fed into a graph neural network (GNN) encoding similar to \cite {gnn_low_energy}. It used a k-nearest neighbor approach to construct the graph followed by edge convolution \cite{edgeconv_reference}. The output of the encoding was then mapped to the normalizing-flow parameters. Here, we improve on that work for the directional reconstruction in several ways.

\begin{figure}[ht]
    \centering
    \begin{subfigure}[b]{0.49\linewidth}
    \centering
    \begin{tikzpicture}[
  img/.style={draw, minimum width=2.5cm, minimum height=1.6cm, align=center},
  box/.style={draw, rounded corners, minimum width=2cm, minimum height=0.5cm, align=center},
  >=latex
]
\matrix (grid) [
  matrix of nodes,
  nodes={img},
  column sep=0.9cm,
  row sep=0.5cm
] {
  {\includegraphics[width=2.5cm]{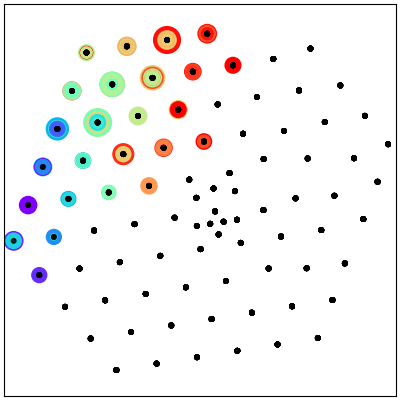}} \\
  {\includegraphics[width=2.5cm]{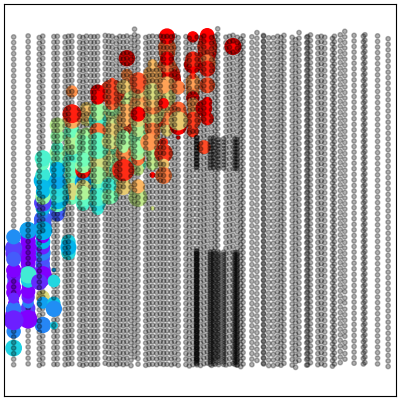}} \\
};

\node[label, above=3pt of grid-1-1] (col-track) {};

\node[label, above=-3pt of grid-1-1,draw, fill=white] (row-top) {top};
\node[label, above=-3pt of grid-2-1,draw, fill=white] (row-side) {side};

\node[
  draw,
  dashed,
  rounded corners,
  inner sep=10pt,
  fit=(grid) (col-track) (row-top) (row-side)
] (topbox) {};

\node[
    anchor=south,
    fill=white,      
    inner sep=1pt,
    font=\bfseries\small
  ] at ([yshift=-0.5ex]topbox.north) {IceCube event (track)};

\node[box, below=1.1cm of grid-2-1] (second) {Transformer};
\node[box, below=0.5cm of second, minimum width=3cm] (final) {Posterior (localized) \\ \includegraphics[width=4cm]{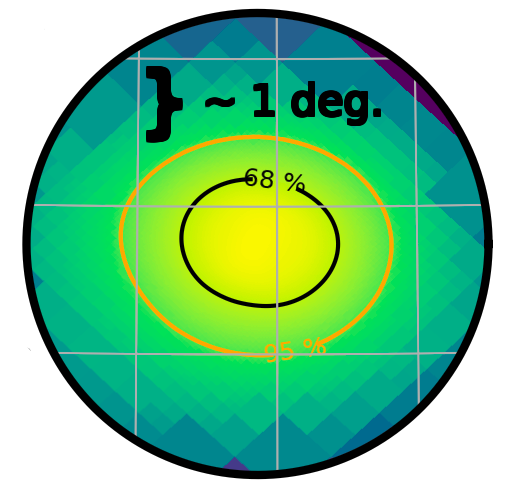}};

\draw[->] ([yshift=-0.45cm]grid-2-1.south) -- (second.north);
\draw[->] (second.south) -- (final.north);

\end{tikzpicture}
\caption{track example}

\end{subfigure}
\begin{subfigure}[b]{0.49\linewidth}
\centering
    \begin{tikzpicture}[
  img/.style={draw, minimum width=2.5cm, minimum height=1.6cm, align=center},
  box/.style={draw, rounded corners, minimum width=2cm, minimum height=0.5cm, align=center},
  >=latex
]
\matrix (grid) [
  matrix of nodes,
  nodes={img},
  column sep=0.9cm,
  row sep=0.5cm
] {
  {\includegraphics[width=2.5cm]{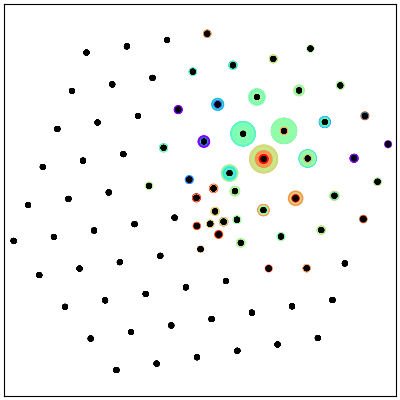}} \\
  {\includegraphics[width=2.5cm]{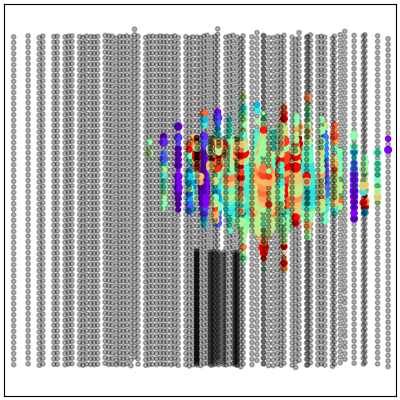}} \\
};


\node[label, above=-3pt of grid-1-1,draw, fill=white] (row-top) {top};
\node[label, above=-3pt of grid-2-1,draw, fill=white] (row-side) {side};

\node[
  draw,
  dashed,
  rounded corners,
  inner sep=10pt,
  fit=(grid) (col-track) (row-top) (row-side)
] (topbox) {};

\node[
    anchor=south,
    fill=white,      
    inner sep=1pt,
    font=\bfseries\small
  ] at ([yshift=-0.5ex]topbox.north) {IceCube event (shower)};
  
\node[box, below=1.1cm of grid-2-1] (second) {Transformer};
\node[box, below=0.5cm of second, minimum width=3cm] (final) {Posterior (all-sky) \\ \includegraphics[width=6.3cm]{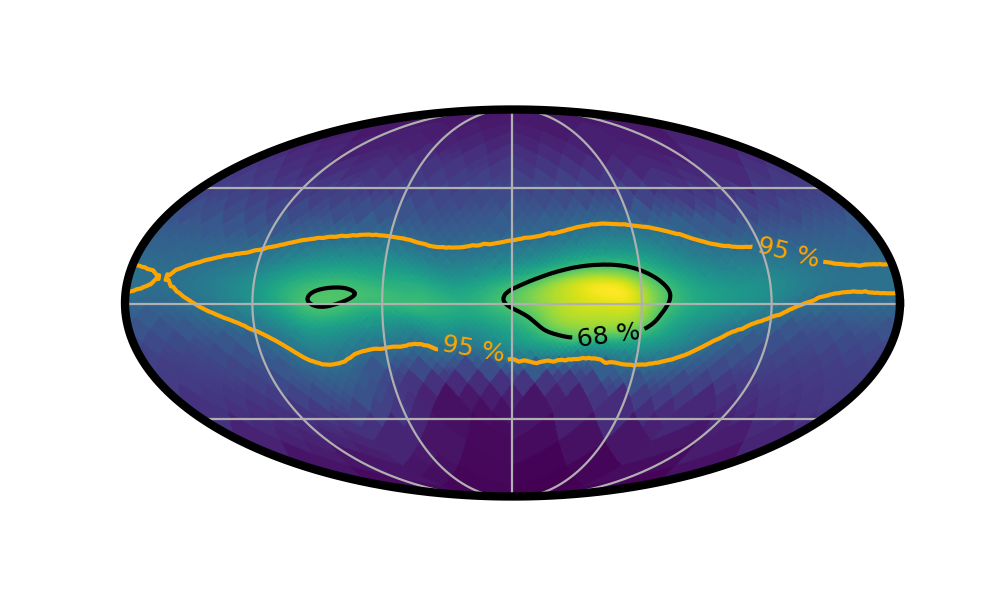}};

\draw[->] ([yshift=-0.45cm]grid-2-1.south) -- (second.north);
\draw[->] (second.south) -- (final.north);

\end{tikzpicture}
\caption{shower example}
\end{subfigure}

    \caption{Schematic overview of transformer-based amortized neural posterior estimation for neutrino event directions in IceCube. Two example posteriors are depicted: a posterior (orthographic projection) localized on a scale of a few degrees for a track (a) and a more extended posterior (Mollweide projection) for a shower (b). The photon data is collected in optical modules that are hexagonally distributed in a cubic-kilometer volume. The color of the modules indicates time, where red is early and blue is late. The transformer maps the event-dependent photon data to the normalizing-flow parameters and thereby encodes all posteriors for events similar to the training dataset at once.}
    \label{fig:overall_overview}
\end{figure}
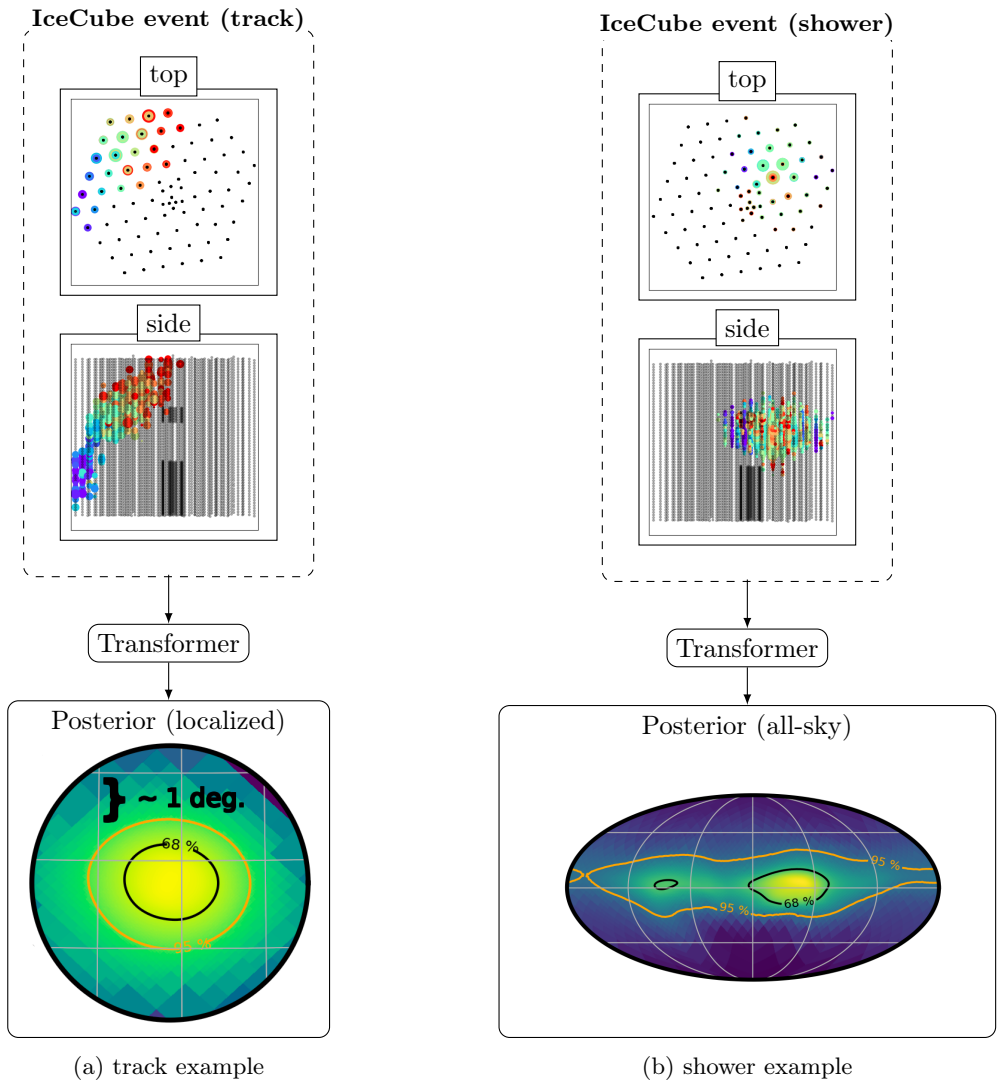

First, we change the GNN to a transformer \cite{transformer_paper} encoding. Secondly, we develop a new efficient manifold normalizing-flow on the 2-sphere which is numerically stable in this conditional setting and is able to handle the extra requirement of modeling several orders of magnitude in scale simultaneously. To achieve this, we extend the recursive construction in \cite{tori_and_spheres} which relies on rational-quadratic splines \cite{neural_spline_flows}. We add a von-Mises-Fisher scaling function, remove the recursive conditioning scheme and make the rational-quadratic spline flow $C^2$-continuous. 
A schematic overview of the proposed algorithm is shown in Fig. \ref{fig:overall_overview}. Depicted are the detected photon counts within the IceCube neutrino detector for a track-like example and a shower-like example event. The detector has a quasi-hexagonal shape and each detection unit or digital optical module (DOM) is indicated by a black dot. DOMs that are detecting photons are color coded with early-hit modules shown in red and late-hit modules shown in blue. The photon data is encoded via a transformer and mapped to the normalizing-flow parameters. More details of the detector and encoding pipeline are given in section \ref{architecture_section}.

The paper is structured as follows. In section \ref{vi_section} we introduce amortized neural posterior estimation. In section \ref{nfs_section} we introduce normalizing flows on the 2-sphere, describe our new manifold normalizing flow and showcase the advantage of a manifold normalizing flow for astronomical skymap creation. In section \ref{architecture_section} we describe the overall architecture, including the data preparation from raw IceCube data. In section \ref{section:transformer}, in particular, we motivate the switch from a GNN to a transformer encoding, introducing the concept of \enquote{Bayesian inductive bias}. In section \ref{training_section} we describe the training procedure and introduce the concept of \enquote{probabilistic regularization}. Finally, in section \ref{results_section} we show results and end with a discussion in section \ref{discussion_section}.

\section{Amortized neural posterior estimation - a form of \enquote{ELBO-free} stochastic variational inference}
\label{vi_section}

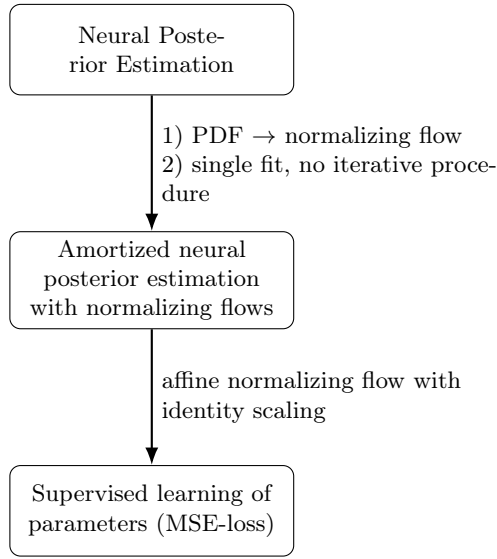
\begin{figure}[ht]
    \centering
    \begin{tikzpicture}[
        node distance=1.8cm,
        align=center,
        box/.style={draw, rounded corners, text width=3.5cm, minimum height=1.2cm, font=\small},
        anno/.style={align=left, text width=4.5cm, font=\small},
        arrow/.style={-{Latex}, thick}
    ]
    
    \node[box] (npe) {Neural Posterior Estimation};
    \node[box] (svicnf) [below=of npe] {Amortized neural posterior estimation with normalizing flows};
    \node[box] (mse) [below=of svicnf] {Supervised learning of parameters (MSE-loss)};
     
    \draw[arrow] (npe) -- node[anno, right] {1) PDF $\rightarrow$ normalizing flow \\2) single fit, no iterative procedure} (svicnf);
    \draw[arrow] (svicnf) -- node[anno, right] {affine normalizing flow with identity scaling} (mse);
    
    \end{tikzpicture}
    \caption{Relation of amortized neural posterior estimation with normalizing flows (used here) to generic neural posterior estimation and standard supervised regression of parameters.}
    \label{fig:vi_overview}
\end{figure}

We can write Bayes' theorem connecting observations $x$, for example data in the IceCube detector, and parameters $\theta$ as 
\begin{equation}
    p(\theta|x)=\frac{p(x|\theta) \cdot p(\theta) }{p(x)},
\end{equation}
where we call $p(\theta|x)$ the posterior and $p(x|\theta)$ the \enquote{data generating distribution} when seen as a probability distribution function (PDF)\footnote{We assume we have continuous data for simplicity and thus call it a \enquote{PDF}. The argument is of course also valid for discrete data which would technically be a probability mass function} over $x$ for fixed $\theta$ or \enquote{likelihood function} when seen as a function $\mathcal{L}(\theta)$ of $\theta$. For IceCube, we have independent photon observations per DOM and per photon, so we can write
\begin{equation}
    \mathcal{L}(\theta)=\prod_{j=1}^{N_{\mathrm{DOM}}} \prod_{i=1}^{N_{p,j}} p_j(x_i|\theta), \label{eq:likelihood}
\end{equation} with number of DOMs $N_{\mathrm{DOM}}$ and number of photons for the $j$th DOM $N_{p,j}$. The likelihood is a product of independent and identically distributed data (IID) factors per DOM.
A standard approach to obtain the posterior from a measurement outcome is via Markov-Chain Monte Carlo (MCMC) sampling \cite{mcmc_evolution} of an objective function that is proportional to the likelihood function. However, this does not scale to high dimensions very well.
In the 1990s, variational inference \cite{vi_1} emerged as an alternative to Markov-Chain Monte Carlo for Bayesian analysis, where the posterior solution is found via optimization instead of sampling. This \enquote{traditional} variational inference approach requires the Evidence Lower Bound (ELBO) to be formulated as a loss function. While this in principle allows the handling of higher dimensional scenarios, it is still inefficient in the sense that it requires a full ELBO optimization for a single measurement and corresponding posterior inference.
In recent years, neural posterior estimation \cite{npe_definition} has emerged as a way to directly approximate the posterior distribution of interest without an ELBO objective - hence one can also call it "ELBO-free" variational inference. It involves an iterative scheme of sampling from a \enquote{proposal prior} $\tilde{p}(\theta)$, running a simulator $\tilde{p}(x|\theta)$ to produce data \enquote{$x$} using the samples from the proposal prior as input parameters, and fitting an approximate posterior $q_\psi(\theta|x)$ parametrized by a neural network via $\psi$. Here and in the following, $x$ refers to all data in an event, not individual photons as in eq. \ref{eq:likelihood}. The procedure is repeated until the approximate posterior converges to the true posterior for a specific event datum $x_o$. It is part of the broader scheme of simulation based inference \cite{sbi_review}.
One can also decide to just run a single training run and fit $q_\psi(\theta|x)$ on samples drawn from the initial proposal prior, and thereby approximate all posteriors of the dataset at once. It assumes the proposal prior is sufficient or its influence becomes negligible once data is observed. This is a form of \enquote{amortized} neural posterior estimation, as all possible posterior predictions are amortized by a neural network with no extra refinement. We adapt this procedure here, since simulation data for IceCube is costly to produce and adaptive re-simulations typically not feasible. The loss function $\mathcal{L}_{\mathrm{NPE}}(\psi)$ in this scheme can be derived from the forward Kullback-Leibler (KL) divergence \cite{kl_divergence_paper} between the joint distribution in the Monte Carlo simulation $\tilde{p}(x,\theta) = \tilde{p}(\theta) \cdot \tilde{p}(x|\theta)$, from which we have samples in the simulation, and an approximation of the joint distribution that involves the posterior $q_\psi(\theta|x)$, over the neural network parameters $\psi$. The KL divergence is a well-known quantity from information theory and is zero if the involved probability distributions are identical and positive otherwise. The implicit minimization of this quantity is performed via a Monte Carlo approximation of the integral and subsequent dropping of constant terms in the optimization parameters $\psi$ as
\begin{align}
\psi^\star 
&= \arg\min_{\psi}\;
D_{\mathrm{KL}}\!\left(
\tilde{p}(x,\theta)
\;\Big\|\;
q(x)\,q_{\psi}(\theta | x)
\right) \label{eq:joint_kl} \\
&= \arg\min_{\psi}\;
\mathbb{E}_{\tilde{p}(x ,\theta_{})}
\!\left[
\log\frac{\tilde{p}(x,\theta)}{q(x)q_{\psi}(\theta| x)}
\right] \\
&= \arg\min_{\psi}\;
\mathbb{E}_{\tilde{p}(x)} \left[ D_{\mathrm{KL}}\!\left(
\tilde{p}(\theta|x)
\;\Big\|\;
q_{\psi}(\theta | x)
\right) \right] + \mathrm{const} \label{eq:posterior_kl} \\
&= \arg\min_{\psi}\;
\mathbb{E}_{\tilde{p}(x,\theta_{})}
\!\left[
-\log q_{\psi}(\theta| x)
\right]
\qquad(\text{drop terms constant in }\psi) \label{eq:cross_entropy} \\
&\approx \arg\min_{\psi}\; \frac{1}{N}\sum_i^{N} -\log q_{\psi}(\theta_{i}|x_i) \equiv \arg\min_{\psi}\;
 \mathcal{L}_{\mathrm{NPE}}(\psi). \label{eq:loss_function}
\end{align}
It can be seen that the loss function is a Monte Carlo approximation (eq. \ref{eq:loss_function}) of the conditional cross entropy (eq. \ref{eq:cross_entropy}) and shares the same minimum with the expected KL-divergence between the implicit simulation posterior $\tilde{p}(\theta;x)$ and the posterior approximation $q_{\psi}(\theta | x)$ (eq. \ref{eq:posterior_kl}) because constant terms can be dropped. Since $N$ is typically in the millions, in practice it is optimized stochastically over batches as in standard neural network training. The target parameters $\theta$ of the posterior can be chosen to be any subset of simulated parameters of interest. In our case we focus on the zenith and azimuth angles, while ignoring other parameters, such as the event position or energy, whose contributions appear as constant terms in eq. \ref{eq:posterior_kl}. Implicitly, these other parameters act as nuisance parameters and are marginalized in the result obtained by optimizing eq. \ref{eq:loss_function}.
The choice for the posterior PDF is not predetermined and can be a mixture model as in the original NPE paper \cite{npe_definition}, for example. However, if a normalizing flow is chosen as the posterior approximation, it can be shown \cite{nfs_systematics} that it not only allows efficient amortized posterior estimation to be performed, it also strictly generalizes supervised neural network training with a Mean-Squared-Error (MSE) loss. The latter can be viewed as a special case of NPE with a restricted affine normalizing flow. We therefore get a direct link between scalable Bayesian analysis and traditional MSE-based supervised regression for parameter prediction. An overview of these relationships is shown in Fig. \ref{fig:vi_overview}.

\section{Conditional normalizing flows on the 2-sphere}
\label{nfs_section}
In order to perform neural posterior estimation for the direction of a neutrino, we need a conditional probability distribution on the 2-sphere. We base this distribution on a manifold normalizing flow.

\subsection{Technical definitions}
A normalizing flow describes a PDF $p_u(\theta)$ using a diffeomorphic mapping $\theta=f_u(z_b)$ with parameters $u$ and so-called \enquote{base distribution} $p_0(z_b)$ with a change-of-variable formula
\begin{equation}
    p_u(\theta)=p_0(f_u^{-1}(\theta))\cdot |\mathrm{det}J_u^{-1}(\theta)|, \label{eq:change_of_variable}
\end{equation}
where $J_u$ is the Jacobian of the mapping $J_u=\frac{df_u(z_b)}{dz_b}$. Here, we call \enquote{$z_b$} the auxiliary base space and \enquote{$\theta$} the target space since we are interested in describing a posterior over parameters $\theta$. Eq. \ref{eq:change_of_variable} then allows access to a valid probability distribution over $\theta$ in closed form.

\subsection{Manifold distributions}
For manifold distributions it makes sense in this context to differentiate between intrinsic coordinates $\theta_{int}$ and embedding coordinates $\theta_{emb}$. For the 2-sphere, the intrinsic coordinates are the zenith and azimuth angles, i.e. $\theta_{int}=\{\theta_{zen}, \theta_{azi}\}$, while the embedding coordinates are the corresponding Euclidean embedding coordinates $\theta_{emb}=\{x,y,z\}$. Eq. \ref{eq:change_of_variable} can then be extended to define a distribution over a non-Euclidean manifold \cite{tori_and_spheres}, for example the $2$-sphere, by a modification of the Jacobian determinant via
\begin{eqnarray}
\label{eq:manifold_nf_rule}
|\mathrm{det}J_u^{-1}(\theta)| &\rightarrow \sqrt{ \mathrm{det}\left( (J_u^{-1} \cdot P)^T \cdot J_u^{-1} \cdot P \right)  }[\theta_{emb}] \\
& \equiv \mathrm{det}\mathcal{J}^{-1}_u[\theta_{emb}].
\end{eqnarray}
Here, $P(\theta_{emb})$ is a projection matrix that consists of the orthogonal vectors in the tangent plane at $\theta_{emb}$, which in the case of the sphere are two orthogonal vectors. The term $J_u^{-1}(\theta_{emb})$ is the inverse Jacobian defined and interpreted in embedding coordinates. One can think of this modified structure as first projecting the Jacobian into the tangent space before taking the determinant. The final determinant then measures the local area change in the tangent space on the manifold. 

\subsection{Conditional PDFs}
We now use a neural network mapping $u=g_\psi(X)$ to predict the flow function parameters $u$ from data $X$ and thereby describe a conditional PDF on the 2-sphere as
\begin{equation}
\label{eq:conditional_nf}
p_\psi(\theta_{emb}|X) =
p_0(f_{g_\psi(X)}^{-1}(\theta_{emb}))\cdot \mathrm{det}\mathcal{J}^{-1}_{g_\psi(X)}[\theta_{emb}] .
\end{equation}
This conditioning scheme requires an efficient normalizing flow that should not have more than $\sim 10^{2}\text{--}10^{3}$ parameters since all of them are predicted by $g_\psi(X)$. The function $g_\psi(X)$ could be any neural network encoding architecture that is appropriate to handle the input data. As described in section \ref{section:transformer} we use a transformer encoding for $g_\psi(X)$, which among other things allows the handling of variable-sized input data and encodes the permutation symmetry of arguments to the data generation distribution, as discussed in section \ref{section:transformer}. For convenience, we will in some cases denote $\theta_{emb}$ simply as $\theta$.

\subsection{A new normalizing flow}
\label{section:used_nf_definition}
The flow function $f_u(z_b)$ we use is motivated by the conditional flow scheme described in \cite{tori_and_spheres} which describes a conditional flow on the 2-sphere by transforming the sphere into a cylinder.
This construction has the advantage that the Jacobian determinant of the mapping from spherical coordinates to cylinder coordinates has determinant 1. It is then enough to just calculate contributions to the determinant from transformations on the cylinder. Let us call this transformation from spherical embedding coordinates to cylindrical embedding coordinates the \enquote{cylinder transformation} $\mathrm{C}(\theta_{emb})$. 
We can write it as
\begin{equation}
\mathrm{C}(\theta_{emb})=\mathrm{C}(x,y,z)=
\left(
\begin{array}{c}
\rho \\
\phi\\
z
\end{array}
\right) = 
\left(\begin{array}{c}
1 \\
\mathrm{atan}(y/x)\\
z
\end{array}
\right).
\end{equation}
The Jacobian determinant of this transformation is 1, even without invoking eq. \ref{eq:manifold_nf_rule}.
Next we apply smooth rational-quadratic splines on the $z$ and $\phi$ coordinates, respectively. Rational-quadratic splines (RQS) \cite{neural_spline_flows} are diffeomorphic functions defined on any interval of choice, and with appropriate boundary conditions can be turned into a diffeomorphism on the circle \cite{tori_and_spheres}. Therefore, they serve as flexible normalizing-flow generators on the cylinder height interval $[-1,1]$ and the cylinder angle $[0,2\pi]$. A problem that arises with those functions is that they are not $C^2$ smooth, which usually leads to unphysical features in the resulting normalizing-flow PDF (see Fig. \ref{fig:spline_visualizations}). We use constrained smooth splines that we derive in appendix \ref{appendix:smooth_spline_derivation} to get rid of such features. In this way we obtain a smooth RQS on the interval $[-1,1]$, $\mathrm{rqs}_I(z)$, which we use for $z$. We also obtain a periodic smooth RQS, $\mathrm{rqs}_A(\phi|z)$, on the angle $[0,2 \pi]$ which we use for $\phi$. The angular RQS is conditioned on $z$. We condition it differently than proposed in \cite{tori_and_spheres} by using a fixed polynomial spline interpolator instead of a neural network (see Fig. \ref{fig:simple_spline_nonmlp_conditioning}). By an appropriate blending toward an identity mapping, this construction avoids singular features near the polar regions, an outcome that smoothing alone cannot accomplish. Further details are given in appendix \ref{appendix:smooth_spline_derivation}. 
Next we apply a scaling transformation that corresponds to the normalizing flow depiction of a von-Mises-Fisher (vMF) distribution. The vMF distribution \cite{vmf_distribution} is a common symmetric distribution on the 2-sphere and can be written as a normalizing flow by starting with a flat base distribution, transforming to the cylinder, and applying the scaling function \cite{stable_vmf_sampling}\footnote{We use $\frac{1+z}{2}$ which is equivalent to a uniform random variable as used in \cite{stable_vmf_sampling} since the flat distribution on the sphere corresponds to a uniform distribution on the height $z$ of the cylinder.}
\begin{equation}
\label{eq:fisher_scaling}
\mathrm{F}(z)=1 + (1/\kappa) \cdot \mathrm{ln}\left(\frac{1 + z}{2} + \left(1 - \frac{1+z}{2} \right) \cdot e^{-2 \kappa} \right)
\end{equation}
on the height $z \in [-1,1]$, followed by an inverse cylinder transformation and a rotation. The scaling transformation $F(z)$, together with the cylindrical mappings, allows one to effectively \enquote{zoom in} on a particular region. For neutrinos, the expected directional posterior regions can span several orders of magnitude down to a fraction of a degree, which is challenging to model without such scaling functions. 
As final steps, we transform back from the cylinder to the sphere with $C^{-1}$ and apply a rotation $\mathrm{R}$ that we parametrize via householder reflections \cite{householder_trafos} (see eq. \ref{eq:rot_matrix_hh} in appendix \ref{appendix:rotation_parametrizations}). 
 The combination of all those flows written in a single block $i$ yields
\begin{equation}
    \label{eq:block_definition}
    f_{\phi_i}(\theta_{emb})=\left[\mathrm{R}_i \ \circ \ \mathrm{C}^{-1} \ \circ \ \underbrace{\mathrm{F}_i \ \circ \ \mathrm{rqs}_{\phi,i} \ \circ \ \mathrm{rqs}_{I,i}}_{\mathrm{on} \ \mathrm{cyl.} \ \mathrm{coords.}} \ \circ \ \mathrm{C}\right](\theta_{emb}) \ ,
\end{equation} where $\phi_i$ denotes all parameters of the involved functions bundled together. It is to be read from right to left in order and can be interpreted as a flexible functional building block that maps coordinates on the sphere $\theta_{emb}=(x,y,z)$ to new coordinates $\theta_{emb}'$ in an invertible manner. The Jacobian determinant of $f_i(\theta_{emb})$ is available in closed form and can be efficiently calculated since the determinant of the Jacobian of $\mathrm{C}$, $\mathrm{C}^{-1}$ and the rotation $\mathrm{R}_i$ is 1, while the Jacobian-determinants of the rational quadratic splines and the scaling function $F(z)$ are absolute values of simple one-dimensional derivatives.
For the final normalizing flow, we use 15 blocks of  $f_{\phi_i}(\theta_{emb})$ (eq. \ref{eq:block_definition}) chained together, i.e.
\begin{equation}
    f_{\mathrm{tot}}(\theta_{emb})=\left[f_{\phi_{15}} \circ \ldots \circ f_{\phi_1}\right](\theta_{emb}) \ ,
\end{equation}
which results in an expressive flow that can be multimodal, can express multiple orders of magnitude in scale due to the vMF scalings, and is efficient, i.e. it is described by a few 100 parameters packaged in $(\Phi_1, \ldots , \Phi_{15})$. We can therefore easily transform it into a conditional PDF by predicting those parameters with a neural network using eq. \ref{eq:conditional_nf}, which amortizes those flow parameters with the neural-network parameters $\psi$ and yields the amortized flow function $f_{\mathrm{tot},\psi}(\theta_{emb})$ (see Fig. \ref{fig:transformer_encoding} and Fig. \ref{fig:final_pdf_prediction} for an illustration). As depicted there, we also add a projection function $f_{\mathrm{proj.}}(z_b)$ in the very beginning in practice, which performs a non-learnable fixed stereographic projection from the 2-d plane to the sphere which has been motivated in \cite{nfs_systematics} and effectively allows starting with a Gaussian base distribution. An alternative is to start with the flat distribution on the sphere as a base distribution and drop this projection. The flow has been implemented in the open-source github package \texttt{jammy\_flows} \cite{jammy_flows}. 

\subsection{Sampling and skymaps}

\begin{figure*}
    \centering
    \begin{subfigure}[b]{0.99\textwidth}
    \resizebox{\linewidth}{!}{
     \begin{tikzpicture}
     
    \node[] (image1)  {\includegraphics[width=6cm]{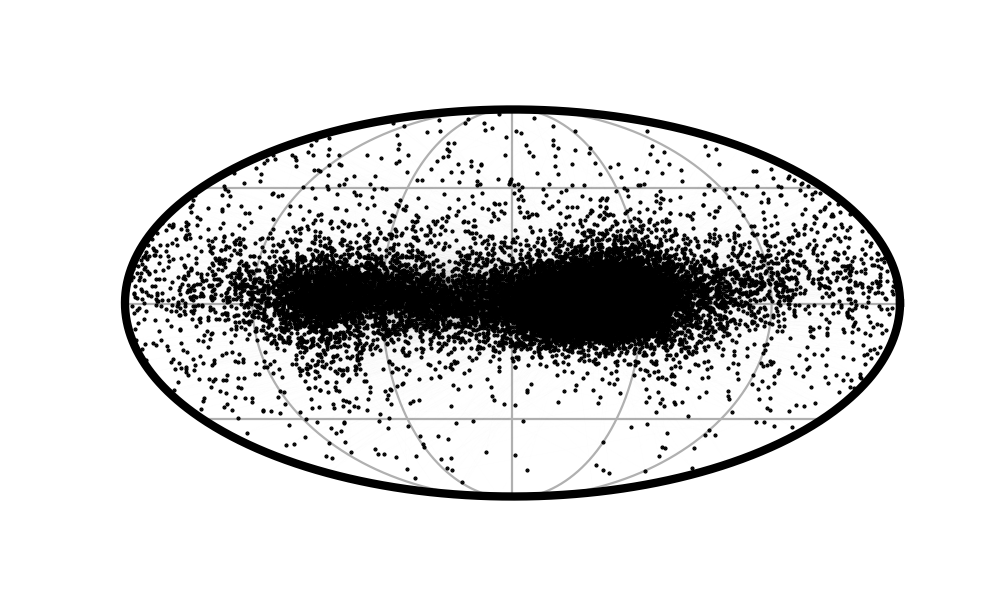} };
    
    \node[right=-0.5cm of image1] (image2)  {\includegraphics[width=6cm]{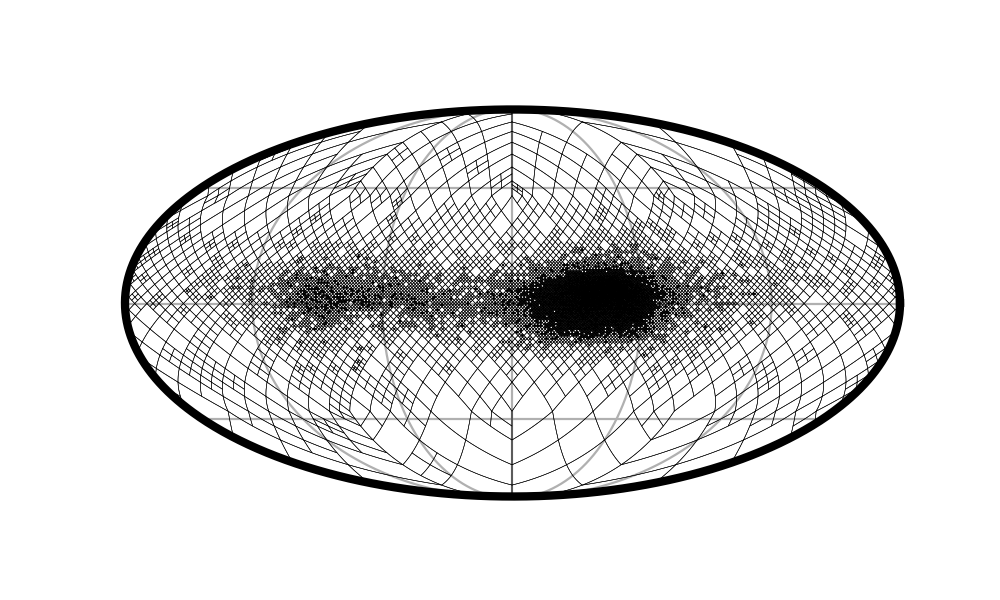} };

    \draw[<-] (image1.east) -- (image2.west);
    
    \node[right=-0.5cm of image2] (image3)  {\includegraphics[width=6cm]{figs/skymap_example_full.png} };

    \draw[<-] (image2.east) -- (image3.west);
    
    \end{tikzpicture} 
    }
    \caption{Full-sky posterior}
    \end{subfigure}

    \centering
    
    \begin{subfigure}[b]{0.99\textwidth}
    \resizebox{\linewidth}{!}{
     \begin{tikzpicture}
     
    \node[] (image1)  {\includegraphics[width=6cm]{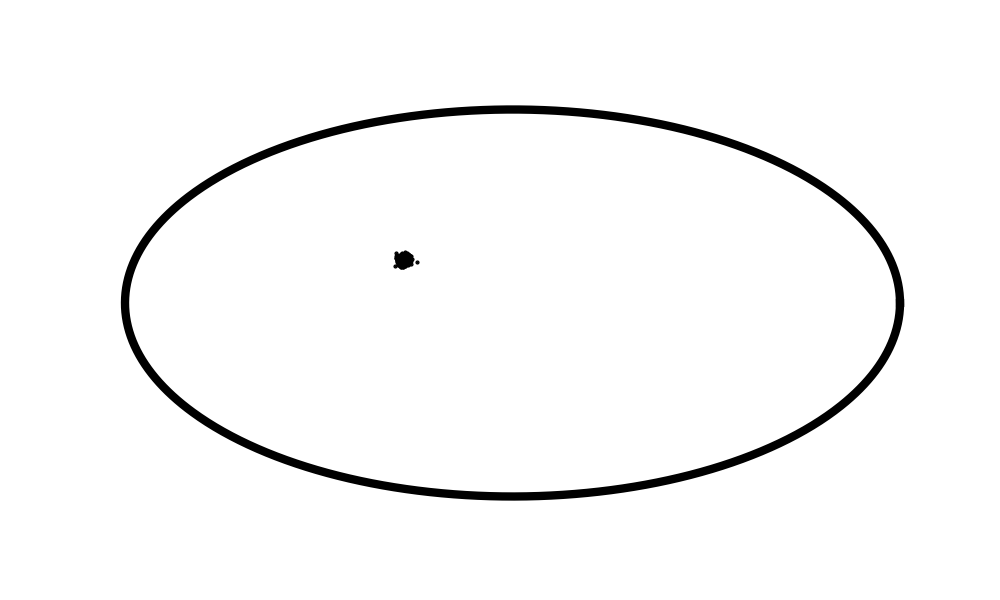} };
    
    \node[right=-0.5cm of image1] (image2)  {\includegraphics[width=6cm]{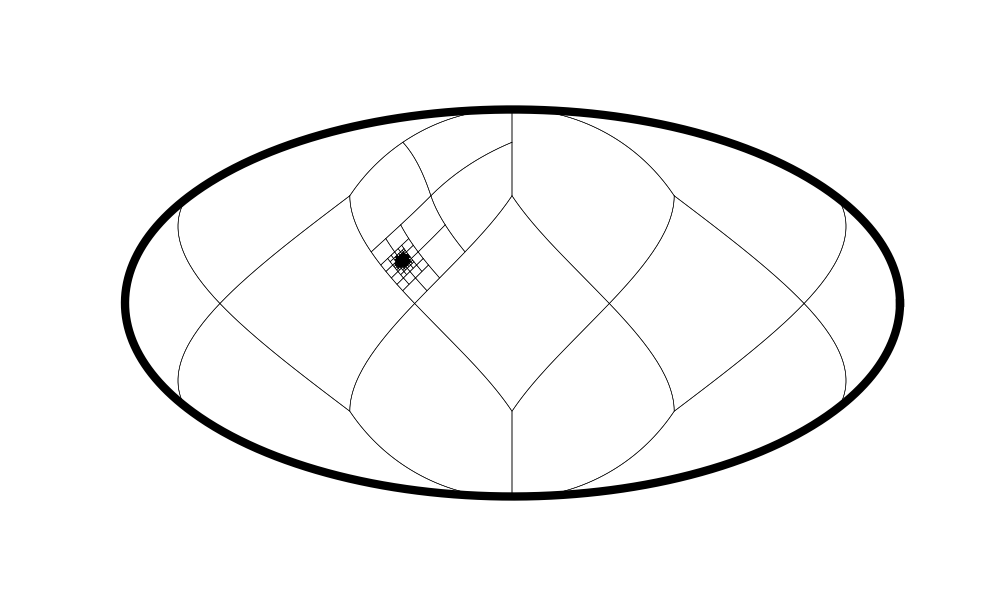} };

    \draw[<-] (image1.east) -- (image2.west);
    
    \node[right=-0.5cm of image2] (image3)  {\includegraphics[width=6cm]{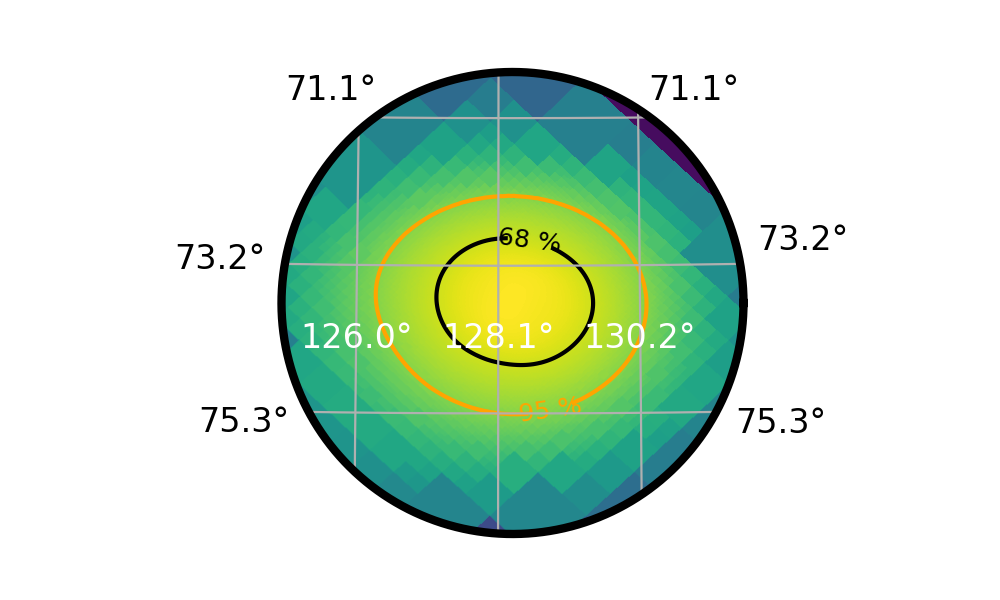} };

    \draw[<-] (image2.east) -- (image3.west);
    
    \end{tikzpicture} 
    }
    \caption{Localized posterior}
    \end{subfigure}
    \caption{Illustration of the constant-time skymap creation using normalizing flows irrespective of shape or size of the posterior. Drawing $N$ samples from the posterior (left) defines adaptive multi-order coverage grid cells (center) via subdivision based on maximally allowing $N_{\mathrm{max}}$ samples per cell. We use $N=10000$ and $N_{\mathrm{max}}=5$. At the cell locations the exact probability can be evaluated which produces smooth PDF maps and contours (right).}
    \label{fig:skymap_illustration}
\end{figure*}

Besides flexible PDF modelling, normalizing flows on the sphere offer a new capability of very efficient astronomical skymap creation. The bijective normalizing flow mapping $f_{tot,\psi}$ transforms samples from the base space to the target space and thereby draws samples from the normalizing-flow distribution on the sphere. These samples can be used to define an adaptive grid on the 2-sphere via an adaptive \texttt{HEALPIX} binning. \texttt{HEALPIX} \cite{healpix_ref} is an equal-binning scheme on the 2-sphere that is widely used in astronomy. It can be adapted for an irregular binning scheme since pixels can be divided into equal-sized smaller pixels on demand, for example via the multi-order coverage \cite{moc_def} (MOC) format as implemented in the package \texttt{mhealpy} \cite{mhealpy} which we utilize in the following. In the normalizing flow context, the irregular MOC binning can be created on the fly from the samples, which will automatically yield finer binning in regions where more detail is necessary. We impose the constraint that a maximum number of samples $N_{\mathrm{max}}$ is allowed in a given pixel, otherwise it is subdivided into 4 sub-pixels. On this irregular grid one can then employ the change-of-variable formula (eq. \ref{eq:change_of_variable}) to obtain the exact PDF value and draw smooth contour lines. The step-by-step procedure from samples to irregular grid and PDF evaluation is indicated in Fig. \ref{fig:skymap_illustration}.
In practice we find that $\sim 10000$ samples with $N_{\mathrm{max}}$ between $5$ and $10$  work well to obtain a smooth PDF in constant time for the normalizing flow described in section \ref{section:used_nf_definition}, independent of the absolute size of the contour. While the binning is subject to variations due to the stochastic sampling process, numerical differences of the drawn contours are not noticable by eye with these settings. The scheme works without iterative brute-force scanning and is in particular useful for contour regions that are orders of magnitude smaller compared to the full sky as well as for irregular PDFs with disconnected regions. 

\section{Combined architecture for IceCube data}
\label{architecture_section}

Neutrino interactions lead to relativistic charged secondary particles like electrons and muons which emit Cherenkov photons \cite{cherenkov_radiation} in transparent matter like the Antarctic ice. IceCube is a cubic-kilometer neutrino detector located at the South Pole. It consists of over 5000 digital optical modules (DOMs) embedded in the ice that detect such incoming Cherenkov photons with photomultiplier tubes (PMTs)  \cite{icecube_instrumentation}. A depiction of the hexagonal layout of these DOMs with two example events is shown in Fig. \ref{fig:overall_overview}.

\subsection{Data pre-processing}

\begin{figure*}

    \begin{subfigure}[b]{1.0\textwidth}
    \begin{tikzpicture}

    \node[anchor=west, scale=0.3] (domimage) at (0,0) 
    {
        \begin{tikzpicture}

        \draw[line width=2mm] (0,0) circle (3cm);
        
        \fill[black] (-3,0) arc[start angle=180,end angle=0,radius=3cm] -- (3,0) -- cycle;
        

        \draw[line width=2mm, draw=gray, ->, decorate, decoration={snake, amplitude=2mm, segment length=10mm}] (-4,-1.5) -- (-1.5,-0.8); 
        \draw[line width=2mm, draw=gray, ->, decorate, decoration={snake, amplitude=2mm, segment length=10mm}] (-4,-2.3) -- (-1.2,-1.5);  
        \draw[line width=2.0mm, draw=gray, ->, decorate, decoration={snake, amplitude=2mm, segment length=10mm}] (-4,-3) -- (-1,-2.3);    
        
        \draw[line width=2mm] (0,3.5) -- (0,3);  
        \draw[line width=2mm] (0,-3.5) -- (0,-3); 
        
        \end{tikzpicture}
    };

    \node (textdom) [above=0.1cm of domimage] {Photons hit DOM};

    \node[right=of domimage, draw] (image)  {\includegraphics[width=10cm]{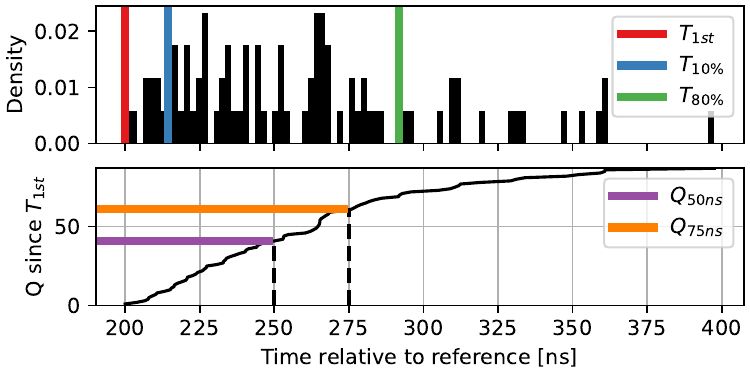} };

    \node (textdom) [above=0.1cm of image] {Reconstructed photon distribution};
    \draw[->] (domimage.east) -- (image.west);

    \node[anchor=west] (line1) at (0,-4) {Normal DOM $\ \ \ \rightarrow T_i=(\tilde{x}, \tilde{y}, \tilde{z}, \tilde{c_x},\tilde{c_y},\tilde{c_z}, \tilde{\bm{\vec{Q}}}, \tilde{T}_{1st},  \tilde{\bm{\vec{T_o}}}, \mathrm{PMT}_{\mathrm{type}}, 1, 0, 0)$};
    \node[anchor=west] (line2) at (0,-4.7) {Saturated DOM $\rightarrow T_i=(\tilde{x}, \tilde{y}, \tilde{z}, \tilde{c_x},\tilde{c_y},\tilde{c_z}, \bm{\vec{0}}, \tilde{T}_{1st},  \bm{\vec{0}}, \mathrm{PMT}_{\mathrm{type}}, 0, 1, 0)$};
    \node[anchor=west] (line3) at (0,-5.4) {Empty DOM $\ \ \ \ \rightarrow T_i=(\tilde{x}, \tilde{y}, \tilde{z}, \tilde{c_x},\tilde{c_y},\tilde{c_z},\bm{\vec{0}}, 0,  \bm{\vec{0}}, \mathrm{PMT}_{\mathrm{type}}, 0, 0, 1)$};

    \node[draw, thick, fit=(line1)(line2)(line3)] (boundingbox) {};
    \draw[->] (image.south) -- (boundingbox.north -| image);

\end{tikzpicture}
        \caption{The top figure shows a photon arrival distribution with example time (\enquote{T}) and charge (\enquote{Q}) summary statistics. The bottom panel shows the different ways that this photon summary statistic is combined with other DOM-specific information into a vector representation $T_i$. The DOM position is denoted by $x,y,z$, the difference of the DOM position to the center of gravity of the event by $c_x,c_y,c_z$, charge quantities of the observed photon hits by $\bm{\vec{Q}}$ and temporal quantities by $T_{1st}$ and  $\bm{\vec{T_o}}$. $\mathrm{PMT}_{\mathrm{type}}$ is either 0 (normal PMT) or 1 (high-QE PMT). The final 3-dimensional one-hot part differentiates between \enquote{normal}, \enquote{saturated} and \enquote{empty} DOMs. The \enquote{$\sim$} above a quantity represents variable-dependent rescaling.}
        \label{fig:time_arrival_distribution}
    \end{subfigure}
    \hfill
    \centering
    
    \begin{subfigure}[b]{0.49\textwidth}
        \centering
        \begin{tikzpicture}[node distance=0.5cm and 0.3 cm, >=Stealth]

    \node (input) [rectangle] {$\ldots$};
    \node (chi2) [draw, rectangle, left=of input] {$T_2$};
    \node (chi1) [draw, rectangle, left=of chi2] {$T_1$};
    \node (chin) [draw, rectangle, right=of input] {$T_{N}$};

    \node (transformer) [draw, rectangle, below= of input, xshift=-0.4cm, align=center] {Transformer Encoder};

    \draw[->] (chi1.south) -- (transformer.north -| chi1);
    \draw[->] (chi2.south) -- (transformer.north -| chi2);
    \draw[->] (chin.south) -- (transformer.north -| chin);

    \node (ldots_placeholder) [rectangle, below=of transformer, xshift=0.4cm, align=center] {$\ldots$};
    \node (t2) [draw, rectangle, left= of ldots_placeholder, align=center] {$\tilde{T}_2$};
    \node (t1) [draw, rectangle, left=of t2,  align=center] {$\tilde{T}_1$};
    \node (tn) [draw, rectangle, right=of ldots_placeholder, align=center] {$\tilde{T}_N$};

    \draw[->] (transformer.south -| chi1) -- (t1.north);
    \draw[->] (transformer.south -| chi2) -- (t2.north);
    \draw[->] (transformer.south -| chin) -- (tn.north);

    \node (aggregation) [draw, rectangle, below=of t2, xshift=0.6cm, align=center] {Aggregate};

    \draw[->] (t1.south) -- (aggregation.north);
    \draw[->] (t2.south) -- (aggregation.north);
    \draw[->] (tn.south) -- (aggregation.north);
    
    \node (mlp) [draw, rectangle, below=of aggregation, align=center] {$\mathrm{MLP}$};
    \draw[->] (aggregation.south) -- (mlp.north);
    
    \node (phi) [below=of mlp, draw, rectangle, align=center] {Flow params: \\ $(\vec{\phi_1}, \vec{\phi_2}, \ldots, \vec{\phi_M})[\psi]$};
    \draw[->] (mlp.south) -- (phi.north);
    
    \node (flowfn) [align=center, draw, below= of phi] {Flow function: \\  $f_{\mathrm{tot},\psi}(z) = [f_{\phi_M} \circ f_{\phi_{M-1}} \circ \ldots \circ f_{\phi_1} \circ f_{\mathrm{proj.}}](z)$};

    \draw[->] (phi.south) -- (flowfn.north);

    \node (parambox) [draw, dashed, rounded corners, thick, orange, fit=(chi1) (chin) (mlp), inner sep=0.3cm] {};
    \node at (parambox.south east) [anchor=south east, xshift=-0.1cm, yshift=0.1cm, orange] {\Large $\psi$};
\end{tikzpicture}
        \caption{Encoding structure from DOM vector representations $T_i$ into the flow function $f_{\mathrm{tot},\psi}(z)$. See section \ref{section:used_nf_definition} for more details on the flow definition.}
        \label{fig:transformer_encoding}
    \end{subfigure}
    \hfill
    \begin{subfigure}[b]{0.49\linewidth}
        \centering
        \resizebox{\linewidth}{!}{
\begin{tikzpicture}

    \node (invisible_node) [align=center] {};
    \node (flowfn_short) [left=3cm of invisible_node] {$f_{\mathrm{tot},\psi}(z)$};
     \node (flowfn_inv_short) [right=3cm of invisible_node] {$f_{\mathrm{tot}, \psi}^{-1}(\theta)$};

    \node (pic1) [above=0.0cm of invisible_node] {\includegraphics[width=2.5cm]{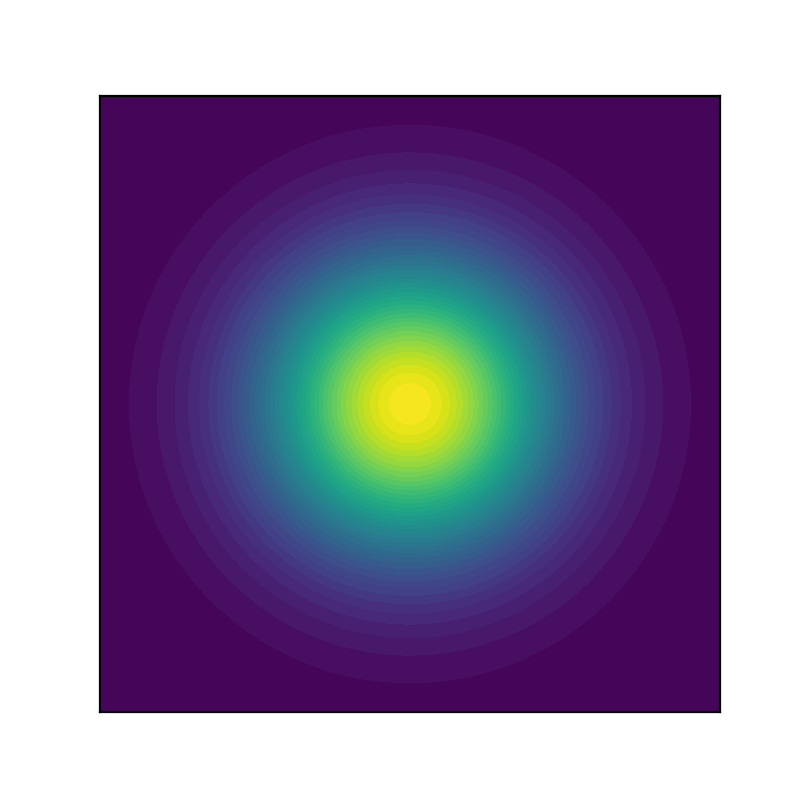}};
    \node (pic1_descr) [above=-0.2cm of pic1] {$p_0(z)$};
    \node (pic1_below) [below=-0.2cm of pic1]{auxiliary base space};
    
    \node (pic2) [below=1.5cm of invisible_node] {\includegraphics[width=5.5cm]{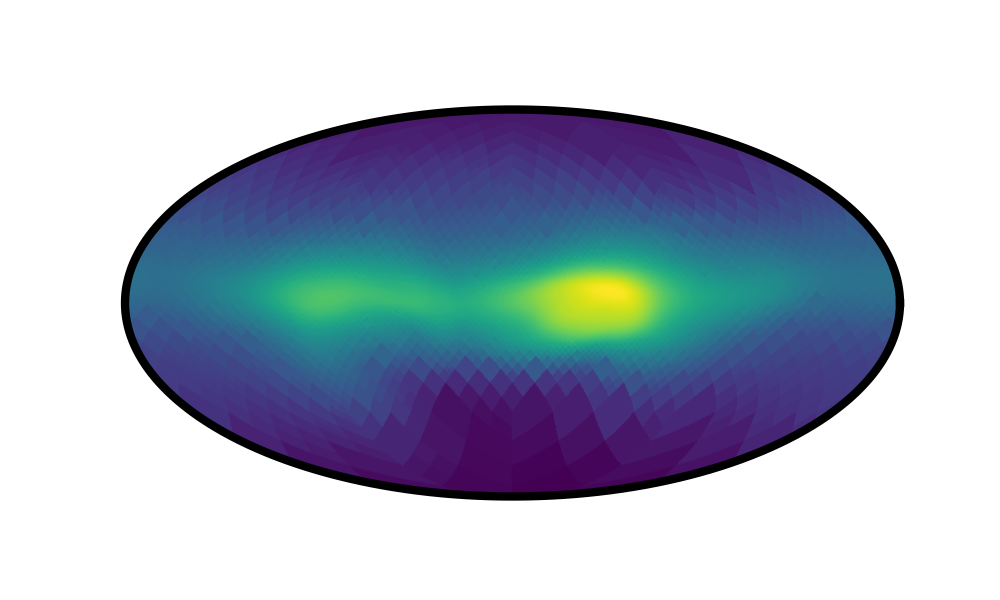}};
    \node (pic2_descr) [draw, above=-0.5cm of pic2, align=center]{Posterior: \\ $p_{\psi}(\theta|\{T_1, T_2, \ldots, T_N\}) $};
    \node (pic2_below) [below=-0.5cm of pic2]{target space \enquote{$\theta$}};
    
    \draw[<-, thick] 
        (pic1.east) .. controls +(1.5,0) and +(1.5,0) .. (pic2.east);

    \draw[<-, thick] 
        (pic2.west) .. controls +(-1.5,0) and +(-1.5,0) .. (pic1.west);

\end{tikzpicture}
}
        \caption{The encoded flow function $f_{\mathrm{tot},\psi}(z)$ is used to define the conditional posterior. It implicitly depends on the neural network parameters $\psi$ and on the DOM summary statistics $T_i$.}
        \label{fig:final_pdf_prediction}
    \end{subfigure}
    \caption{The data encoding pipeline from photon hits to posterior prediction.}
    \label{fig:sidebyside}
\end{figure*}
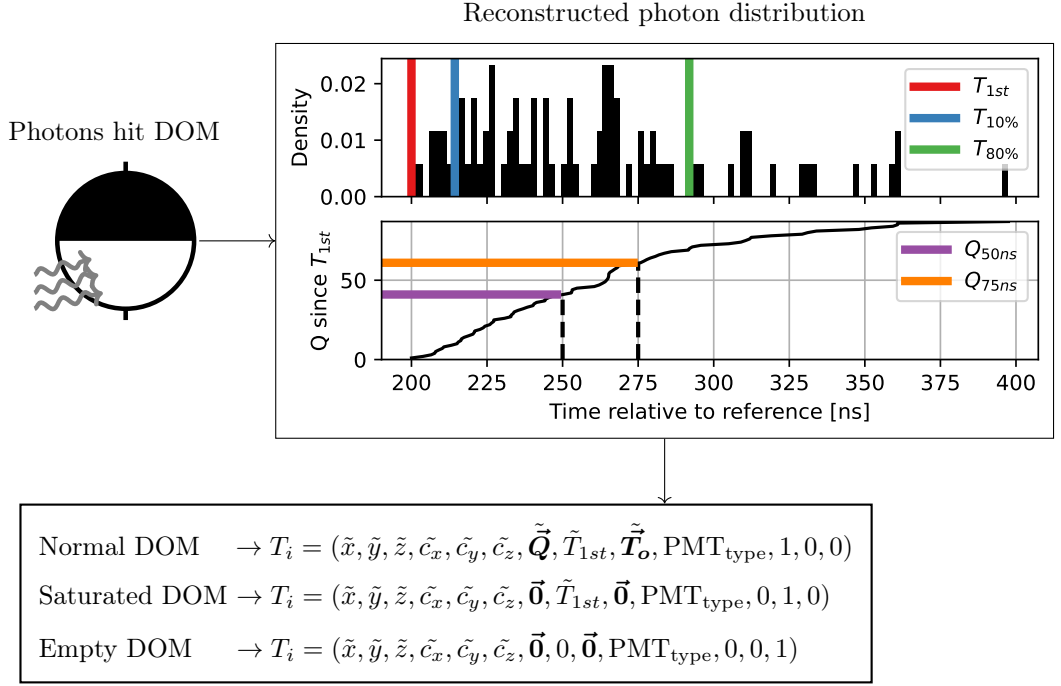
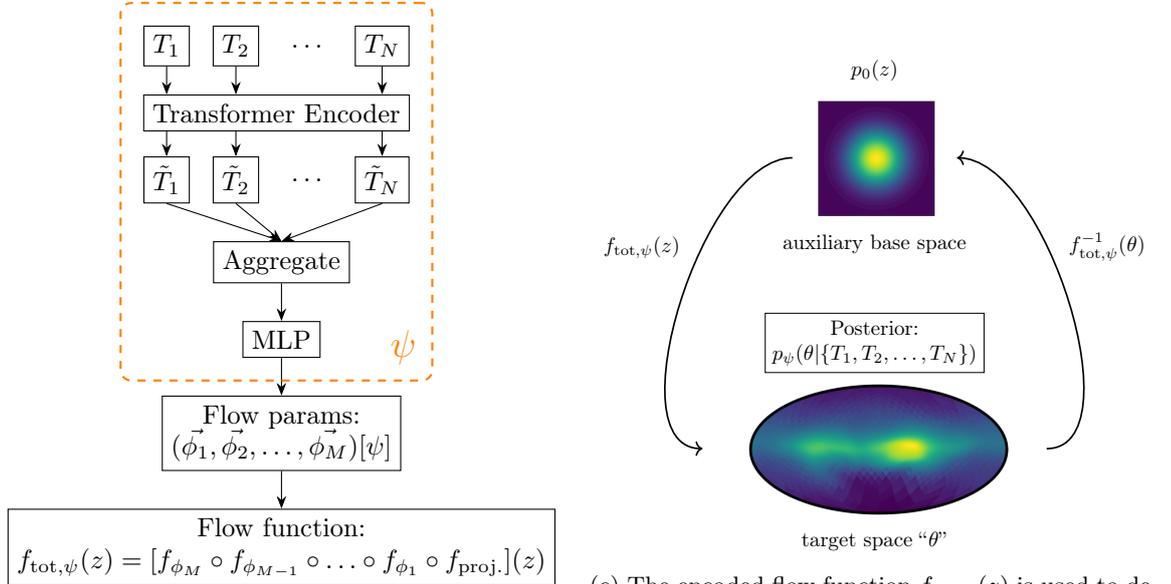

The photons that reach a DOM have a specific time arrival distribution as depicted in Fig. \ref{fig:time_arrival_distribution}. This distribution technically corresponds to reconstructed \enquote{effective photons} which come from a least-squares unfolding algorithm that involves the DOM detector responses \cite{least_squares_unfolding}. In the following discussion we will not differentiate between reconstructed photons and true physical photons for simplicity. The shape of the distribution and the number of photons depend on event parameters like the flavor, energy or direction of the incoming neutrino. In a first step, noise is cleaned with an established cleaning algorithm. DOMs with detected photons that cannot causally be connected to a main \enquote{cluster} of DOMs are removed in this step. It is followed by a PMT afterpulse cleaning in which we only keep photons in a given DOM that arrive at most 5 microseconds after the first hit in the same DOM - this excludes late after pulses that typically arrive on a timescale of 6 microseconds or later after a physical photon. We found in various trainings consistently better results by applying such a combined cleaning first versus taking the raw effective photon data as input for the neural network. We then encode the surviving cleaned observed photon distribution in a fixed-length summary statistic following the ansatz originally described for a convolutional neural network (CNN) encoding \cite{icecube_cnn_paper}. On top of time quantiles and charge quantiles \cite{icecube_cnn_paper}, we also add the absolute DOM position and the DOM position relative to the charge-weighted \enquote{center of gravity} position of the event. We furthermore add an identifier whether a PMT has normal or high quantum efficiency and differentiate between normal, unhit and saturated PMT DOMs via an extra 3-d one-hot encoding. In unhit or \enquote{empty} DOMs we set all the charge and timing information to zero. These are to be differentiated from the few percent of DOMs that are malfunctioning and even in principle would not be able to detect any photons, which are altogether excluded here. In saturated DOMs we remove all the charge and timing information since saturation is not well modeled in our simulation. We only keep the time of the first photon hit, which is reliable for saturated DOMs. Time entries are shifted to be aligned relative to the median time of the event and normalized by $1/10000.0$. Charge entries are calculated as $\tilde{Q}=\mathrm{ln}(1+Q) \cdot 0.2$ which automatically maps zero to zero. Position entries and relative position entries are normalized by $1/500.0$, e.g. $\tilde{x}=\frac{1}{500} \cdot x$. Examples of summary statistics are given in Fig. \ref{fig:time_arrival_distribution}. In total, this per-DOM vector is 27-dimensional in all experiments and we denote it with $T_i$ for DOM $i$.

\subsection{Transformer encoding}
\label{section:transformer}
The transformer architecture \cite{transformer_paper} is a machine learning model that has gained traction over the past years. Originally used in neural language processing \cite{transformer_paper} for sequential language data in an encoder-decoder setting, its use case has expanded to all other data science due to its universality. Here, the transformer encoding is used as a compression algorithm for the IceCube data. The transformer encoding consists of a number of building blocks or layers $N_L$, where we use $N_L=20$ in all experiments. It operates on a set of vectors instead of a single vector, which in our case are the DOM embeddings $T_i$. By default, every layer uses the vanilla architecture with \enquote{pre}-layer normalization \cite{pre_norm_transformer}, followed by a self-attention block that exhanges information between tokens $i$ and $j$ and a final multi-layer perceptron (MLP) without dropout, that is applied per token $i$. For each self-attention block, and given the input vector $X_i$ for each DOM $i$ at a given layer L, we first define the linear query (Q), key (K) and value (V) embeddings for each token, sometimes also called \enquote{QKV embedding}:
\begin{equation}
q_i = X_i \cdot W_Q,\quad k_i = X_i \cdot W_K,\quad v_i = X_i \cdot W_V .
\end{equation}
The self-attention update between query $i$ and key $j$ is then given as
\begin{equation}
\alpha_{ij}=\mathrm{softmax}_j\!\left(\frac{q_i k_j^\top}{\sqrt{d_k}}\right)
\end{equation}
and the output is multiplied by the value embeddings as
\begin{equation}
X_i' = \left[\sum_{j=1}^n \alpha_{ij}\, v_j \right ] \cdot W_O 
\end{equation}
with a final out-projection matrix $W_O$. The result is further processed per token via an MLP. Further details like the addition of an explicit positional encoding can be found in \cite{transformer_paper}. We also test several extensions to the vanilla transformer which are described in section \ref{sec:hyperparams_detail}. Initially, the per-DOM summary statistics $T_i$ are fed into the transformer encoder as depicted in Fig. \ref{fig:transformer_encoding}. Here, all the internal transformer layers are jointly denoted as \enquote{Transformer Encoder}, and transform tokens into transformed versions $\tilde{T}_i$ whose content is now an abstract encoding. In a final step tokens are aggregated by permutation invariant functions, which by default consist of forming the mean and standard deviation for every output feature over all the tokens and then mapping the result via an MLP to predict the normalizing-flow parameters. We also study alternative aggregation schemes with learnable aggregation tokens as described in section \ref{sec:hyperparams_detail}. Finally, the normalizing flow parameters define the target posterior $p_{\psi}(\theta|T_1, \ldots, T_N)$, where all parameters of the transformer and the final MLP are denoted as $\psi$. The posterior depends on the observed data summary statistics of each of the  $N$ DOMs in the event, $T_1, \ldots, T_N$ (Fig. \ref{fig:final_pdf_prediction}), as well as the network parameters $\psi$ and therefore structurally resembles an explicit posterior distribution with the mapping from data to parameters facilitated by the transformer. There are two probabilistic reasons in favor of transformers for variational inference using IceCube data and we call these \enquote{Bayesian inductive biases} in the following.

\subsection{Bayesian inductive bias 1: Permutation invariance}
The first Bayesian inductive bias comes from the transformer permutation equivariance per token, which leads to invariance with an appropriate aggregation process. This is illustrated in Fig. \ref{fig:illustration_perm_invariance}. In general, Bayesian inference does not care in which order one looks at input parameters in the data distribution - they can be permuted, and the inference result would not change. This is even more explicit if the data is IID, which is the case for IceCube (see section \ref{vi_section}). An encoding mechanism that respects this invariance to ordering, like a transformer with appropriate aggregation, therefore has some level of inbuilt advantages over other encoding mechanisms that do not respect it, like encoding schemes via a recurrent neural network (RNN) \cite{rnn_review}, for example. In early inference tests with normalizing flows we have used RNN encodings based on gated recurrent units \cite{gru_paper}, both using random DOM orderings at training and inference time or fixed orderings by mean time per DOM, and those consistently yielded worse results than GNNs or transformers. A GNN encoding based on k-nearest neighbors and edge convolutions \cite{edgeconv_reference} which has been used in IceCube before \cite{gnn_low_energy} can be made permutation invariant in the same manner as a transformer, but seems to lack generalizability to handle variable-sized inputs to the same extent. This will be discussed in the next section. 

\subsection{Bayesian inductive bias 2: dropping data factors}

\begin{figure*}
    \centering

\begin{subfigure}[b]{1.0\textwidth}
\centering
\begin{tikzpicture}

\tikzset{
    layersquare/.style={
        draw=black, 
        line width=1.2pt, 
        rounded corners=10pt, 
        minimum width=2.5cm, minimum height=2.5cm, 
        fill=white 
    },
    topsquare/.style={
        draw=black, 
        line width=1.5pt, 
        rounded corners=10pt, 
        minimum width=2.5cm, minimum height=2.5cm, 
        fill=white, 
        text centered 
    },
    titlebox/.style={
        draw=black, 
        fill=white, 
        font=\bfseries, 
        align=center, 
        text width=3.5cm 
    },
    outerbox/.style={
        draw=black, 
        line width=1.2pt, 
        rounded corners=5pt, 
        inner sep=10pt 
    }
}

\foreach \i in {1,...,2} {
    \node[layersquare] (gnn\i) at  (\i*0.2, -\i*0.2) {};
}

\foreach \i in {1,...,2} {
    \node[layersquare] (trafo\i) at (\i*0.2, -\i*0.2-5) {};
}

\node[topsquare] (ts_gnn) at (0.6,-0.6) {\begin{tikzpicture}
    \tikzset{
        node style/.style={circle, draw, minimum size=0.15cm, inner sep=0, text centered, font=\bfseries},
        arrow style/.style={->, thick},
        square node style/.style={rectangle, minimum size=0.15cm,sharp corners},
    }

    \node[square node style,fill=black] (As) {};
    \node[square node style, fill=black, right=0.3cm of As] (Bs) {};
    \node[square node style, fill=black, draw,right=0.3cm of Bs] (Cs) {};
    \node[square node style, fill=black, draw, right=0.3cm of Cs] (Ds) {};
    
    \node[node style, fill=orange,below=0.3cm of As] (A) {};
    \node[node style, fill=blue, below=0.3cm of Bs] (B) {};
    \node[node style, fill=red, below=0.3cm of Cs,] (C) {};
    \node[node style, fill=black, text=white, below=0.3cm of Ds] (D) {};

    \draw[arrow style] (As) -- (Bs);
    \draw[arrow style] (Bs) -- (Cs);
    \draw[arrow style] (Cs) -- (Ds);

    \draw[arrow style] (A) -- (As);
    \draw[arrow style] (B) -- (Bs);
    \draw[arrow style] (C) -- (Cs);
    \draw[arrow style] (D) -- (Ds);
    
    
    \draw[arrow style] (Ds) -- (2.1, 0); 

\end{tikzpicture}}; 

\node[above=of ts_gnn,draw, fill=white] (descr_encoder) {Encoder};

\node[above=of descr_encoder] (showcase_bayes) {$
p(\theta|\bold{x}) = \frac{\textcolor{orange}{p(x_1;\theta)} \cdot  \textcolor{blue}{p(x_2;\theta)} \cdot \textcolor{red}{p(x_3;\theta)} \cdot \ldots  \cdot p(\theta) }{p(\bold{x})}
 =\frac{\textcolor{orange}{p(x_1;\theta)}  \cdot \textcolor{red}{p(x_3;\theta)} \cdot  \textcolor{blue}{p(x_2;\theta)} \cdot \ldots  \cdot p(\theta) }{p(\bold{x})}$
};

\coordinate (start) at ($(showcase_bayes.north west) + (3.3cm, 0)$); 
    \coordinate (end) at ($(showcase_bayes.north west) + (4cm, 0)$); 

    \draw[<->, bend left=30] (start) to node[above, pos=0.5, sloped] {permute}   (end) ;


\node[draw, rounded corners, thick, inner sep=10pt, fit=(showcase_bayes), minimum height=2cm,yshift=0.2cm] (rounded) {};

  \node[draw, rounded corners, fill=white, anchor=south, text centered, minimum width=3cm, above=of showcase_bayes, yshift=-0.3cm]  {Bayes' theorem with I.I.D. data};

\node[above=0.01cm of ts_gnn, yshift=-0.3cm,draw, fill=white] (title_gnn) {RNN};

\node[right=of gnn1] (post_example_gnn)  {\includegraphics[width=5.5cm]{figs/skymap_example.png}};

\node[left=of gnn1] (data_input_gnn)  {\includegraphics[width=3cm]{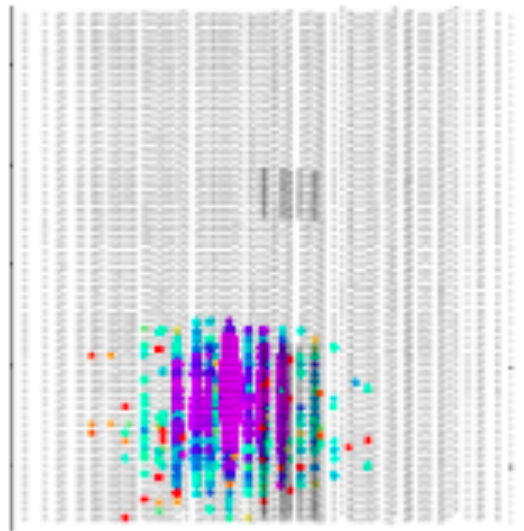}};

\draw[->] ($(data_input_gnn.east) + (-0.1, 0)$) -- ($(gnn1.west) + (-0.1, 0)$);
\draw[->] ($(gnn1.east) + (0.7, 0)$) -- ($(post_example_gnn.west) + (0.7, 0)$);

\node[above=of data_input_gnn,draw, yshift=-1cm, fill=white, align=center] (descr_data) {Data input:};

\node[above=of post_example_gnn,draw, fill=white, align=center, yshift=-1.5cm] (descr_posterior) {Posterior\\$p(\theta|\bold{x})$};

\node[below=of data_input_gnn, align=center, yshift=1cm] (descr_data_2) {\textcolor{orange}{DOM 1} $\rightarrow$ \textcolor{orange}{$x_1$}\\\textcolor{blue}{DOM 2} $\rightarrow$ \textcolor{blue}{$x_2$}\\$\ldots$};

\node[topsquare] (ts_trafo) at (0.6,-0.6-5) {\includegraphics[width=2cm]{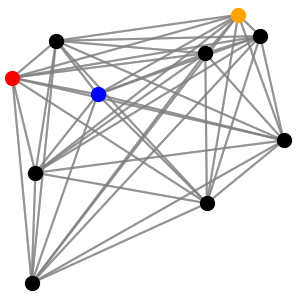}}; 

\node[above=0.01cm of ts_trafo, yshift=-0.3cm,draw, fill=white] (title_gnn) {Transformer};

\node[right=of trafo1] (post_example_trafo)  {\includegraphics[width=5.5cm]{figs/skymap_example.png}};

\node[left=of trafo1] (data_input_trafo)  {\includegraphics[width=3cm]{figs/det_view.png}};

\draw[->] ($(data_input_trafo.east) + (-0.1, 0)$) -- ($(trafo1.west) + (-0.1, 0)$);
\draw[->] ($(trafo1.east) + (0.7, 0)$) -- ($(post_example_trafo.west) + (0.7, 0)$);

\end{tikzpicture}
\caption{Illustration of permutation invariance of data arguments $x_i$ to the posterior. For independent data the data factors in Bayes' theorem can be swapped explicitly. Permutation invariance is respected in a transformer encoding with appropriate aggregation, which gives inductive bias for posterior estimation. Encoding architectures that are not permutation invariant, like RNN encodings, do not have this inductive bias.}
        \label{fig:illustration_perm_invariance}
\end{subfigure}

\begin{subfigure}[b]{1.0\textwidth}
\centering
\begin{tikzpicture}

\tikzset{
    layersquare/.style={
        draw=black, 
        line width=1.2pt, 
        rounded corners=10pt, 
        minimum width=2.5cm, minimum height=2.5cm, 
        fill=white 
    },
    topsquare/.style={
        draw=black, 
        line width=1.5pt, 
        rounded corners=10pt, 
        minimum width=2.5cm, minimum height=2.5cm, 
        fill=white, 
        text centered 
    },
    titlebox/.style={
        draw=black, 
        fill=white, 
        font=\bfseries, 
        align=center, 
        text width=3.5cm 
    },
    outerbox/.style={
        draw=black, 
        line width=1.2pt, 
        rounded corners=5pt, 
        inner sep=10pt 
    }
}

\foreach \i in {1,...,2} {
    \node[layersquare] (gnn\i_normal) at  (\i*0.2, -\i*0.2) {};
}

\foreach \i in {1,...,2} {
    \node[layersquare] (trafo\i_normal) at  (\i*0.2+3.0, -\i*0.2) {};
}

\node[topsquare] (gnn_ts_normal) at (0.6,-0.6) 
{
\includegraphics[width=2cm]{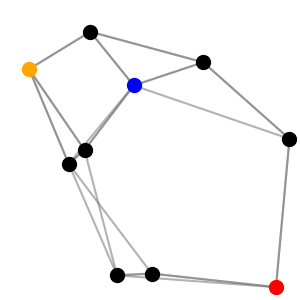}
};

\node[topsquare] (trafo_ts_normal) at (0.6+3.0,-0.6) 
{
\includegraphics[width=2cm]{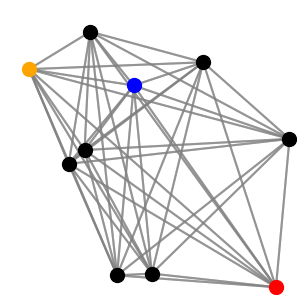}
};

\node[above=0.01cm of gnn_ts_normal, yshift=-0.3cm,draw, fill=white] (title_gnn_normal) {GNN};

\node[above=0.01cm of trafo_ts_normal, yshift=-0.3cm,draw, fill=white] (title_trafo_normal) {Transformer};

\node[right=of gnn1_normal,xshift=-0.7cm,yshift=1cm] (phantom_left) {};

\foreach \i in {1,...,2} {
    \node[layersquare] (gnn\i_removed) at  (\i*0.2+7.0, -\i*0.2) {};
}

\foreach \i in {1,...,2} {
    \node[layersquare] (trafo\i_removed) at  (\i*0.2+10.0, -\i*0.2) {};
}

\node[topsquare] (gnn_ts_removed) at (0.6+7.0,-0.6) 
{
\includegraphics[width=2cm]{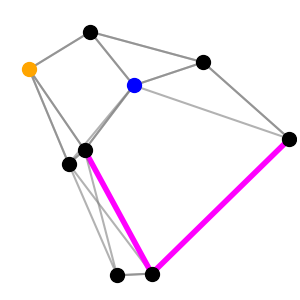}
};

\node[above=0.01cm of gnn_ts_removed, yshift=-0.3cm,draw, fill=white] (title_gnn_removed) {GNN};

\node[topsquare] (trafo_ts_removed) at (0.6+10.0,-0.6) 
{
\includegraphics[width=2cm]{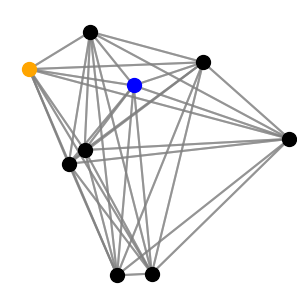}
};

\node[above=0.01cm of trafo_ts_removed, yshift=-0.3cm,draw, fill=white] (title_trafo_removed) {Transformer};

\node[right=of gnn1_removed,xshift=-0.7cm,yshift=1cm] (phantom_right) {};

\node[right=of gnn1_removed,xshift=-0.7cm,yshift=-1.5cm] (phantom_right_lower) {};

\node[right=of gnn1_normal,xshift=-0.7cm,yshift=-1.5cm] (phantom_left_lower) {};

\node[above=0.7cm of phantom_left, align=center] (data_left) {Data factors:  \textcolor{orange}{$x_1$}, \textcolor{blue}{$x_2$}, \textcolor{red}{$x_3$}, $x_4$, $\ldots$};
\draw[->] (data_left.south) to   (phantom_left.north) ;

\node[above=0.7cm of phantom_right, align=center] (data_right) {Data factors:  \textcolor{orange}{$x_1$}, \textcolor{blue}{$x_2$}, $x_4$, $\ldots$};
\draw[->] (data_right.south) to   (phantom_right.north) ;

\node[below=0.7cm of phantom_left_lower, align=center] (posterior_left) {$p(\theta|\bold{x}) = \frac{\textcolor{orange}{p(x_1;\theta)} \cdot  \textcolor{blue}{p(x_2;\theta)} \cdot \textcolor{red}{p(x_3;\theta)} \cdot p(x_4;\theta) \cdot \ldots  \cdot p(\theta) }{p(\bold{x})}$};
\draw[->] (phantom_left_lower.south) to   (posterior_left.north) ;

\node[below=0.7cm of phantom_right_lower, align=center] (posterior_right) {$p(\theta|\bold{x}) = \frac{\textcolor{orange}{p(x_1;\theta)} \cdot  \textcolor{blue}{p(x_2;\theta)} \cdot p(x_4;\theta) \cdot \ldots  \cdot p(\theta) }{p(\bold{x})}$};
\draw[->] (phantom_right_lower.south) to   (posterior_right.north) ;

\end{tikzpicture}
\caption{Illustration of removing likelihood data factors in Bayes' theorem. On the left side we use all data, including $x_3$, while on the right side we drop $x_3$ and predict a posterior without that data point. In contrast to the all-to-all connectivity of the transformer, the k-nearest neighbor edge-forming algorithm in the GNN creates new edges (pink edges) which more strongly affects the rest of the graph.}
\label{fig:illustration_input_generalizability}
\end{subfigure}
\caption{Inductive biases of a transformer encoding for a) the data encoding in posterior estimation and b) the selective removal of data factors from the encoding.}

\end{figure*}

Every neutrino interaction produces a different number of detected photons in the detector and the number of hit modules varies from event to event. The encoding scheme therefore has to handle variable-sized inputs. In the Bayesian picture (Fig. \ref{fig:illustration_input_generalizability}), variable sized data inputs correspond to a varying number of data factors in the likelihood function. In IceCube, we sometimes have faulty DOMs so we additionally need to be able to remove inputs dynamically. A transformer can naturally handle variable-sized inputs, but so can a GNN or an RNN. However, when removing data factors in Bayes' theorem we conjecture the transformer can handle those situations better than k-nearest-neighbor based GNNs because unlike the GNN no new spurious edge connections are created when removing data as illustrated in Fig. \ref{fig:illustration_input_generalizability}. This might partially explain the performance difference in section \ref{results_section}. We can additionally support prediction of the posterior for a varying number of inputs by probabilistically dropping inputs during training. We employ this as a form of data augmentation and probabilistic regularization during training (see section \ref{training_section}).

\section{Training details}
\label{training_section}

We focus on two types of events for training: charged-current $\nu_e$-interactions (showers) and charged-current $\nu_\mu$-interactions (tracks). An overview of the training datasets is given in table \ref{table:datasets}. The track dataset for training includes both throughgoing and starting tracks, so a single track model can handle both types. For showers, a $\nu_e$ charged-current model is also expected to work for neutral-current interactions or sub-PeV $\nu_\tau$ charged-current interactions. The lepton direction is taken as the target label, which above a few TeV is co-aligned with the neutrino direction due to kinematics. One can also train on the neutrino direction directly, but then one is more dependent on the spectral assumptions of the training dataset for low energies.

The training datasets contain simulated neutrinos with energies ranging between $100$ GeV to $100$ PeV for both tracks and showers. For weighting purposes, we further calculate the deposited energy (see section \ref{results_section} for details on the different weighting schemes, which are effectively treated as hyperparameters). For tracks, we define this as the sum of energy losses deposited within an \enquote{active} region which extends up to $350$m outside of the detector boundary. However, we only include tracks in training that cross within $50$m of the detector boundary at their point of closest approach. Similarly for showers, we require a shower to be within $50$m of the detector boundary. We found that not requiring such a cut would bias the training data too strongly towards events lying far outside the instrumented volume at high energies. Regular simulated neutrino events can contain coincident muons from air showers. Such events were removed to ensure pure signal events for training. The track training dataset contains all track event morphologies, including starting and throughgoing tracks, to obtain one unified track model after training. For testing, we use separate \enquote{throughgoing} and \enquote{starting} test datasets to later evaluate this trained model on those morphologies independently. 
In total, the shower training set comprises roughly 7 million events while the track training set comprises roughly 12 million events. For validation, we use $10000$ events each with similar structure as the respective training datasets. All data is simulated using the newest ice model description FTP-v3 \cite{ftp_paper} and the default hole ice parametrization for FTP-v3 which corresponds to the central \enquote{Flasher unfolding} datapoint in the unified hole ice parametrization \cite{hole_ice_paper}.

\renewcommand\tabularxcolumn[1]{>{\raggedright\arraybackslash}m{#1}}

\newcolumntype{L}[1]{>{\raggedright\arraybackslash}m{#1}}
\newcolumntype{C}[1]{>{\centering\arraybackslash}m{#1}}

\newcommand{\cutlist}[1]{%
  \begin{tabular}[c]{@{}l@{}}#1\end{tabular}%
}

\begin{table}[t]
\centering

\begin{tabularx}{\linewidth}{L{0.17\linewidth} C{0.2\linewidth} X}
\toprule

\multicolumn{3}{c}{\textbf{Training / validation}}\\
\midrule
\addlinespace[0.25em]
Dataset & Num.\ events & Specifications \\
\midrule

showers
& $\approx 7$ million (train)  $10000$ (val)
& \cutlist{%
  \textbullet\ only $\nu_e$ charged-current interactions\\
  \textbullet\ interaction max. 50m outside hexagon\\
  \textbullet\ no cosmic-ray air showers present\\
  \textbullet\ events pass filter for showers\\
  \textbullet\ $10^2 \ \mathrm{GeV} < E_\nu < 10^8 \ \mathrm{GeV}$
}
\\
\midrule

tracks
& $\approx 12$ million (train)  $10000$ (val) 
& \cutlist{%
  \textbullet\ only $\nu_\mu$ charged-current interactions\\
  \textbullet\ all morphologies (starting/throughgoing) \\
  \textbullet\ dep. energy calculated within hexagon + 350m\\
  \textbullet\ no cosmic-ray air showers present \\
  \textbullet\ events pass filter for muons \\
  \textbullet\ $10^2 \ \mathrm{GeV} < E_\nu < 10^8 \ \mathrm{GeV}$}
\\
\addlinespace[0.75em]
\midrule
\multicolumn{3}{c}{\textbf{Testing}}\\
\midrule
\addlinespace[0.25em]
Dataset & Num.\ events & Specifications \\
\midrule 
showers
& $\approx 10000$
& \cutlist{%
  \textbullet\ well-contained showers, similar selection as \cite{monopod_update}\\
  \textbullet\ contains both neutral and charged-current events\\
  \textbullet\ $10^4 \ \mathrm{GeV} < E_\nu < 10^7 \ \mathrm{GeV}$}
\\
\midrule
starting tracks
& $\approx 20000$
& \cutlist{%
  \textbullet\ geometrically selected: starting within the hexagon\\
  \textbullet\ at least 300m of track length within the hexagon\\
  \textbullet\ similar to train dataset otherwise}\\
\midrule
throughgoing tracks
& $\approx 20000$
& \cutlist{%
  \textbullet\ geometrically selected as passing through the hexagon\\
  \textbullet\ at least 300m of track length within the hexagon\\
  \textbullet\ similar to train dataset otherwise}\\
\addlinespace[0.75em]
\midrule
\multicolumn{3}{c}{\textbf{Datasets for data and systematics checks (section \ref{sec:data_systematics})}}\\
\midrule
\addlinespace[0.25em]
Dataset  & & Specifications \\
\midrule 
showers
&  & selection as developed for \cite{gal_plane_paper} with slightly more stringent final-level cuts
\\
\midrule
tracks
&  & final-level selection as developed for \cite{ngc_1086} \\

\end{tabularx}
\caption{Dataset specifications for the model training loop, split in training, validation and testing. Also shown are datasets used for data / Monte-Carlo comparisons and systematics checks. The \enquote{hexagon} refers to the quasi-hexagonal instrumented volume of IceCube. All Monte Carlo datasets use the FTP-v3 ice description \cite{ftp_paper}.}
\label{table:datasets}
\end{table}

A typical model with the described architecture from section \ref{architecture_section} has a few million parameters, depending in detail on the chosen hyperparameters. Because a model training takes about one to two months on a RTX-3090 GPU machine in our setup, we decided to do manual hyperparameter variations based on outcomes of a set of parallel training runs. In total we trained about 60 models split over three consecutive training \enquote{periods}, where each period contains parallel runs. The hyperparameters from the best-performing run in each period were carried forward as \enquote{defaults} for the next period, with further variations introduced manually. For most trainings, we fix the normalizing flow to the one described in section \ref{section:used_nf_definition} which has been found to be flexible enough to describe most relevant posterior shapes. The exception are runs that use a simple von-Mises Fisher distribution instead of a complex flow. We focus most of the hyperparameter variations on the exact architecture of the Transformer block as the encoding scheme has been found to be a major bottleneck in neural posterior estimation performance previously \cite{nfs_systematics}. For run 1, we train some models on showers and some on tracks. For run 2, we focus solely on tracks. And for run 3, we again train on showers and tracks. The optimization process is performed via stochastic gradient descent (SGD) with ADAM \cite{adam_paper} using default settings. We required flash attention v2 \cite{dao2024flashattention} to be able to train efficiently since we have a strongly varying token length between tens to thousands per batch item. For all training runs, we apply cosine-annealing \cite{cos_annealing_paper} on the learning rate whose amplitude we reduce at fixed intervals of 200k steps. We found cosine-annealing to help with instabilities of previously used normalizing flows. The novel flow  described in this paper likely does not require it anymore, but we retained it anyway. Additionally, some runs have a burn-in duration of 50k steps where the learning rate is linearly increased from zero to the initial learning rate, before starting with the scheduling. We anneal the learning rate towards a final learning rate $\eta_f$ that is defined as $\eta_f=2\cdot \frac{B}{N}$, with $B$ the batch size and $N$ the dataset size. This is motivated by the interpretation of SGD as stochastic variational inference over the weight posterior \cite{mandt_2016}. The batch size $B$ is 80 through all trainings. We split each batch into 4 sections. In the first 20 items, we include all DOMs, in the second 20 items, we remove empty DOMs, in the third 20 items, we remove saturated DOMs, and in the last 20 items, we remove both empty and saturated DOMs. Additionally, with a $50\%$ chance we further remove a random amount of DOMs from the input for each batch item. This helps as a form of \enquote{probabilistic regularization} as the model learns to mimic the structure of Bayes' theorem by training on all combinations of IID data factors, which is especially efficient for transformers via Bayesian inductive bias (see Fig. \ref{fig:illustration_input_generalizability}). Dropping data randomly can also be seen as a form of data augmentation. Every time the same event is passed to the model during training, its input data to the model might look different.  We perform stochastic weight averaging (SWA) \cite{swa_paper} during the whole training run on the side to always have an averaged model available. For showers, we typically exit the run early  due to overfitting at the highest energies (see a discussion in section \ref{section:validation_curves} or Fig. \ref{fig:validation_curves}), and the model used for testing is then the averaged model available at the best validation loss. For tracks, we also perform a few more 100k iterations of SWA at the end of training with the final learning rate $\eta_f$ and the test model is then taken at the end of that. More details on how a model is selected for testing based on the validation loss are given in section \ref{section:validation_curves}.

\subsection{Hyperparameters}
\label{sec:hyperparams_detail}
Our baseline transformer model in training period 1 uses 20 layers, pre-layer normalization, an internal embedding dimension of $96$ and a single head. The MLP part uses a hidden dimension of size 512. We aggregate at the end using the mean and diagonal variances of the final tokens and map the output to a \enquote{bottleneck} dimension of size 32. This representation is then mapped to the normalizing flow with a single MLP and hidden dimension 128. We vary this baseline multiple times which comprises training period 1. For the further periods we took the best models, varied them again, and in this way trained period 2 and 3. The hyperparameter extensions include a variation of the bottleneck dimension, an increase of the embedding dimension and the number of heads. We also checked standard sinusoidal positional encoding based on the $(x,y,z)$ DOM coordinates and relative value positional encoding \cite{shaw_et_al} in several variations, the latter one having been used in the recent IceCube Kaggle challenge \cite{kaggle_paper}. For relative positional encoding, we had to cap the maximum allowed number of input tokens to 200 due to memory constraints. We further explored using two simultaneous residual streams \cite{residual_citation} and nonlinear QKV embedding, which potentially captures correlations between the Q,K and V tokens during the embedding stage. As an alternative to the standard aggregation scheme we test three different variants of class tokens, similar to their usage in vision transformers \cite{vision_transformers}. In order to test the influence of a fully-fledged normalizing flow compared to a simpler distribution, we also train standard vMF distributions with two different rotation parametrizations. Lastly, we compare against a non-transformer model based on GNNs with an architecture that has been described in \cite{gnn_low_energy}.
More details on the hyperparameters are given in appendix \ref{appendix:hyperparameters}.

\subsection{Validation curves}
\label{section:validation_curves}

\begin{figure}
    \begin{subfigure}[t]{0.5\textwidth}
    \centering
    \includegraphics[width=0.99\linewidth]{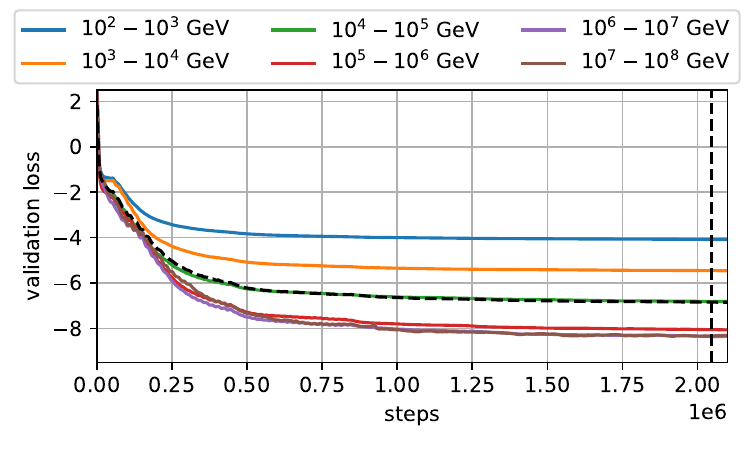} 
    \caption{Best track model.}
    \label{fig:validation_curves_a}
    \end{subfigure}
    \begin{subfigure}[t]{0.5\textwidth}
        \centering
         \includegraphics[width=0.99\linewidth]{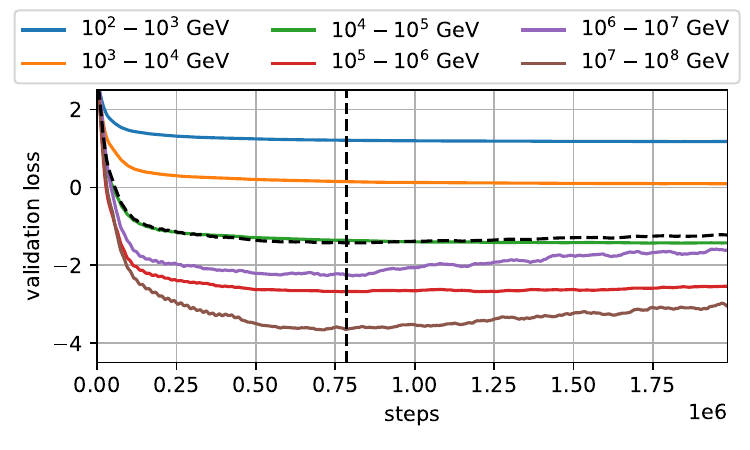} 
        \caption{Best shower model}
        \label{fig:validation_curves_b}
    \end{subfigure}
    \caption{Validation-loss curves for the best track and shower model. The x-axis shows optimization steps. Curves are shown for different energy ranges, along with the equally weighted average validation loss (dashed). The vertical dashed line marks the step at which the model is frozen for testing.}
    \label{fig:validation_curves}
\end{figure}

The validation curves for tracks and showers look qualitatively slightly different as seen in Fig. \ref{fig:validation_curves} which shows the validation curves for the the best throughgoing track model and the best shower model. The validation is split up into curves for different deposited energy in log space. An average validation loss, which is calculated by weighting each energy-decade loss curve equally, is also shown. The model for further testing is selected based on the minimum of this average validation loss, as illustrated by the vertical dashed line. For tracks, we have more training data available at the highest energies, and there is minimal to no overfitting. The model for testing is typically obtained after SWA has run for an extended period late in the training run (see Fig. \ref{fig:validation_curves_a}). For showers, above a PeV there are only a few hundred thousand events in the training sample, while we have millions of events available around a TeV. During training, we train on all events equally weighted per deposited logarithmic energy, which means the high-energy events will be picked more often in batches and overfit earlier than at low-energy (Fig. \ref{fig:validation_curves_b}). For the same reason, the validation dataset also has less events at higher energies, and the loss for 100 TeV to 1 PeV (red curve) is actually below the loss for events between 1 TeV to 10 PeV (purple curve), while for enough statistics one should always expect an ordering based on energy since the associated posteriors become more compact. We accept this slight imprecision in validation and for the future plan to use a larger admixture at high energies. In general, there is always slight compromise for the tested shower models, where the low energy events could gain more from further training, while the high-energy events require an earlier cutoff to decrease the overfitting. This can be remedied for future studies with increasing the available training data at high energies and balancing out the statistics.

\section{Results}
\label{results_section}

We test the shower models on a test dataset that has been originally introduced in \cite{monopod_update} and consists of roughly 10000 well-contained showers between 10 TeV and 10 PeV. For tracks we have two different test datasets: roughly 20000 events sampled equally in log-energy that are starting in the detector and roughly 20000 tracks that are throughgoing. This separation was not made during training, where all track morphologies were used to train the models. All test datasets use the same ice description as in training, FTP-v3 \cite{ftp_paper}. More details are given in table \ref{table:datasets}. To give an impression of some of the predicted posteriors, example event views are shown in appendix \ref{appendix:example_events}.

\subsection{Test results}

\begin{figure*}
    \centering
   \begin{subfigure}[b]{0.99\textwidth}
   \centering
    \includegraphics[width=0.99\textwidth]{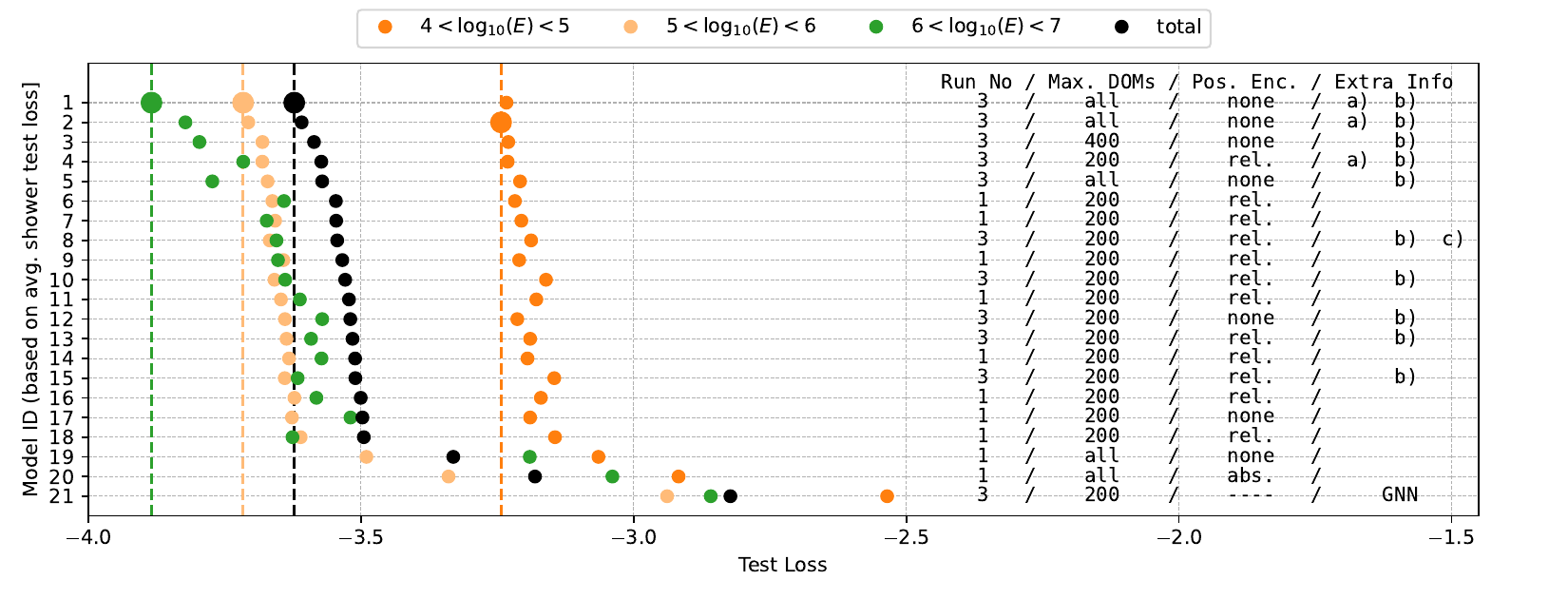} 
    \caption{Shower test results}
    \label{subfig:testresults_cascades}
    \end{subfigure}
    \par\bigskip
    \begin{subfigure}[b]{0.99\textwidth}
    \centering
    \includegraphics[width=0.99\textwidth]{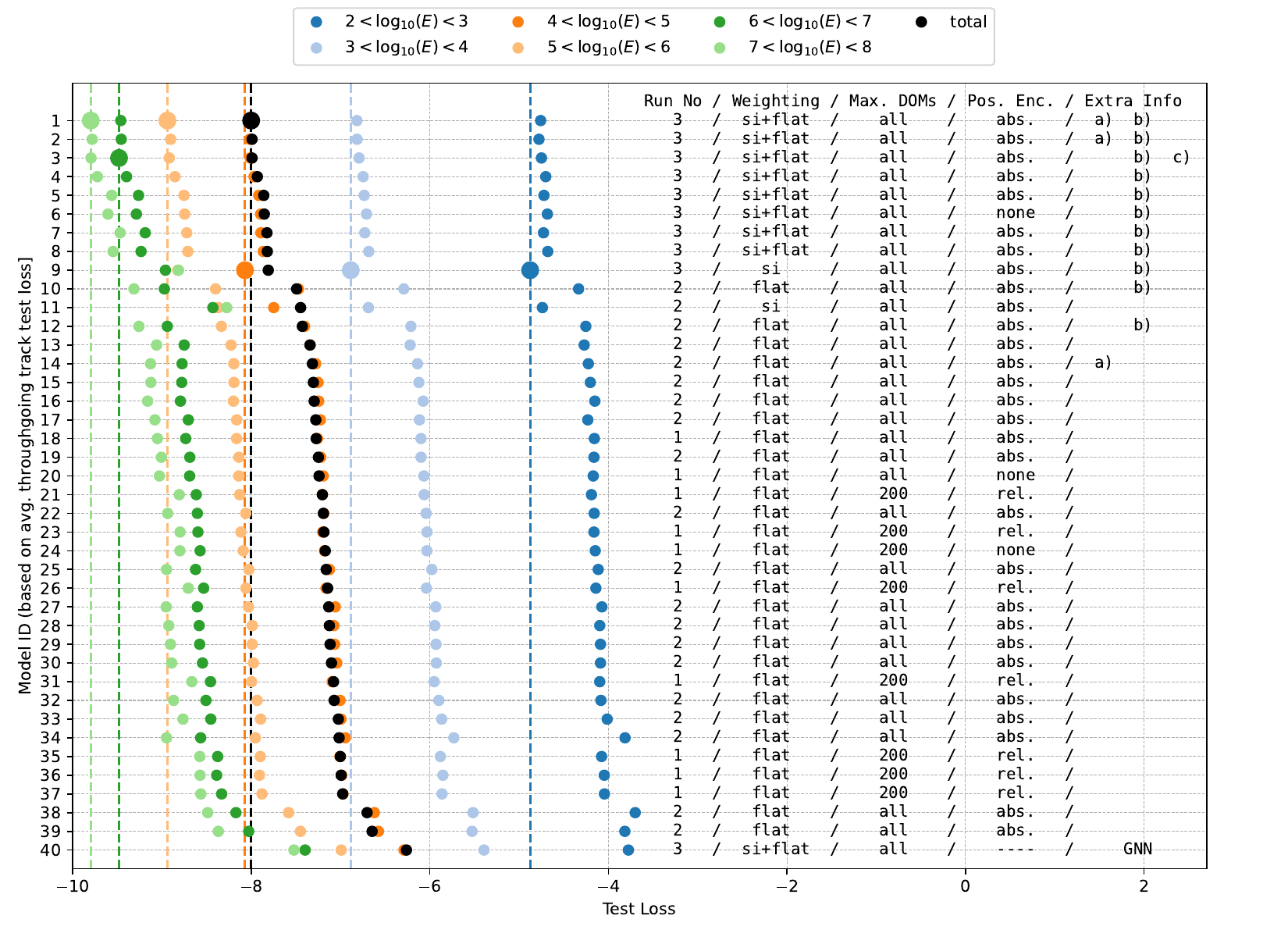} 
    \caption{Throughgoing track test results}
    \label{subfig:testresults_throughgoing}
    \end{subfigure}
    \caption{Test results for different hyperparameters for showers and throughgoing tracks. Models are sorted by average total test loss (black). The best test loss per energy range is highlighted by a larger marker and a corresponding vertical cashed line to compare to other models.  Options a), b) and c) (\enquote{Extra Info}) are described in the text in section \ref{results_section}. A detailed description of all parameters of each model is given in table \ref{tab:cascade_options} and \ref{tab:track_options}. }
    \label{fig:test_results}
\end{figure*}

 In Fig. \ref{fig:test_results} we summarize the test losses for showers and throughgoing tracks for all models. The results for starting tracks are summarized in Fig. \ref{fig:test_results_starting} in appendix \ref{appendix:hyperparameters}. The models are sorted along the $y$-axis based on performance of the average total test loss. For starting tracks, that sorting uses the throughgoing track test results in order to see if the order changed dramatically, which it does not. Along the $x$-axis the test loss is split into different deposited energy regions. The best loss within a given energy loss is highlighted with a slightly larger marker. This is done to differentiate cases where the average model loss might be good for a specific model, but maybe another model has a better low energy performance. This is the case for model 9 and 11 for tracks, for example, as they were trained using a pure spectral weighting which over-emphasizes low energy events to the neglect of the PeV events and beyond. On the right of the plot we summarize relevant hyperparameter settings to give a quick overview in addition to the precise summary in tables \ref{tab:track_options} and \ref{tab:cascade_options}. Important settings include the weighting scheme during training, the maximum number of DOMs given to the algorithm, and also the positional encoding for the transformer. One weighting scheme used is a weighting to a spectrum with spectral index of $-1.8$ in neutrino energy (we call this \enquote{si} weighting in Fig. \ref{fig:test_results}), which was observed to approximately lead to a flat distribution in observed energy. In practice, however, it actually slightly overemphasized low-energy events in early training runs. Therefore, we introduced a second weighting that explicitly constructs weights based on a flat distribution in observed deposited energy (\enquote{flat}) - a scheme that turned out to produce an overemphasis on high-energy events while low-energy performance was lacking. Therefore, for training session 3, we also used weights that are a mixture of both weighting schemes (\enquote{si+flat}). This weighting turned out to give a good performance over all energies from $100$GeV to $100$ PeV. For relative value positional encoding , we had to limit the number of allowed input DOMs to be 200 - ordered by observed charge - due to memory constraints. The column \enquote{Extra Info} summarizes three specific hyperparameters that we found had the overall biggest influence on performance improvements. These are (a) nonlinear in-projection to the QKV tensor, (b) dual residual connections (\enquote{ResiDual}) \cite{residual_citation} and (c) aggregating with an \enquote{improved class token}, a modified version of the class token in vision transformers \cite{vision_transformers} that interacts via cross-attention, as indicated in Fig. \ref{fig:test_results}. A more detailed list of the used parameters and their description is found in table \ref{tab:track_options} and \ref{tab:cascade_options} in appendix \ref{appendix:hyperparameters}. 

 For shower models, absolute position encoding (model 20) or no positional encoding (model 19) did not yield good results in training period 1 compared to relative value positional encoding with maximally 200 DOM input tokens. It seemed the information could not be adequately processed, although more training runs are necessary to settle this. However, switching to dual residual connections (model 5) changes this, even though a limitation of input tokens is then still favored (model 3). Adding nonlinear QKV projection in addition (model 1 and 2) gives another big boost which then allows the information of the full detector to be utilized without any input restriction. In the future it might be worthwhile to also try standard positional encodings again with these improvements.

 For track models, relative absolute or no positional encoding with the full DOM token input was always slightly better (model 18 or 20) compared to restriction in input tokens and relative value positional encoding (model 21). Again, using dual residual connections in combination with nonlinear QKV projection yields top performances (model 1 and 2). However, dual residual connections and a separate class aggregation gives nearly similar performance in this case (model 3). It might be interesting to combine all three of them in the future. In both cases of showers and tracks, a GNN encoding following the structure from \cite{gnn_low_energy} ranks last in model performance compared to all tested transformer architectures.

\subsection{Angular resolution and coverage}

\begin{figure*}
    \centering
   \begin{subfigure}[b]{0.95\textwidth}
   \centering
    \includegraphics[width=0.95\textwidth]{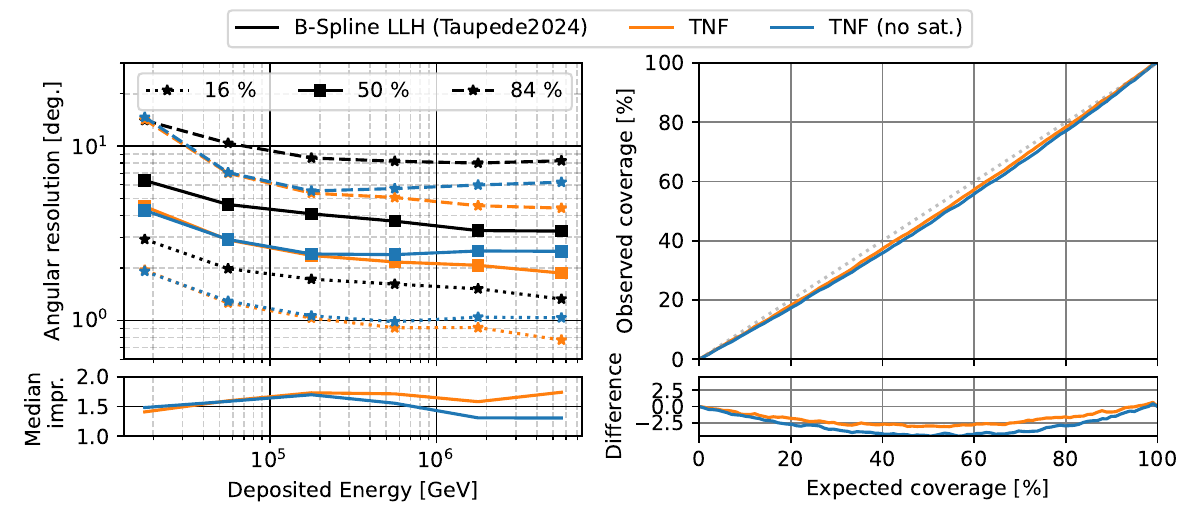} 
    \caption{Neutrino-induced showers}
    \label{subfig:cascades}
    \end{subfigure}
    \par\bigskip
    \begin{subfigure}[b]{0.95\textwidth}
    \centering
    \includegraphics[width=0.95\textwidth]{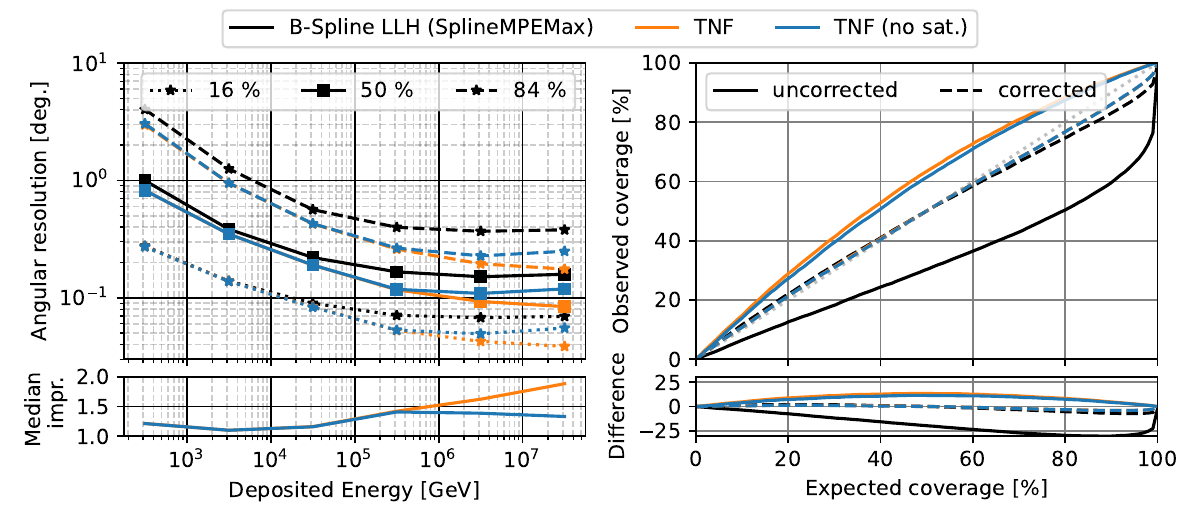} 
    \caption{Throughgoing tracks (at least 300m distance within detector)}
    \label{subfig:throughgoing_300}
    \end{subfigure}
    \par\bigskip
    \begin{subfigure}[b]{0.95\textwidth}
    \centering
    \includegraphics[width=0.95\textwidth]{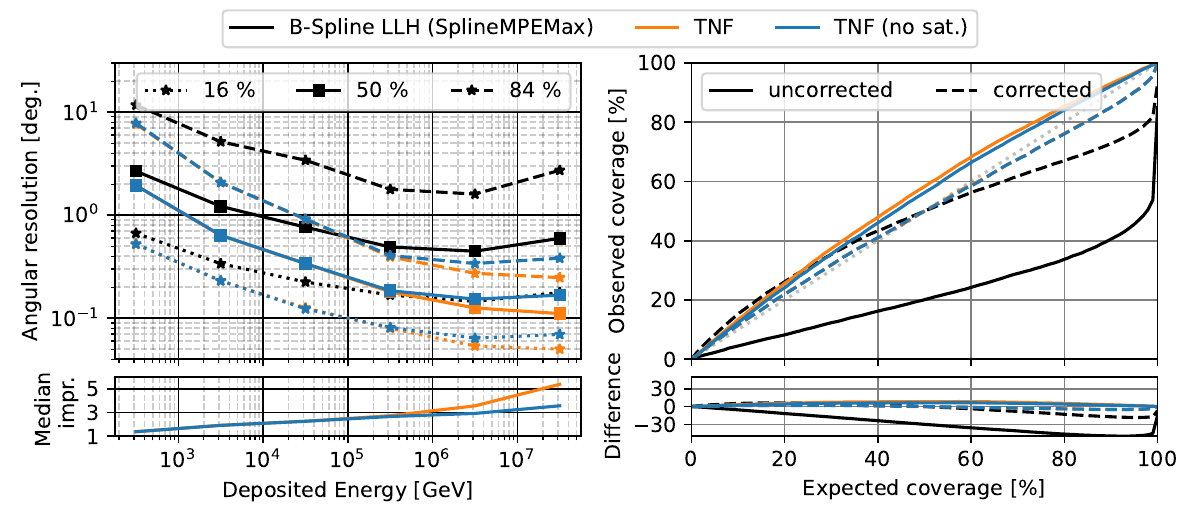} 
    \caption{Starting tracks (at least 300m distance within detector)}
    \label{subfig:starting_300}
    \end{subfigure}
    \caption{On the left, angular resolution ($16\%$, $50\%$, $84\%$ quantiles) of the transformer normalizing flow (TNF) with and without saturated DOMs (TNF no sat.) compared to the state-of-the-art respective likelihood method based on B-splines, \textit{SplineMPEMax} \cite{schatto_thesis} for tracks and \textit{Taupede2024} \cite{monopod_update} for showers. The improvement of median angular resolution of the new method versus the respective likelihood method is shown in the lower left figures. The right part shows expected versus observed coverage for all coverage probabilities. The corrected coverage for TNF assumes the von-mises-Fisher approximation for its calculation.}
    \label{fig:resolution_all}
\end{figure*}

\begin{figure*}
    \centering
   \begin{subfigure}[t]{0.49\textwidth}
   \centering
    \includegraphics[width=\linewidth]{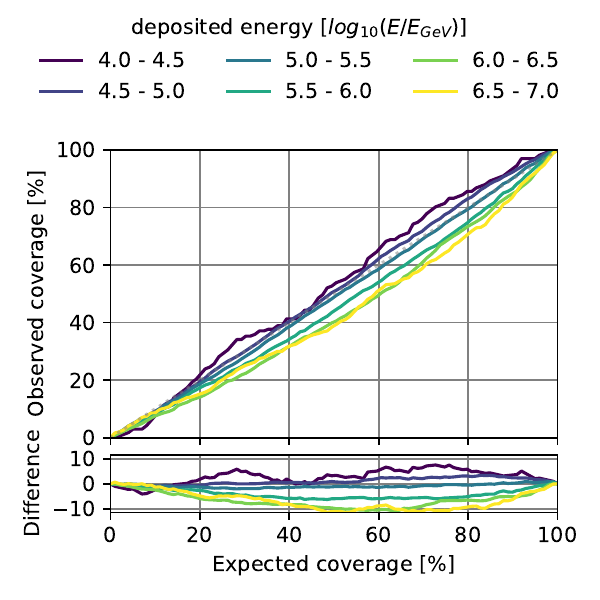} 
    \caption{Neutrino-induced showers}
    \label{subfig:coverage_energies_cascades}
    \end{subfigure}
    \begin{subfigure}[t]{0.49\textwidth}
    \centering
    \includegraphics[width=\linewidth]{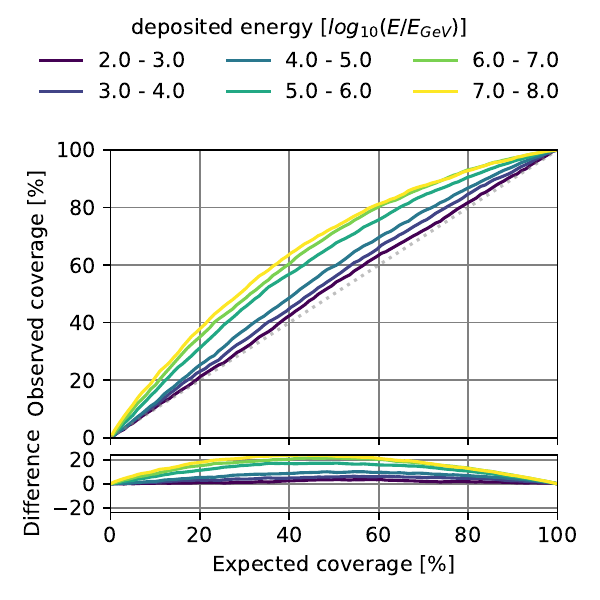} 
    \caption{Throughgoing tracks (at least 300m distance within detector)}
    \label{subfig:coverage_tracks}
    \end{subfigure}
    \caption{Coverage split up in different deposited energy bands.}
    \label{fig:coverage_energy_dependent}
\end{figure*}

In the following we calculate angular resolutions using the maximum of the posterior (MAP) estimate of the posterior estimate in order to compare to the respective maximum likelihood estimator. The transformer normalizing flow is abbreviated as \enquote{TNF} for simplicity. Figure \ref{subfig:cascades} shows the median angular resolution for showers using TNF improves by up to a factor of two compared to the existing best B-spline based likelihood reconstruction \textit{Taupede2024} \cite{monopod_update}. It should be noted that the \textit{Taupede2024} shower reconstruction by default excludes saturated DOMs and DOMs with a charge higher than 15 times the mean charge, while the TNF reconstruction has no exclusions. 
This has to be considered for the comparison mostly above 100 TeV, while below 100 TeV there are few saturated DOMs or high-charge outliers and the comparison can be interpreted as running on the exact same input. There is typically no uncertainty contour associated to the B-spline construction for showers, mostly because a profile-likelihood map is costly and can take several hours on a cluster (see Fig. \ref{fig:runtimes}) and an alternative via the Hessian has not been established in terms of numerical stability. In analyses applications \cite{gal_plane_paper}, we typically use separate uncertainty predictors trained with high-level features for these reasons, but here we are interested in a self-consistent method comparison. That is why we do not include a coverage result for the B-spline applied to showers. The coverage for the transformer-based normalizing flow is calculated using the exact contours of the full-sky probability maps. It seems to perform well overall, but when split among energies (Fig. \ref{subfig:coverage_energies_cascades}) it becomes evident that it is good at low energies, but starts to undercover from a few 100 TeV upwards. This has to do with the lack of training statistics at high energies which shows up as earlier overfitting as discussed in section \ref{section:validation_curves}. Training on more simulations should alleviate this problem. 

For throughgoing tracks (Fig. \ref{subfig:throughgoing_300}) the median angular resolution improves over the whole energy range compared to the standard method based on B-splines, \textit{SplineMPEMax} \cite{schatto_thesis}, that has been in use for most track-based neutrino point source analyses in the past, e.g. in the detection of NGC 1068 \cite{ngc_1086}. An improvement or gain by a given factor means the angular resolution decreases by that factor and allows better localization of a neutrino source. In the relevant energy range from 1 to 100 TeV, the gain in angular resolution is about a factor of $1.1$ to $1.3$. This is the first time a machine-learning reconstruction yields consistently better results anywhere over the whole energy range since the introduction of \textit{SplineMPEMax} in 2014. Additionally, at high energies there is no sign of a strong resolution floor. The standard B-spline reconstruction flattens out, since it does not fully take into account stochastic losses which dominate at PeV energies. The transformer-based normalizing flow, on the other hand, seems to perform well also in this regime. For starting tracks the performance gain in angular resolution is larger than for throughgoing tracks. The gain over the B-spline reconstruction reaches a factor of $2.5$ at 100 TeV (Fig. \ref{subfig:starting_300}) and goes beyond a factor of $3$ above a PeV. One reason is that the B-spline ansatz uses a chain of seeding reconstructions \cite{segmented_spline_paper} which often gets stuck in local optima for starting tracks due to the extra hadronic shower from the neutrino interaction and shorter overall track lengths. The transformer-based normalizing flow seems to adapt to different morphologies automatically, as it was trained jointly on both throughgoing and starting tracks.
In constrast to showers, for tracks we use a standard profile likelihood fit \cite{paraboloid_paper} to obtain an uncertainty estimator for \textit{SplineMPEMax}. Typically, an energy-dependent correction is applied to widen the contours. Even after this correction, the coverage properties of the B-spline reconstruction are undercovering for the tails of the distribution, in particular for starting tracks. The TNF reconstruction for tracks is overcovering, which is conservative. We analyse the energy dependence of this behavior later in this section.

It should be said that the track B-spline method \cite{schatto_thesis} uses an outdated ice model for numerical reasons, and efforts are underway to improve it with a newer ice description in the near future. In all cases we observed that the inclusion of saturated DOMs as described in Fig. \ref{fig:transformer_encoding} helps at high energies for angular resolution. The inclusion of empty DOMs did not have a big influence, so we did not include it in the plots. For energy reconstruction, on the other hand, we expect the inclusion of empty DOMs to have an non-negligible contribution to the performance at all energies.

Looking at the energy dependence of the track coverage for TNF (Fig. \ref{subfig:coverage_tracks}), it has proper coverage for lower energies, but starts to have too large contours above a few 100 TeV. This effect is the opposite of the behavior for showers. The training statistics for tracks at PeV energies is a factor of a few larger than for showers, and this shows here. The over-coverage likely comes from the fact that the absolute resolution goes down to sub-degrees at high energies and several scales of resolution are combined during training. Events with small angular contours dominate the loss function, and start fluctuating more compared to lower energy events. We suspect further improvements with the exact parametrization of the flow, in particular the rotation parametrization, and a better tuned optimization routine might alleviate this issue in the future. It should be noted however, that over-coverage means the predicted uncertainty contours are too large, which can be acceptable and is in general conservative.

\subsection{Application on real data and systematics}
\label{sec:data_systematics}
In order to test the robustness of the algorithm, we apply it on data and Monte Carlo datasets with various systematic uncertainties. 

\begin{figure*}
    \centering
    \begin{subfigure}[t]{0.85\textwidth}
   \centering
    \includegraphics[width=\linewidth]{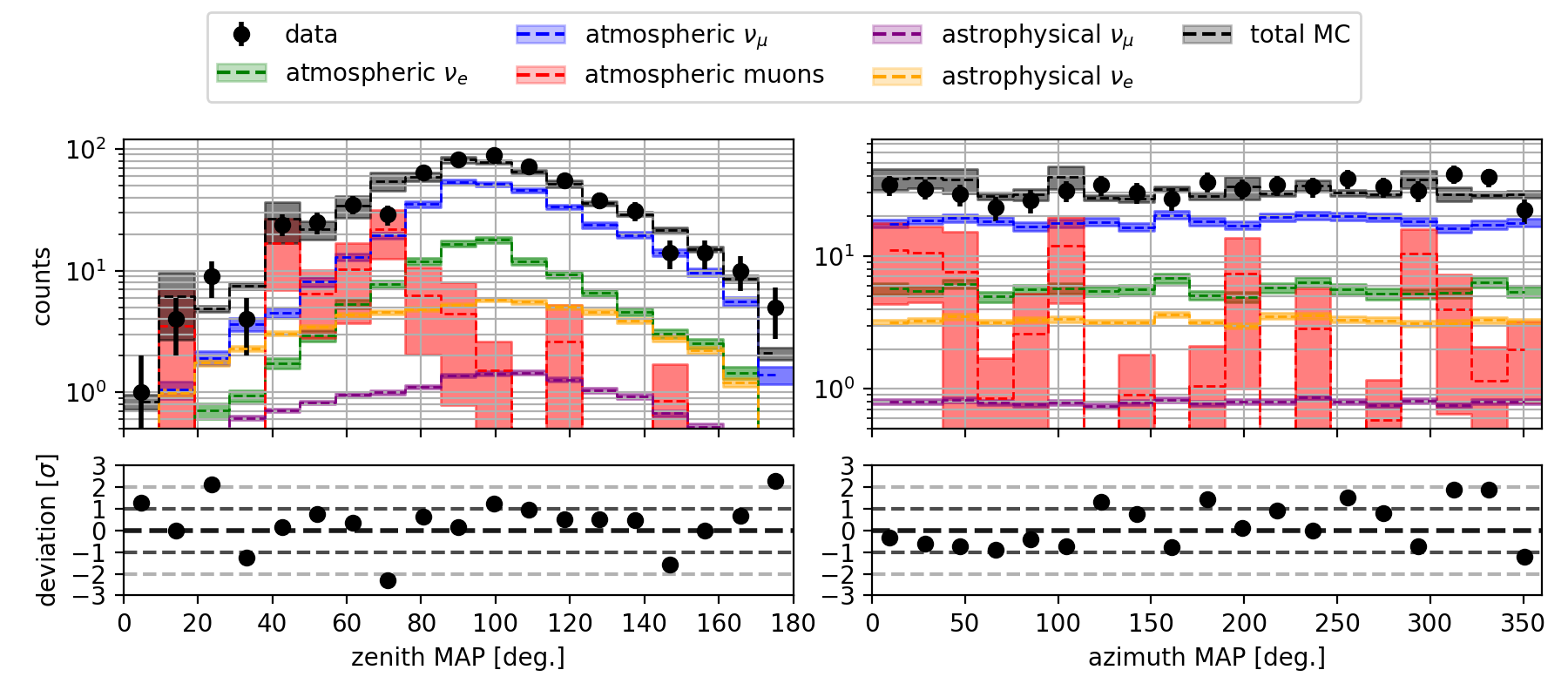} 
    \caption{}
    \end{subfigure}
    \par\bigskip
   \begin{subfigure}[t]{0.49\textwidth}
   \centering
    \includegraphics[width=\linewidth]{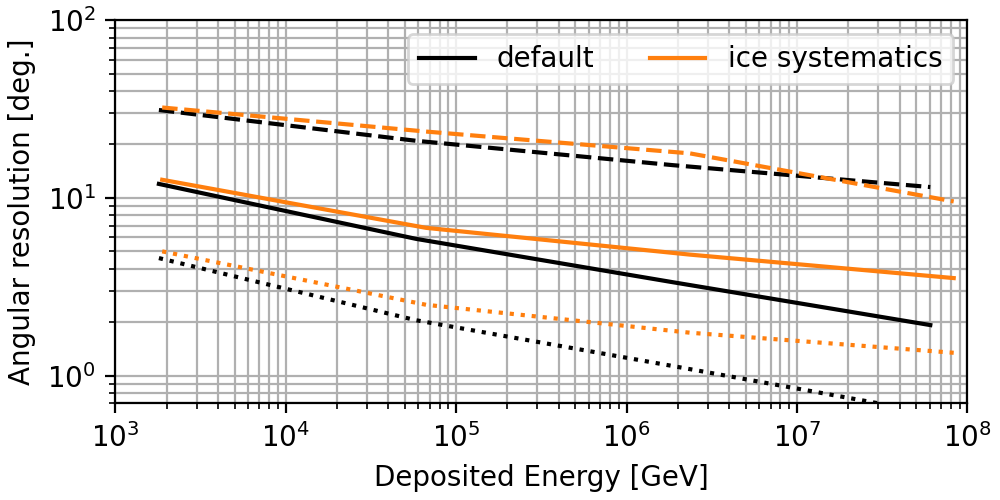} 
    \caption{}
    \end{subfigure}
    \begin{subfigure}[t]{0.49\textwidth}
    \centering
    \includegraphics[width=\linewidth]{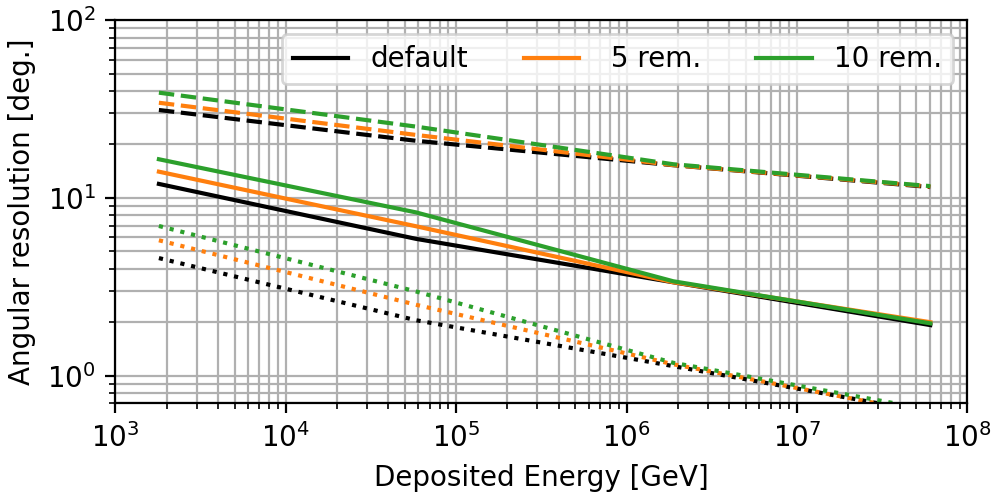} 
    \caption{}
    \end{subfigure}
    \par\bigskip
    \begin{subfigure}[t]{0.49\textwidth}
   \centering
    \includegraphics[width=\linewidth]{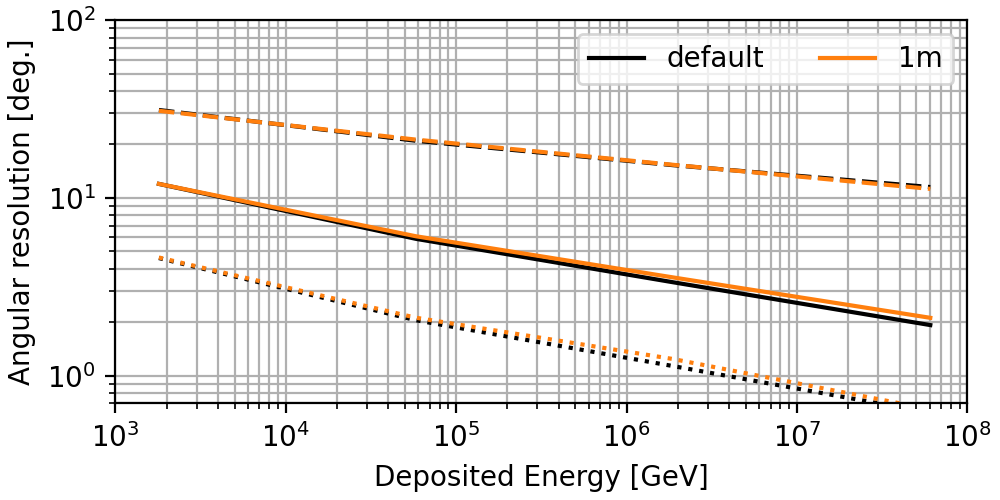} 
    \caption{}
    \end{subfigure}
    \begin{subfigure}[t]{0.49\textwidth}
    \centering
    \includegraphics[width=\linewidth]{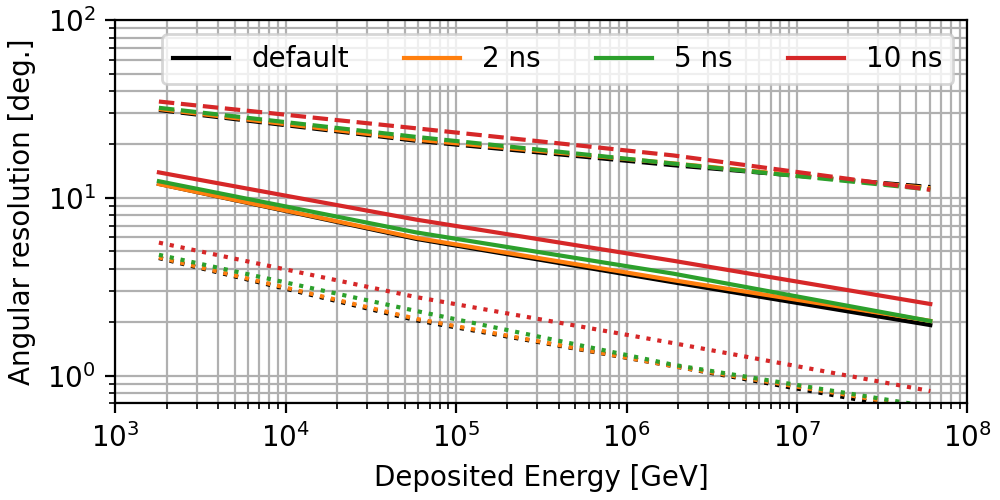} 
    \caption{}
    \end{subfigure}
    \caption{Data / Monte Carlo comparison of the zenith and azimuth of the maximum of the posterior (MAP) (a) and effects of various systematics on angular resolution (b-e) for a slightly modified selection of the one used in \cite{gal_plane_paper}. The systematics checks are visualized as $16 \%$ (dotted), $50 \%$ (solid) and $84 \%$ (dashed) quantiles to indicate the distributions. They include (b) ice model systematics, (c) the removal of \enquote{$x$} highest charge DOMs, (d) random variation (in terms of standard deviation) in $(x,y,z)$ coordinates of individual strings and (e) random variation (in terms of standard deviation) of absolute time offset in each DOM.}
    \label{fig:shower_data_mc}
\end{figure*}

For showers, we apply it on a small test dataset of 3 months livetime using a final level shower selection similar to the one used in \cite{gal_plane_paper} (see also table \ref{table:datasets}). Fig. \ref{fig:shower_data_mc} a) shows the shower data/MC comparison of the MAP of the zenith and azimuth posterior after matching the total Monte Carlo rate to the data rate. The Poisson uncertainty is indicated for visualization purposes by error bars on the data, while the uncertainty from finite weighted Monte Carlo statistics is indicated as colored bars for the respective Monte Carlo prediction using a Gamma approximation following \cite{say_llh}. The deviation in the lower part of the figure utilizes the combined uncertainty via an appropriate Gamma-Poisson Mixture. Cosmic-ray interactions in the atmosphere produce \enquote{atmospheric} neutrinos and muons which show up as irreducible background events in this selection. Muons are irreducible here because they enter the detector between strings or skim a corner of the detector so they appear as showers. The atmospheric neutrino fluxes assume a primary cosmic-ray flux model from \cite{gaisser_2012} and atmospheric interactions following the conventional model in \cite{sybill23_paper}. Also shown are \enquote{astrophysical} diffuse neutrino flux predictions assuming the measurement results from \cite{global_fit_paper}. A few outliers are visible which is likely due to the low Monte Carlo statistics of the cosmic-ray muon background and unmodeled ice systematics. Overall the data/MC agreement is similar to expectations from comparisons with equivalent B-spline reconstructions. It is noteworthy that most of the remaining events from downgoing muons are correctly reconstructed as downgoing, even though the network was only trained on electron neutrino charged-current interactions and has never seen muons during training. Furthermore, Fig. \ref{fig:shower_data_mc} b)-e) show various systematics checks, where the angular resolution of the MAP is compared to the no-systematics baseline. The \enquote{ice systematics} curve corresponds to an average of different effects which comprise photon absorption, photon scattering, effective light yield and angular PMT acceptance related to the local ice structure. Overall the impact on the angular resolution by ice-related systematics is a little larger than detector related systematics, which is known behavior in existing shower reconstructions. The removal of the 5 or 10 highest charge DOMs per event affects mostly low-energy events, which can be explained by those events already having a low number of DOMs to begin with, and the removal of a fixed amount of high charge DOMs has a larger relative impact in this case.

\begin{figure*}
    \centering
    \begin{subfigure}[t]{0.85\textwidth}
   \centering
    \includegraphics[width=\linewidth]{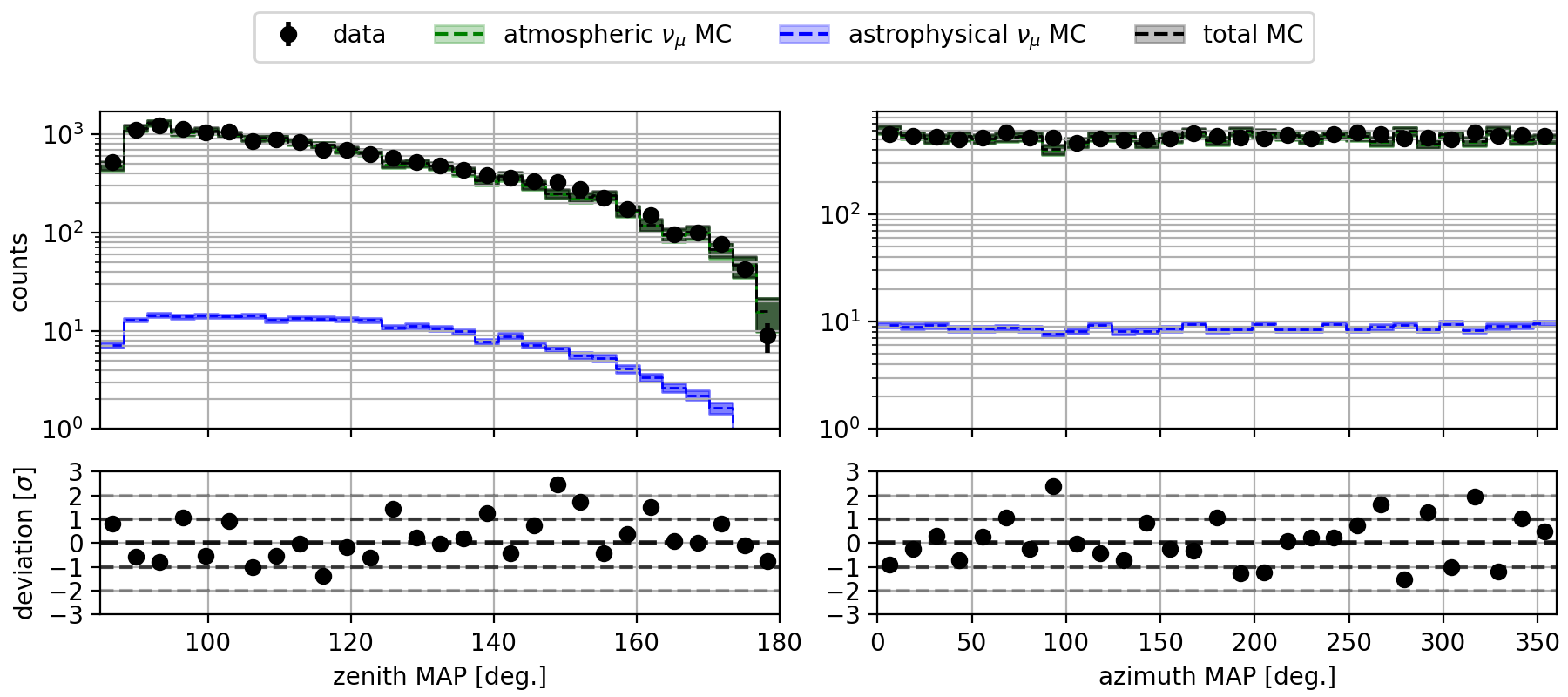} 
    \caption{}
    \label{subfig:casc_ice}
    \end{subfigure}
    \par\bigskip
   \begin{subfigure}[t]{0.49\textwidth}
   \centering
    \includegraphics[width=\linewidth]{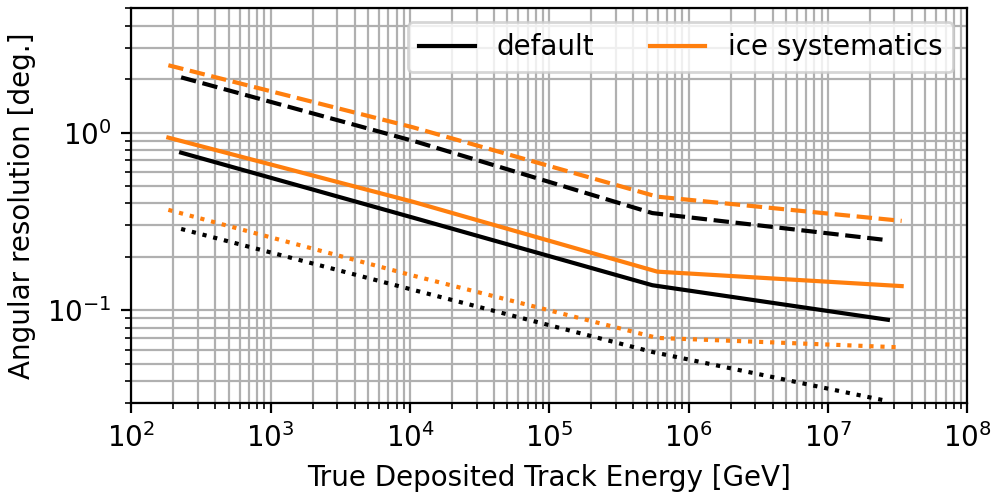} 
    \caption{}
    \end{subfigure}
    \begin{subfigure}[t]{0.49\textwidth}
    \centering
    \includegraphics[width=\linewidth]{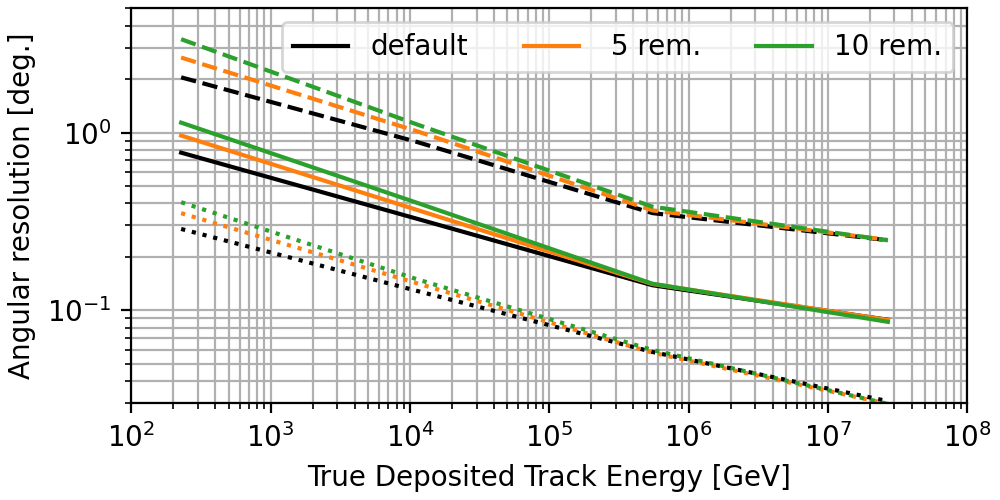} 
    \caption{}
    \end{subfigure}
    \par\bigskip
    \begin{subfigure}[t]{0.49\textwidth}
   \centering
    \includegraphics[width=\linewidth]{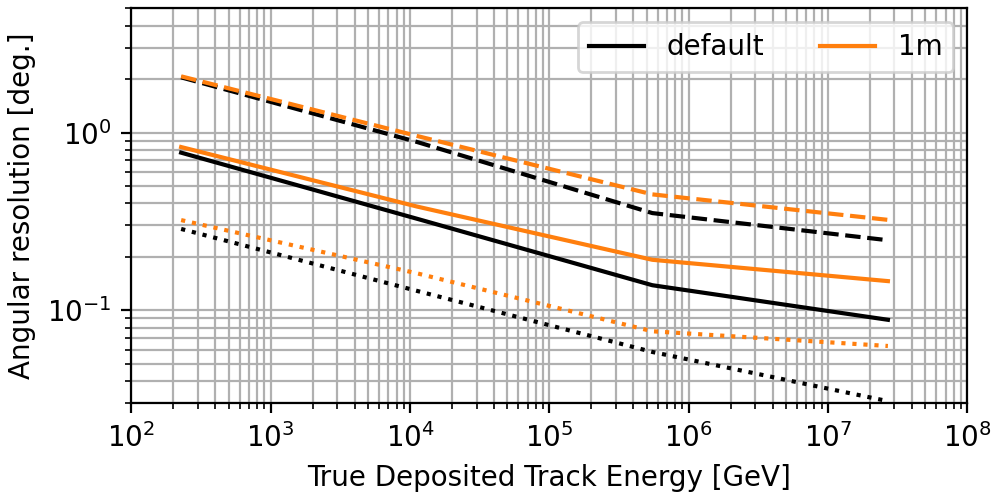} 
    \caption{}
    \end{subfigure}
    \begin{subfigure}[t]{0.49\textwidth}
    \centering
    \includegraphics[width=\linewidth]{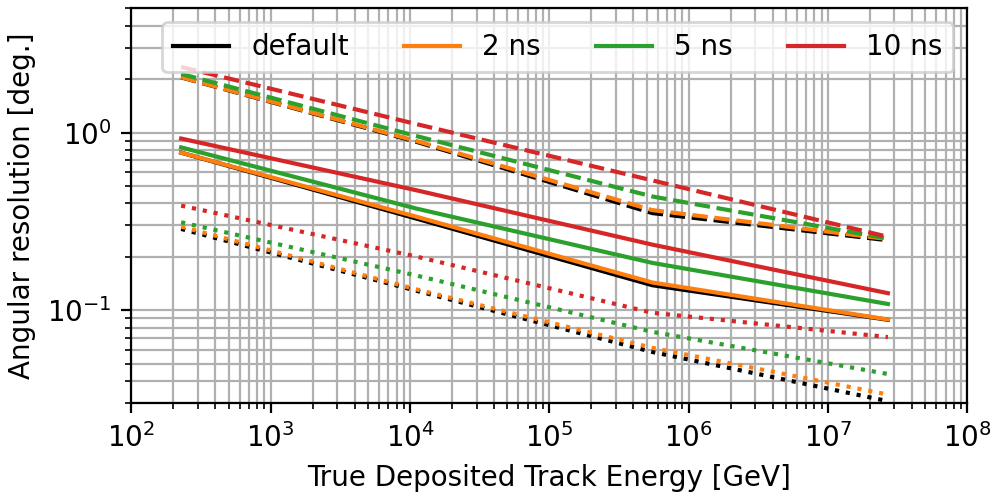} 
    \caption{}
    \end{subfigure}
    \caption{Data / Monte Carlo comparison of the zenith and azimuth of the maximum of the posterior (MAP) (a) and effects of various systematics on angular resolution (b-e) for the final track selection used in \cite{ngc_1086}. The systematics checks are visualized as $16 \%$ (dotted), $50 \%$ (solid) and $84 \%$ (dashed) quantiles to indicate the distributions. They include (b) ice model systematics, (c) the removal of \enquote{$x$} highest charge DOMs, (d) random variation (in terms of standard deviation) in $(x,y,z)$ coordinates of individual strings and (e) random variation (in terms of standard deviation) of absolute time offset in each DOM.}
    \label{fig:tracks_data_mc}
\end{figure*}

For tracks, we use 2.5 months of livetime for data using the final sample selection developed in \cite{ngc_1086}. Fig. \ref{fig:tracks_data_mc} shows similar data/MC and systematics comparisons as for showers, again after matching the overall Monte Carlo rate to the data rate. The Data/MC comparison looks good, similar to existing likelihood reconstructions. Overall, the relative ice-systematic effect is slightly smaller than for showers, while geometry effects play a bigger role. This makes sense, since track events on average contain more unscattered Cherenkov light which can reduce the impact of ice effects. On the other hand, the angular resolution of a few tenths of a degree is precise enough such that position and timing of the modules becomes more relevant compared to showers whose resolution is about an order of magnitude worse.

For specific analyses, it can make sense to include systematics during training. This can be achieved either by an explicit inclusion in the model, or by training on an ensemble of different systematics realizations and thereby effectively learn the marginalized posterior as demonstrated in \cite{nfs_systematics}. We leave this for future work.

\subsection{Timing}

\begin{figure*}
    \begin{subfigure}[t]{0.99\textwidth}
    \centering
    \includegraphics[width=\linewidth]{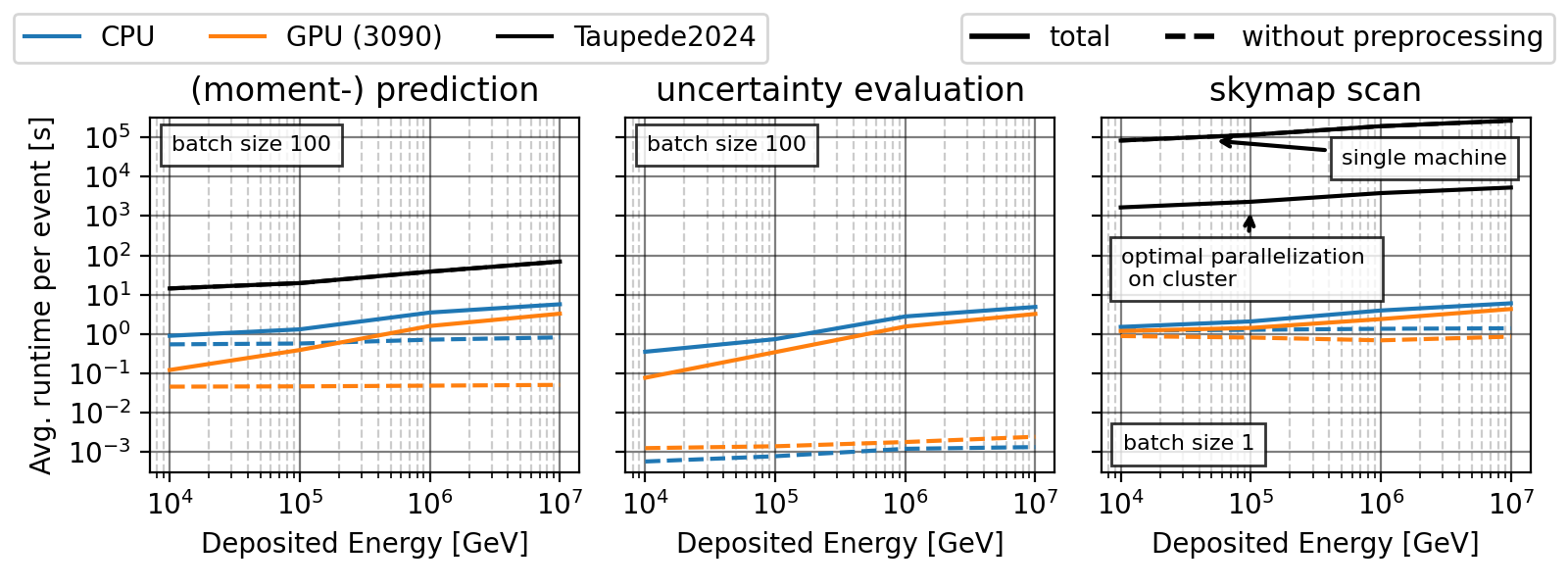} 
    \caption{Timing comparison showers}
    \label{subfig:runtimes_cascades}
    \end{subfigure}
    \begin{subfigure}[t]{0.99\textwidth}
        \centering
     \includegraphics[width=\linewidth]{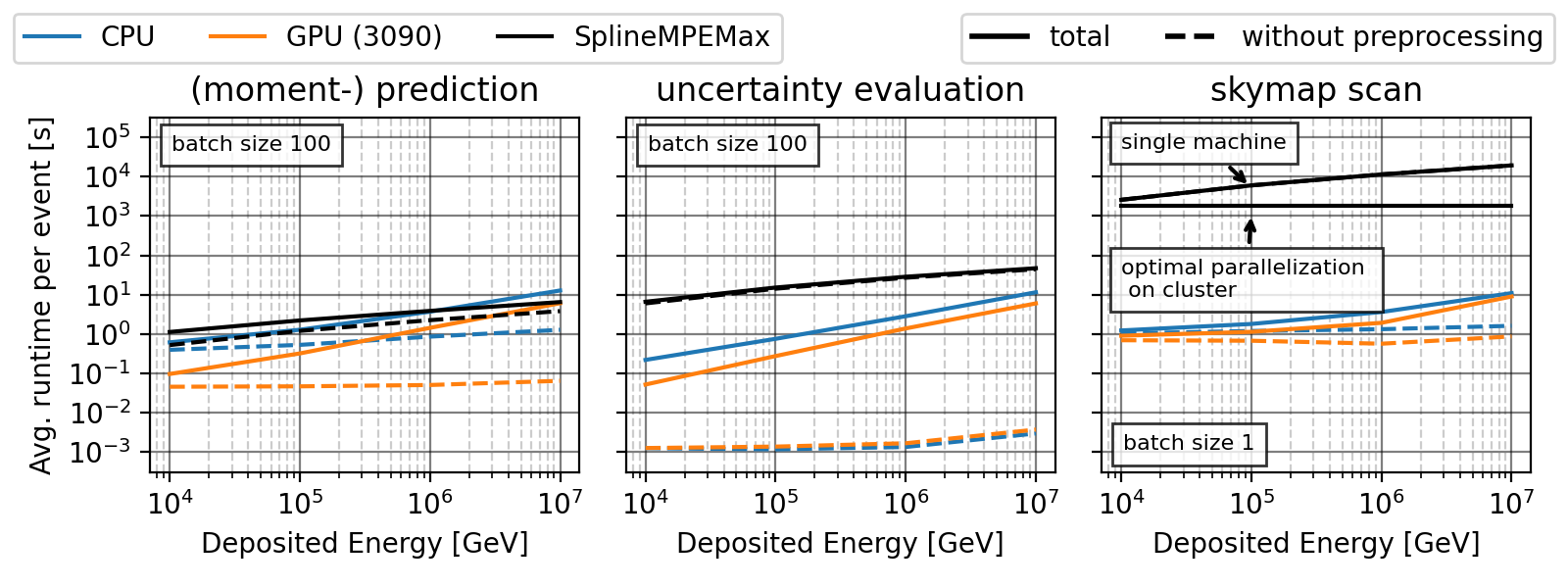} 
        \caption{Timing comparison tracks}
        \label{subfig:runtimes_tracks}
    \end{subfigure}
    \caption{Runtimes of TNF for CPU and GPU in comparison to the B-spline method which runs on CPU. For TNF we allow for 4 CPU cores (CPU) with access to an additional RTX3090 (GPU). The total time is shown in solid, the time without preprocessing and encoding in dashed.}
    \label{fig:runtimes}
\end{figure*}

Figure \ref{fig:runtimes} shows the runtimes of TNF compared to the standard B-spline construction for both showers (Fig. \ref{subfig:runtimes_cascades}) and tracks (Fig. \ref{subfig:runtimes_tracks}). We split runtime calculations in three parts. The first part is moment prediction, which involves sampling from the PDF (by default we sample 10000 times) and calculating the first moments. We obtain the mean and the kappa parameter of the vMF approximation that closest matches the samples via a likelihood fit. The corresponding step for the B-spline is to obtain the full uncertainty contour. The second part is uncertainty evaluation, which just performs a single PDF evaluation --- a quantity that can be important for re-evaluation of uncertainties during an analysis, for example a point source likelihood analysis. The third part is a skymap scan, which involves the skymap creation strategy described in section \ref{section:used_nf_definition}. It starts with 10000 samples, which then guides the creation of the \texttt{HEALPIX} evaluation grid which is on the order of 5000 pixels. For the B-spline skymap-scan, the whole sky is typically scanned with a refinement strategy and parallelized on a cluster. All the checks for TNF are performed for a pure CPU application with up to 4 cores and additionally with a Geforce GTX 3090 for the GPU check. Additionally we separate into total time and time without pre-processing. The preprocessing time is the time it takes to convert IceCube specific data to a format that feeds into the transformer or GNN encoder, and then also pass it through that encoding part, but before we activate the normalizing flow. For the respective B-spline reconstruction, pre-processing involves all reconstructions in the seeding chain that come before it. For moment prediction and uncertainty evaluation, we calculate results in batches and quote the time per batch item. For the skymap scan, we perform it for a single batch item at a time.

For showers, the moment prediction of TNF is faster by at least a factor 10 than the likelihood counterpart. For uncertainty prediction the go-to algorithm for the B-spline is an all-sky profile-likelihood scan, so we can directly compare the skymap-scan running times, since they are roughly equal for the TNF in both cases as they are dominated by preprocessing time. Here we see TNF is faster by several orders of magnitude compared to the B-spline reconstruction, even if we fully utilize a cluster of several 100 compute nodes. For tracks the time spent in moment prediction is roughly on par between TNF and the B-spline based method. However, once we go to uncertainty evaluation, which in the B-spline case relies on a profile-likelihood scan around the best fit point \cite{paraboloid_paper}, we have a speed up by more than an order of magnitude. For the skymap scan, the speed up is again several orders of magnitude as for showers.

We can see that in all cases the proportion of time spent in preprocessing rises with energy. This could potentially be optimized in the future as it currently relies in parts on python implementations that could be ported to C/C++.

\section{Discussion and Outlook}
\label{discussion_section}

In this paper, we combined a spherical normalizing flow with a transformer encoding to learn a model of the posterior distribution of the direction of leptons induced by high energy neutrinos in IceCube - an instance of amortized neural posterior estimation and ELBO-free variational inference. The posterior model takes as input the summary statistics of the raw photon data in each DOM, passes it as per-DOM data tokens through a transformer encoding, aggregates the result in a single vector, and in a final step maps that vector to a normalizing-flow posterior over the direction.
To this end, we developed a novel spherical normalizing flow that combines smooth rational quadratic splines, scale transformations and rotations to learn a flexible conditional posterior that can span several orders of magnitude in extent for different events. For the rational quadratic splines in particular, the $C^2$ smoothness constraint together with a simplistic yet flexible spline-based conditioning were crucial for the stability of the learning process.

We argue that the transformer encoding provides Bayesian inductive bias in two ways: a) it encodes the permutation invariance of the labeling of the input dimensions of the data PDF in Bayes' theorem and b) it provides generalization capabilities when IID data factors (DOMs) are dropped and information is removed. 

After hyperparameter optimization of the transformer architecture in several iterative training sessions, we find the leading model for charged-current electron neutrinos (showers) and charged-current muon neutrinos (tracks) to be architecturally similar up to positional encoding. In both cases, the leading model mostly benefits from nonlinear QKV projection in each attention layer and dual residual pipelining (\enquote{ReSiDual}) compared to our vanilla transformer baseline. For tracks, a specific weighting scheme was additionally important to balance out low- and high energy tracks during the learning process.

We compare the leading models to state-of-the-art B-spline-based likelihood reconstructions. The angular resolution is better than the B-spline-based likelihood reconstructions over the whole tested energy ranges. The gain is between a factor of $1.1$ for 1 TeV throughgoing tracks up to a factor of $1.5$ or $2.5$ for 100 TeV showers or starting tracks, respectively. Processing times are also favorable, in particular for all-sky scans which are sent out in realtime alerts where the processing time can be pushed down from hours to seconds. This is achieved by synergies of the normalizing flow properties together with irregular MOC \texttt{HEALPIX} maps. The coverage for the trained posteriors is typically also better than the B-spline based counterparts, but not perfect. There is slight undercoverage at higher energies for showers, which stems from a lack of training data, and overcoverage at high energies for tracks, which stems from the interaction of the final large learning rate together with the specific flow parameterization. This prevents the neural-network prediction for high energy events with very small contour sizes to settle into local weight minima and effectively broadens their predicted contours compared to what they could be otherwise.
In the near future, we expect these remaining issues to disappear with more training data and some finetuning of either the late-stage learning-rate scheduling or the precise flow parametrization. Furthermore, dedicated training on systematics datasets can incorporate complex systematic uncertainties and should make the application to real data robust. 

As the leading model for showers and tracks shares the same architecture - up to positional encoding - it is conceivable we will soon converge on hyperparameters that also agree in totality. It will then be possible to do a joint training of all neutrino interactions, including tau neutrinos which sit morphologically between showers and tracks. This would harness extra training statistics while being confident that neither individual class loses performance due to the model architecture. 

Looking further into the future, the algorithm is suited to face the challenges of heterogeneous detectors consisting of different types of optical modules like the IceCube Upgrade \cite{upgrade_paper} and planned IceCube-Gen2 \cite{gen2_paper} with minimal changes.

\ack{The IceCube collaboration acknowledges the significant contributions to this manuscript from Thorsten Gl\"usenkamp. The authors gratefully acknowledge the support from the following agencies and institutions:
USA {\textendash} U.S. National Science Foundation-Office of Polar Programs,
U.S. National Science Foundation-Physics Division,
U.S. National Science Foundation-EPSCoR,
U.S. National Science Foundation-Office of Advanced Cyberinfrastructure,
Wisconsin Alumni Research Foundation,
Center for High Throughput Computing (CHTC) at the University of Wisconsin{\textendash}Madison,
Open Science Grid (OSG),
Partnership to Advance Throughput Computing (PATh),
Advanced Cyberinfrastructure Coordination Ecosystem: Services {\&} Support (ACCESS),
Frontera and Ranch computing project at the Texas Advanced Computing Center,
U.S. Department of Energy-National Energy Research Scientific Computing Center,
Particle astrophysics research computing center at the University of Maryland,
Institute for Cyber-Enabled Research at Michigan State University,
Astroparticle physics computational facility at Marquette University,
NVIDIA Corporation,
and Google Cloud Platform;
Belgium {\textendash} Funds for Scientific Research (FRS-FNRS and FWO),
FWO Odysseus and Big Science programmes,
and Belgian Federal Science Policy Office (Belspo);
Germany {\textendash} Bundesministerium f{\"u}r Forschung, Technologie und Raumfahrt (BMFTR),
Deutsche Forschungsgemeinschaft (DFG),
Helmholtz Alliance for Astroparticle Physics (HAP),
Initiative and Networking Fund of the Helmholtz Association,
Deutsches Elektronen Synchrotron (DESY),
and High Performance Computing cluster of the RWTH Aachen;
Sweden {\textendash} Swedish Research Council,
Swedish Polar Research Secretariat,
Swedish National Infrastructure for Computing (SNIC),
and Knut and Alice Wallenberg Foundation;
European Union {\textendash} EGI Advanced Computing for research;
Australia {\textendash} Australian Research Council;
Canada {\textendash} Natural Sciences and Engineering Research Council of Canada,
Calcul Qu{\'e}bec, Compute Ontario, Canada Foundation for Innovation, WestGrid, and Digital Research Alliance of Canada;
Denmark {\textendash} Villum Fonden, Carlsberg Foundation, and European Commission;
New Zealand {\textendash} Marsden Fund;
Japan {\textendash} Japan Society for Promotion of Science (JSPS)
and Institute for Global Prominent Research (IGPR) of Chiba University;
Korea {\textendash} National Research Foundation of Korea (NRF);
Switzerland {\textendash} Swiss National Science Foundation (SNSF).}

\appendix          

\section{Example events}
\label{appendix:example_events}

\begin{figure*}
    \centering
    \begin{subfigure}[t]{0.95\textwidth}
   \centering
    \includegraphics[width=\linewidth]{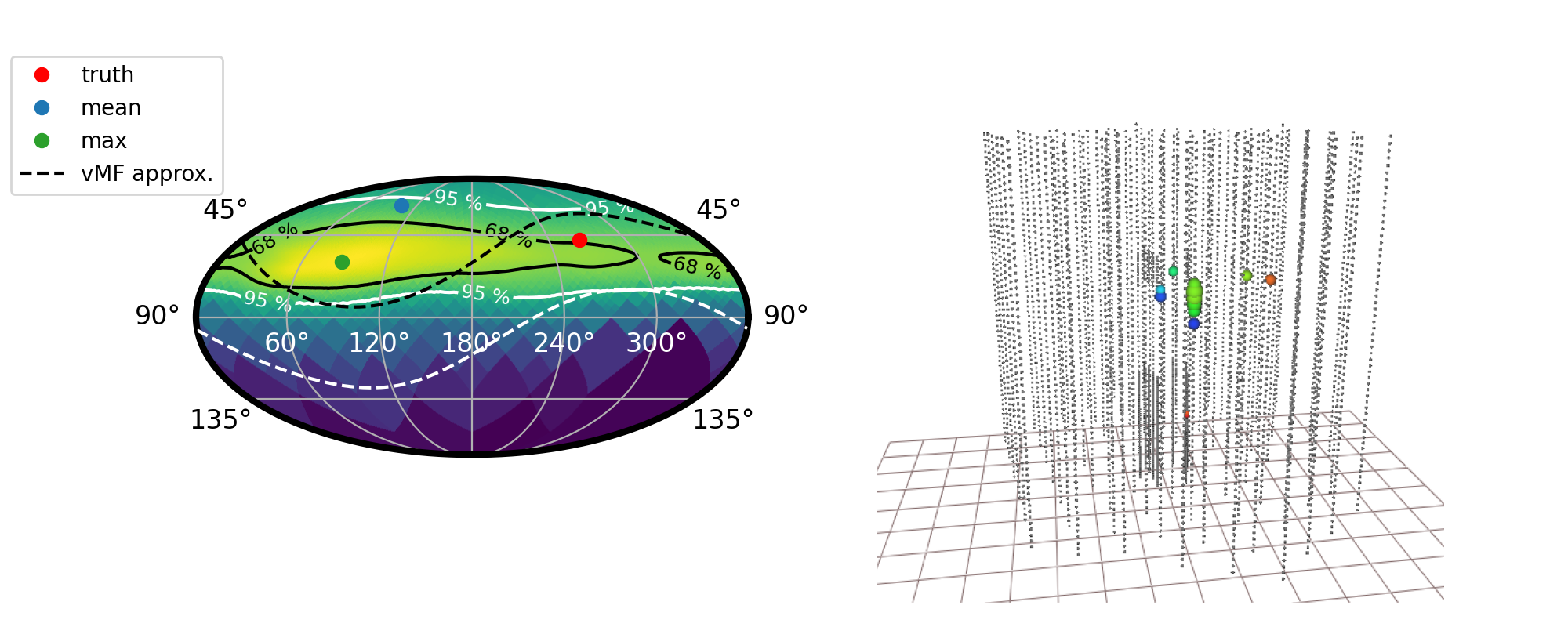} 
    \caption{A low-energy shower that is mostly visible in a single string.}
    \end{subfigure}
    \begin{subfigure}[t]{0.95\textwidth}
   \centering
    \includegraphics[width=\linewidth]{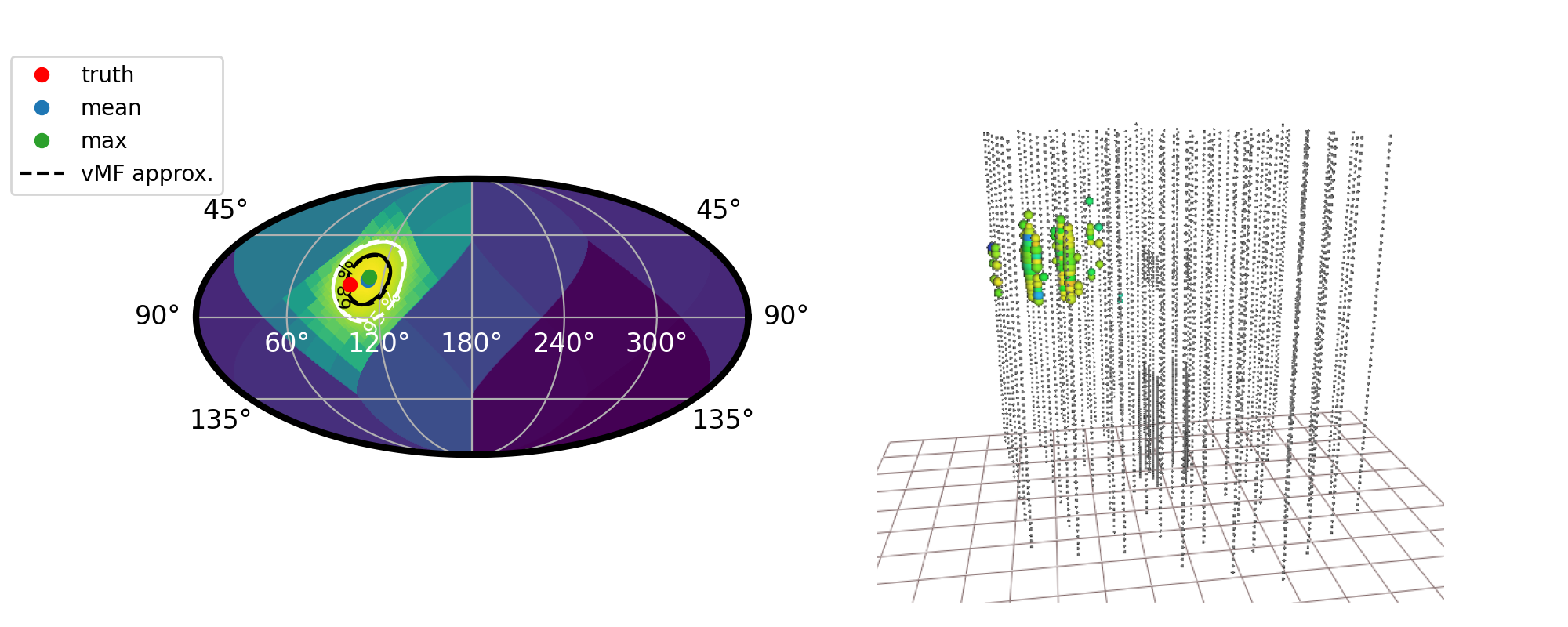} 
    \caption{A shower which is visible in several neighboring strings.}
    \end{subfigure}
    \begin{subfigure}[t]{0.95\textwidth}
   \centering
    \includegraphics[width=\linewidth]{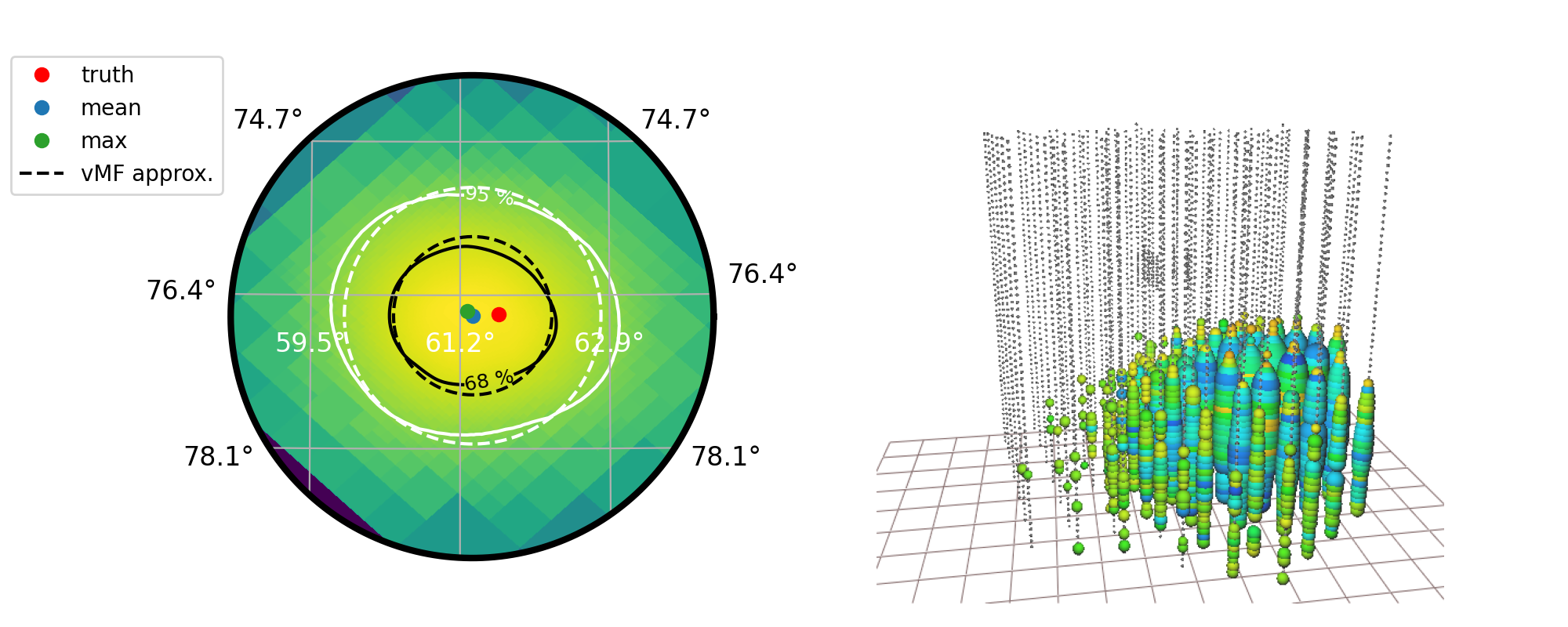} 
    \caption{A multi-PeV shower in the clear ice.}
    \end{subfigure}
    \caption{Contours (left) and event view (right) for example shower events. The contours are indicated in solid for $68 \%$ (black) and $95 \%$ (white) contained probability mass. The corresponding contours of the von-Mises Fisher approximation, i.e. the second moment, are shown with dashed lines. The colors in the event view indicate time, with red being early and blue being late.}
    \label{fig:example_events_showers}
\end{figure*}

At low energies, showers are often single-string dominated, which introduces azimuthal degeneracy into the posterior (Fig. \ref{fig:example_events_showers} a). At higher energies, this degeneracy is typically broken and posteriors are more Gaussian (Fig. \ref{fig:example_events_showers} b)+c)).

\begin{figure*}
    \centering
    \begin{subfigure}[t]{0.95\textwidth}
   \centering
    \includegraphics[width=\linewidth]{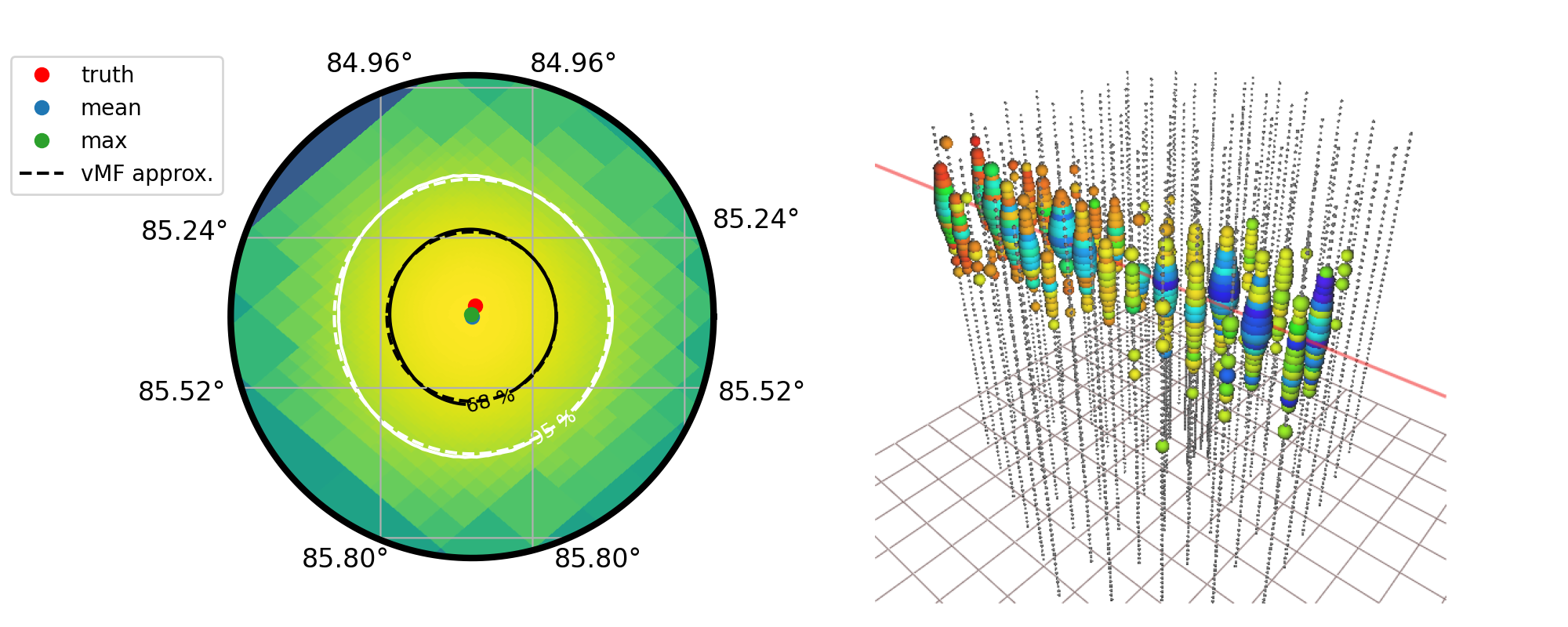} 
    \caption{Bright horizontal track.}
    \end{subfigure}
    \begin{subfigure}[t]{0.95\textwidth}
   \centering
    \includegraphics[width=\linewidth]{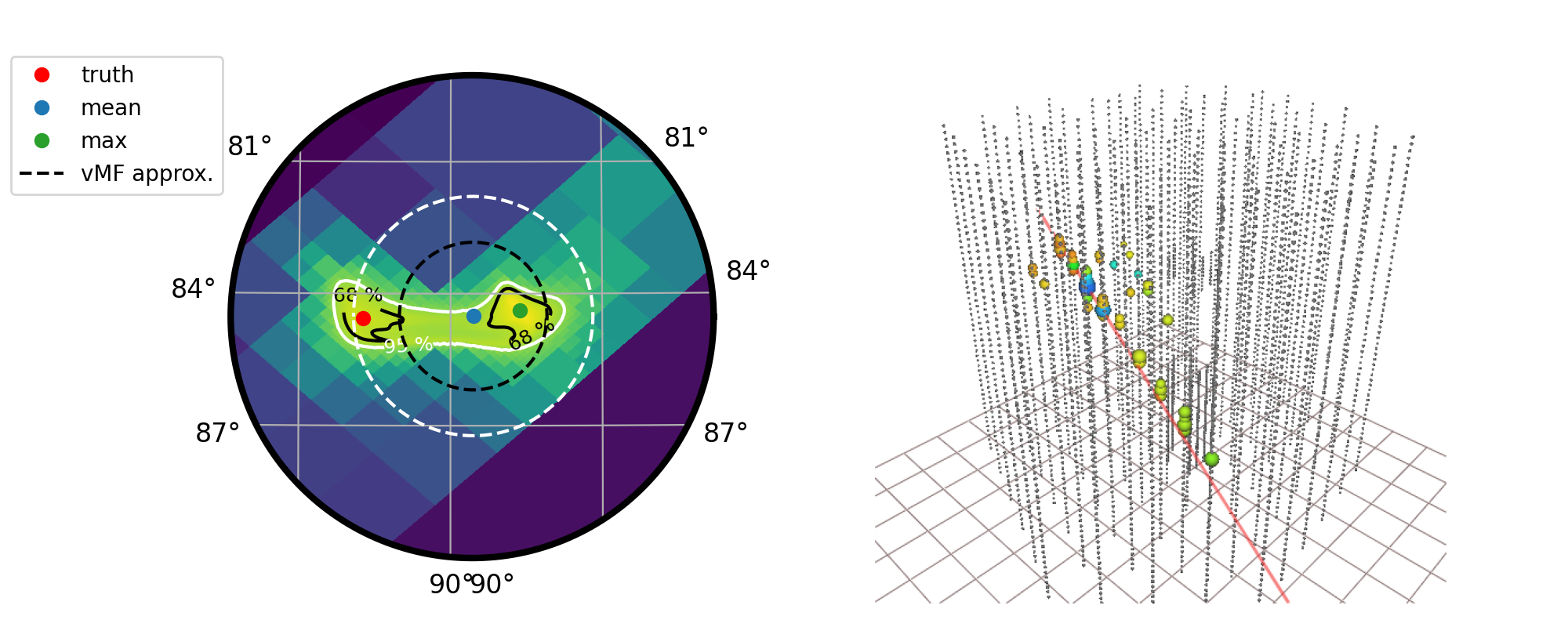} 
    \caption{Muon track along string lines.}
    \end{subfigure}
    \begin{subfigure}[t]{0.95\textwidth}
   \centering
    \includegraphics[width=\linewidth]{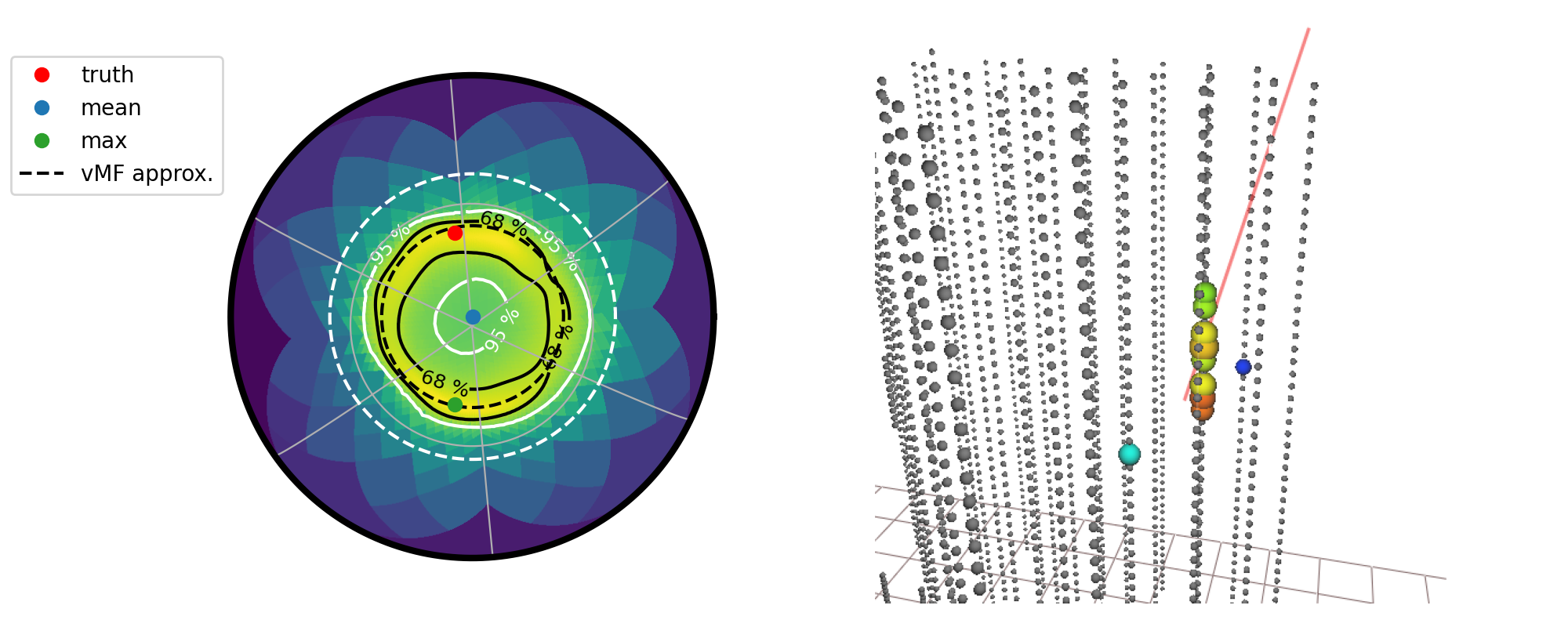} 
    \caption{Low energy contained upgoing track along a single string line.}
    \end{subfigure}
    \caption{Contours (left) and event view (right) for example track events. The contours are indicated in solid for $68 \%$ (black) and $95 \%$ (white) contained probability mass. The corresponding contours of the von-Mises Fisher approximation, i.e. the second moment, are shown with dashed lines. The colors in the event view indicate time, with red being early and blue being late.}
    \label{fig:example_events_track}
\end{figure*}

Most track events are visible in multiple strings that are not lying within a plane and have fairly Gaussian posteriors, as in the high-energy example shown in Fig. \ref{fig:example_events_track} a). However, due to the geometry there are outliers that are co-aligned with symmetry axes. Fig. \ref{fig:example_events_track} b) shows an event that directly passes along string lines that all lie within the same plane, which leads to a slight bimodal degerenacy in azimuth. Fig. \ref{fig:example_events_track} c) shows a low-energy contained muon that passes upwards close to a single string whose posterior is ring-like.

\section{Smooth rational-quadratic spline derivation}
\label{appendix:smooth_spline_derivation}

\begin{figure*}
    \begin{subfigure}[t]{0.24\textwidth}
    \centering
    \includegraphics[width=\linewidth]{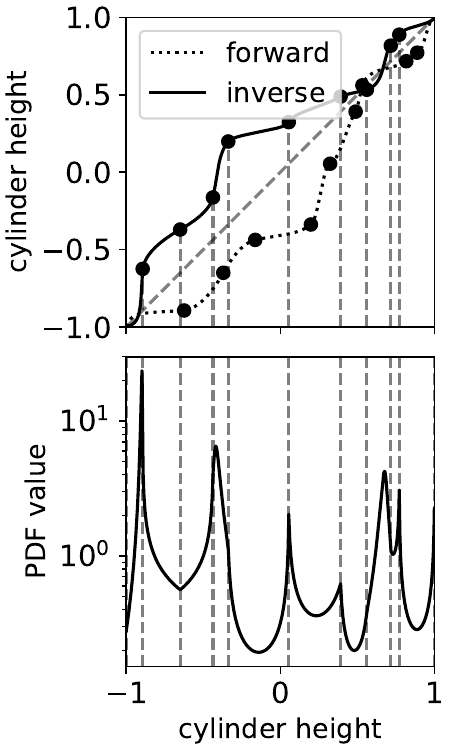} 
    \caption{Unconstrained rational quadratic splines.}
    \label{subfig:unconstrained_splines}
    \end{subfigure}
    \begin{subfigure}[t]{0.24\textwidth}
        \centering
         \includegraphics[width=\linewidth]{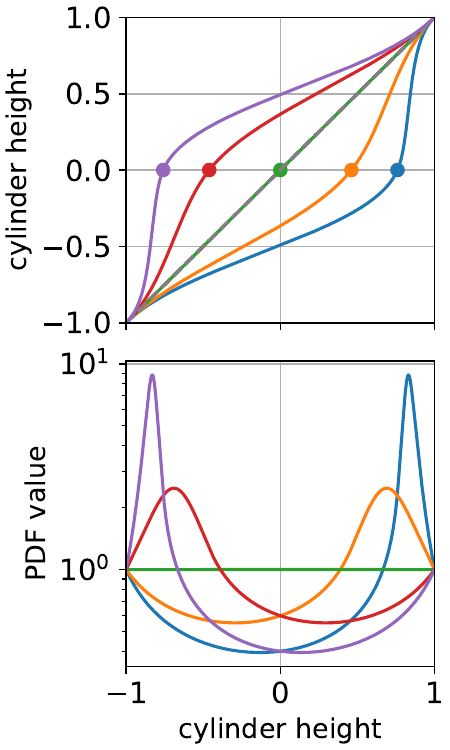} 
        \caption{Smooth splines with two knot intervals on cylinder height.}
        \label{subfig:constrained_splines_height_1}
    \end{subfigure}
    \begin{subfigure}[t]{0.24\textwidth}
        \centering
     \includegraphics[width=\linewidth]{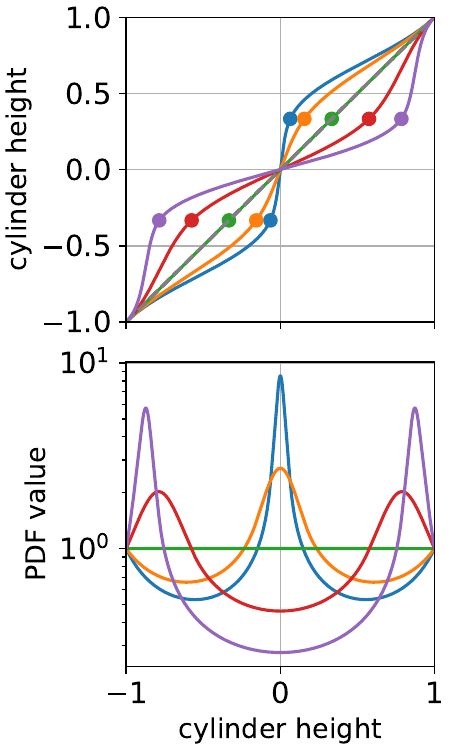} 
        \caption{Smoth splines with three knot intervals on cylinder height.}
        \label{subfig:constrained_splines_height_2}
    \end{subfigure}
    \begin{subfigure}[t]{0.24\textwidth}
        \centering
     \includegraphics[width=\linewidth]{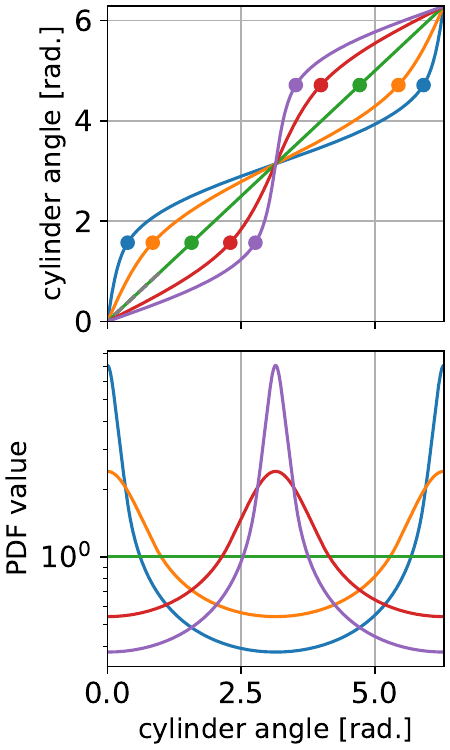} 
        \caption{Smooth periodic splines  with two knot intervals. One interval wraps the boundary.}
        \label{subfig:periodic_shifted_splines}
    \end{subfigure}
    \caption{Depictions of rational quadratic splines with various constraints. Knots are shown as circles. In general the upper plot shows the inverse spline function and the corresponding normalizing flow PDF that follows from eq. \ref{eq:change_of_variable} is shown in the lower part. The exception is a) which also depicts the forward transformation. All colored functions denote a real-1-parameter spline parametrization, $\mathrm{rqs}_I(z)_a$ for the cylinder height $z$ and $\mathrm{rqs}_A(\phi)_a$ for the periodic angle $\phi$, where the parameter $a$ is a left-over parametric degree of freedom that varies from negative values (i.e. blue) to positive ones (i.e. purple). Setting $a=0$ equals an identity mapping (green).}
    \label{fig:spline_visualizations}
\end{figure*}

The forward function $f(z)$ represented as a monotonic rational-quadratic spline is defined per knot segment $k$ between knot points $k$ and $k+1$ and given by \cite{neural_spline_flows}
\begin{equation}
f[\xi(z)]=y_k+ \frac{(y_{k+1}-y_k)[s_k \cdot \xi^2+\delta_k \cdot \xi (1-\xi)]}{s_k+[\delta_{k+1}+\delta_k-2s_k] \cdot \xi(1-\xi)},
\end{equation}
where $x_k$ and $x_{k+1}$ are the knot x-positions, $y_k$ and $y_{k+1}$ the knot y-positions, $\delta_k$ and $\delta_{k+1}$ the derivatives at the knots and $\xi(z)=\frac{z-x_k}{x_{k+1}-x_k} \in [0,1]$ a relative coordinate between the knots.
We also define the width $w_k=x_{k+1}-x_k$ and height $h_k=y_{k+1}-y_k$ of segment k and $s_k=\frac{h_k}{w_k}$ the linear slope between the knots in the segment.
If the derivatives  $\delta_k$ at the knot points are positive the resulting function is monotonically increasing \cite{monotonic_spline_defs}. The exact derivative at any point between the knots is given by \cite{neural_spline_flows}
\begin{equation}
\frac{df}{dz}[\xi(z)]=\frac{s_k^2[\delta_{k+1}\xi^2+2 s_k \cdot \xi(1-\xi)+\delta_k(1-\xi)^2]}{[s_k+[\delta_{k+1}+\delta_k-2s_k] \cdot \xi(1-\xi)]^2}
\end{equation} and is by construction positive due to the underlying monotonicity.
Additionally, the exact analytic inverses of both these functions exist \cite{neural_spline_flows} but we omit to write them down here for brevity. Using these definitions and multiple such segments we can define a flexible normalizing flow on an interval using eq. \ref{eq:change_of_variable}.

In this default parametrization, the derivatives at the knot points $\delta_k$ are fitted and can be small or large independent of the knot positions, which in general results in the function only being $C^1$-smooth. The jumps in the second derivative lead to sharp features in the PDF via eq. \ref{eq:change_of_variable} (see Fig. \ref{subfig:unconstrained_splines}). 

In the following, we impose that the second derivative at the knot points should be equal between two consecutive knot segments which leads to $C^2$ smoothness. Forming the second derivative of segment $k$ and evaluating it at its upper knot point $k+1$ yields
\begin{equation}
\label{eq:smoothness_upperbound_lowersegment}
\frac{df_{k}}{d^2z}[z=x_{k+1}]  =\frac{2 \left[\delta_{k+1} (\delta_k + \delta_{k+1})-s_k(\delta_{k+1}+s_k) \right]  }{s_k w_k}.
\end{equation}
Forming the second derivative of segment $k+1$ and evaluating it at its lower knot point $k+1$ yields
\begin{eqnarray}
\label{eq:smoothness_lowerbound_uppersegment}
\frac{df_{k+1}}{d^2z}[z=x_{k+1}]=\frac{2 s_{k+1} (\delta_{k+1}+s_{k+1}) }{s_{k+1} w_{k+1}} 
-\frac{ 2\delta_{k+1}(\delta_{k+1}+\delta_{k+2}) 
 }{s_{k+1} w_{k+1}}.
\end{eqnarray}
The term $w_k$ denotes the width of segment $k$, $w_k=x_{k+1}-x_k$. The two equations can be set equal to each other to impose continuity on the second derivative which constrains the inner first derivative parameter $\delta_k+1$, i.e.
\begin{equation}
\label{eq:smoothness_constraint}
\frac{df_{k}}{d^2z}[z=x_{k+1}]=\frac{df_{k+1}}{d^2z}[z=x_{k+1}].
\end{equation}
For $N$ segments, we have $N-1$ such constraints which are quadratic in the respective inner derivatives, and the equations are coupled since the derivative $\delta_{k+2}$ or $\delta_k$ in the constraint \ref{eq:smoothness_constraint} also appear in constraints involving neighboring segments $k+2$ or $k-1$.
For N such segments chained after each other, we have in total $4N$ knot position values $x_k, x_{k+1}, y_{k}, y_{k+1}$ and $2N$ derivatives at the segment boundaries. We also have $2(N-1)$ constraints for the segments to be continuous in function value (for both $x$ and $y$ coordinates to agree) and another $N-1$ constraints to be continuous to 1st order. This leads to $6N-3(N-1)=3N+3$ free parameters in the default parametrization. If we fix the outer knot positions of the left-most and right-most segment, which on the cylinder and the circle has to be done, we can subtract $4$ again to obtain $3N-1$ free parameters. We have not imposed the $2nd$-order constraint (eq. \ref{eq:smoothness_constraint}) yet. In the following we differentiate two cases: application on the interval $z \in [-1,1]$ and on the circle $\phi$.
\subsection{Interval on cylinder height $z \in [-1,1]$}
For the normalizing flow we describe in section \ref{section:used_nf_definition} the sphere is deformed to the cylinder, and for non-singular behavior the spline flow on the cylinder height $z$ should be equal to an identity mapping at $z=-1$ and $z=1$, which correspond to the poles in the spherical representation. This can be achieved by enforcing the two outer-most knot derivatives to be equal to 1. If we now additionally apply a 2nd-order smoothness constraint using eqs. \ref{eq:smoothness_constraint} on all inner $N-1$ inner knots connecting segments, we have overall 
\begin{eqnarray}
\label{eq:generic_constraint}
p_{\mathrm{free}}=3N-1-(N-1)-2=2N-2
\end{eqnarray} free parameters for the total spline flow. Since the constraints in eq. \ref{eq:smoothness_constraint} lead to $N-1$ coupled quadratic equations in $N-1$ variables, they are not trivial to be solved analytically in all generality. In the following we solve them for $N=2$ and for $N=3$ with an extra symmetry constraint.
\paragraph{Case $N=2$:}
For 2 segments we have a single quadratic equation in the intermediate derivative $\delta_1$. The single usable solution is given by 
\begin{equation}
\delta_{1}=-\frac{p}{2}+\sqrt{\left(\frac{p^2}{4}-q \right)}
\end{equation}
with
\begin{eqnarray}
p=  \frac{h_0 \cdot (s_1-1) + h_1\cdot (s_0-1)}{h_0+h_1}, \\
q=-h_0 h_1 \left(\frac{h_0}{ (h_0 + h_1) w_1^2} +  \frac{h_1}{ (h_0 + h_1) w_0^2}\right).
\end{eqnarray}
Since $q$ is negative, the second solution for $\delta_{1}$ is always negative which violates the increasing monotonicity of the resulting spline function and it can therefore be omitted. We have 2 free parameters left to define the shape (eq. \ref{eq:generic_constraint}), which we choose to be the intermediate knot $x$ and $y$ values, $x_1$ and $y_1$. Since the lower and upper knot positions are fixed to be $x_0=-1$ and $x_2=1$ on the cylinder height, they are contained in the width and height parameters. The width parameters, for example, are defined as $w_0=x_1-x_0=x_1-(-1)=x_1+1$ and $w_1=x_2-x_1=1-x_1$. For numerical stability, we found it useful to use a 1-parameter parametrization, where $w_0=w_1=1$ and the height is parametrized by a real parameter $a$ in log space such that $h_0=h_1=1$ if $a=0$. This allows the transformation to have a smooth 1-parameter set of curves that defines the normalizing flow (see Fig. \ref{subfig:constrained_splines_height_1}) and results in an identity for $a=0$.

\paragraph{Case $N=3$, symmetric:}
For $N=3$ segments there are four free parameters to fit according to eq. \ref{eq:generic_constraint}. In addition, from  the two 2nd-order constraints we have two quadratic coupled equations in 2 variables, the two intermediate knot derivatives $\delta_1$ and $\delta_2$. In general, those two coupled equations are hard to solve analytically. However, we can impose two extra symmetry conditions that the first and third segment dimensions are equal, i.e. $w_0=w_2$ and $h_0=h_2$. In this case, there are four solutions for the two derivatives. Only two of those solutions are positive for both derivatives at the same time, and only one of those two solutions has the same result for the two derivatives, which is what we expect from symmetry considerations. This unique solution for the two coupled quadratic equations turns out to be
\begin{equation}
\delta_{1}=\delta_{2}=-\frac{p}{2}+\sqrt{\left(\frac{p^2}{4}-q \right)}
\end{equation}
with
\begin{eqnarray}
p &=&  \frac{h_1 (w_0 w_1 - h_0(w_0+w_1))}{(2 h_0+h_1)w_0 w_1}, \\
q&=&-h_0 h_1 \frac{h_1 w_0^2 + h_0 w_1^2}{(2h_0 + h_1) w_0^2 w_1^2 } .
\end{eqnarray}
In addition, the symmetry constraint eliminates two from the four overall free parameters, and we end up with two free parameters to determine the shape of the flow function, which can for example be $w_0$ and $h_0$. Again, it is useful to fix the width $w_0$, this time to a third of the overall interval via $w_0=2.0/3.0$, and model the height $h_0$ as a relative real 1-parameter curve $a$ in log space such that $a=0$ leads to the identity mapping, as depicted in Fig. \ref{subfig:constrained_splines_height_2}. In contrast to the 2-interval solution (Fig. \ref{subfig:constrained_splines_height_1}) such transformations lead to symmetric PDFs on $z$.

\subsection{Interval on cylinder angle $\phi \in [0, 2\pi]$}
The cylinder angle $\phi$ is periodic. Instead of fixing the two boundary derivatives to $1$ as was done for the height $z$, there is only the single constraint that the derivatives have to agree. This results in one more free parameter. On the other hand, also the second derivative must agree on the boundaries, a constraint that did not exist for the height, which results in one less free parameter. In total the number of free parameters therefore is similar to the situation for cylinder height $z$ and equals also $2N-2$ with slightly different constraint equations.  For $N=2$, which corresponds to 3 knot positions, we can solve for the derivative at the inner knot $\delta_1$ and one of the outer knots, i.e. $\delta_0$. Using the first constraint the two outer knot derivatives are the same, so $\delta_0=\delta_2$. Using the two second derivative constraints at the inner and one outer knot we obtain two solutions, of which only one is always positive. The two derivative solutions for $\delta_0$ and $\delta_1$ are always equal for the unique positive solution, similar to the situation with three segments and 4 knots on the cylinder height, but with slightly different dependency on the widths and heights of the segments. Since $\delta_0$ = $\delta_2$, the result for all derivatives is
\begin{equation}
    \delta_0 = \delta_1 = \delta_2 \equiv \delta = -\frac{p}{2}+\sqrt{\left(\frac{p^2}{4}-q \right)}
\end{equation}
with
\begin{eqnarray}
p=  -\frac{h_0 h_1 (w_0+w_1)}{2 (h_0+h_1) w_0 w_1}, \\
q= -\frac{h_0 h_1 (h_1 w_0^2+h_0 w_1^2)}{2 (h_0+h_1) w_0^2 w_1^2}.
\end{eqnarray}
We fix $w_0=w_1=\pi$ and define the height again as a relative 1-parameter curve in log-space parametrized by $a$, where $a=0$ leads to the identity. The solution turns out not to be symmetric over the interval $[0,2\pi]$. We can however introduce an extra adaptive \enquote{shift} $s$ to effectively turn the transformation into a symmetric one where the peak does not move. The total transformation is
\begin{equation}
    \mathrm{rqs}_A(\phi)= \{f_{\mathrm{base}}(\phi)+s\} \bmod 2\pi,
\end{equation}
 with shift $s$,
\begin{eqnarray}
    s=2 \pi - h_0 
    -\frac{h_1 (\frac{1}{2} w_0 -\pi) (h_0 (\frac{1}{2} w_0 -\pi)-\delta  w_0 (\frac{1}{2} w_0+w_1-\pi))} {(h_0 w_1^2+2 (\frac{1}{2} w_0-\pi) (h_0-\delta  w_0) (\frac{1}{2} w_0+w_1 -\pi))},
\end{eqnarray}
where $f_{\mathrm{base}}(\phi)$ is the standard smooth rational quadratic spline without shift. Note that we have not added an extra knot, so there are in total still three knots or two intervals at use. One of the intervals however wraps around the $0/2\pi$ boundary with the included shift. We use this form in all experiments, as it is more stable for optimization due the independence of the peak position with the parametrized degree of freedom. The final transformation for the angle including this shift is illustrated in Fig. \ref{subfig:periodic_shifted_splines}, which shows the 1-real-parameter parametrization $\mathrm{rqs}_A(\phi)_a$  of this transformation in dependence of the free degree of freedom $a$. 

\begin{figure}
    \centering
     \includegraphics[width=0.6\linewidth]{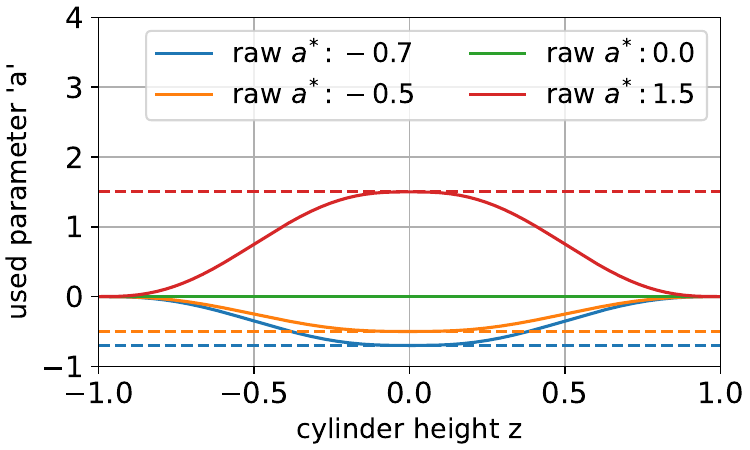} 
    \caption{Conditioning function $a=f_{a^{*}}(z)$ on the cylinder height $z$ to produce a regularized parameter $a$ to be used for the azimuthal angle flow function, $\mathrm{rqs}_A(\phi)_a$, depicted in Fig. \ref{subfig:periodic_shifted_splines}. At the poles ($z=-1$ and $z=1$) the output is always $a=f_{a^{*}}(z=-1/z=1)=0$ which enforces an identity mapping as $\mathrm{rqs}_A(\phi|z)_{a=0}=\phi$. The horizontal dashed lines correspond to specific values of $a^{*}$.}
    \label{fig:simple_spline_nonmlp_conditioning}
\end{figure}

In \cite{tori_and_spheres} the azimuthal transformation is defined in dependence on the cylinder height $z$, by conditioning on $z$ using an MLP. We found this can lead to numerical issues in the polar regions. In order to mitigate this, but still have some form of conditioning scheme implemented, we use a simple spline-based non-MLP conditioning that blends towards an identity mapping at the polar regions, as depicted in Fig. \ref{fig:simple_spline_nonmlp_conditioning}. 
The function is just parametrized by single parameter $a^{*}$, its overall normalization, and towards the poles, i.e. in the cylinder representation at $z=-1$ and $z=1$, always tends towards $0$ by construction. The parameter $a^{*}$ acts as a surrogate parameter for the actual parameter $a=f_{a^{*}}(z)$ which is then plugged into the parametrization in Fig. \ref{subfig:periodic_shifted_splines}. The end result is the angular flow transformation $\mathrm{rqs}_A(\phi|z)_{a^{*}}$ that depends on the cylinder height $z$ via the single parameter $a^{*}$. At the poles the mapping $a=f_{a^{*}}(z=-1/z=1)=0$ and the corresponding polar angle transformation $\mathrm{rqs}_A(\phi)_{a=0}=\phi$ is the identity (green curve in Fig. \ref{subfig:periodic_shifted_splines}). This scheme is quite effective in turning an extra $\phi$ transformation on or off if desired and numerically stable since it smoothly blends to an identity at the poles irrespective of $a^*$. The explicit form of $f_{a^{*}}(z)$ is a fifth-order polynomial spline in two intervals which was derived enforcing the necessary boundary conditions,
\[
f_{a^*}(z)=
\left\{
\begin{array}{ll}
a^{*}\,(6z^5+15z^4+10z^3+1), & z\le 0,\\
a^{*}\,(-6z^5+15z^4-10z^3+1), & z>0~,
\end{array}
\right.
\]
i.e. it contains appropriate smoothness constraints at $z=0$ and fulfills $f_{a^{*}}(z=-1/z=1)=0$.

\section{Rotation Parametrizations}
\label{appendix:rotation_parametrizations}
The following are parametrizations of the rotation matrix $\mathrm{R}$ discussed in section \ref{section:used_nf_definition}. The baseline rotation parametrization is based on householder reflections \cite{householder_trafos}. An individual householder reflection in 3-dimensional space can be defined by a 3-dimensional vector $v$ as 
\begin{equation}
    \mathrm{R}_{hh,i}=\mathbb{1}-2 v \times v^{T}/|v|^2.
\end{equation}
For the full rotation we chain three such transformations as
\begin{equation}
   \label{eq:rot_matrix_hh}
    \mathrm{R}_{hh}=\mathrm{R}_{hh,1} \circ \mathrm{R}_{hh, 2} \circ \mathrm{R}_{hh,3},
\end{equation}
which has in total $9$ real parameters and technically ends up in an improper rotation matrix with determinant $-1$, since each householder reflection has negative parity. For pure von-Mises-Fisher distributions we use another representation which reads

\begin{equation}
 \label{eq:rot_matrix_original_fisher}
\mathrm{R}_F=
\begin{pmatrix}
1 - \mu_1^2/(1 + \mu_3) & -\mu_1\mu_2/(1 + \mu_3) & \mu_1 \\
-\mu_1\mu_2/(1 + \mu_3) & 1 - \mu_2^2/(1 + \mu_3) & \mu_2 \\
-\mu_1 & -\mu_2 & \mu_3
\end{pmatrix}
\end{equation}
and is parametrized by a normalized mean vector $\vec{\mu}$. As briefly mentioned in section \ref{section:used_nf_definition}, we can describe a von-Mises-Fisher distribution via the change-of-variables formula which makes it a normalizing flow. In particular, the determinant of the Jacobian of the inverse of the block transformation in eq. \ref{eq:block_definition} where the smooth spline transformations are identities, i.e. $\mathrm{rqs}_\phi^{-1}(x)=x$ and $\mathrm{rqs}_I(x)=x$, yields
\begin{eqnarray}
    \mathrm{det}\left(\mathcal{J}_F^{-1}\right) &=&\left(\frac{d}{dz}\mathrm{F}^{-1}\right)[R_F^{-1} \cdot \vec{x}]_z\\ 
    &=& \frac{\kappa}{\mathrm{sinh}(\kappa)}e^{\kappa \cdot  [R_F^{-1} \cdot \vec{x}]_z} \\
    &=& \frac{\kappa}{\mathrm{sinh}(\kappa)}e^{\kappa \cdot  \vec{\mu} \cdot \vec{x}}\label{eq:det_of_fisher}.
\end{eqnarray}
since the Jacobian determinant of the rotation matrix and the cylinder transformations from eq. \ref{eq:block_definition} are all equal to one, and the only term remaining after some algebra is $\frac{d}{dz}\mathrm{F}^{-1}(z)$, evaluated at the z-component of $R_F^{-1} \cdot \vec{x}$. Using eq. \ref{eq:change_of_variable} we can combine this density update with a flat base distribution, $p_0=1/{4\pi}$, to obtain the known von-Mises-Fisher density. We use this rotation parametrization only to describe the standard vMF density. In practice we found it to be unstable when used in longer normalizing-flow chains.

\section{Additional test results and model hyperparameter settings}
\label{appendix:hyperparameters}

\begin{figure*}
    \centering
   \begin{subfigure}[b]{0.99\textwidth}
   \centering
    \includegraphics[width=0.99\textwidth]{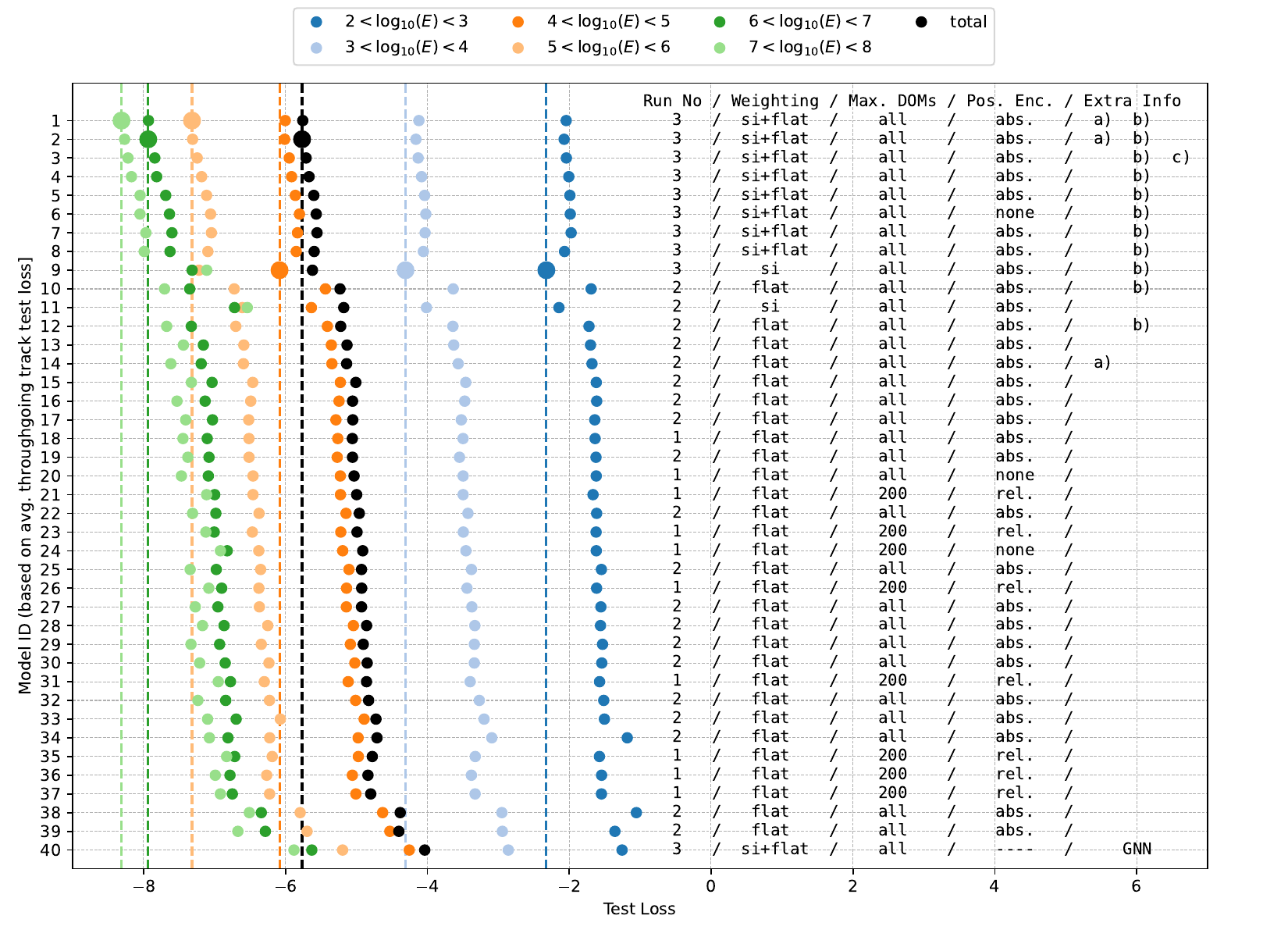} 
    \end{subfigure}
    \caption{Test results for different hyperparameters for starting tracks. Models are sorted by average total test loss from throughgoing tracks as shown in Fig. \ref{subfig:testresults_throughgoing} which defines the \enquote{model ID}. The best test loss per energy range is highlighted by a larger marker and a corresponding vertical cashed line to compare to other models. Options a), b) and c) (\enquote{Extra Info}) are described in the text in section \ref{results_section}. A detailed description of all parameters of each model is given in table \ref{tab:track_options}.}
    \label{fig:test_results_starting}
\end{figure*}

This section lists the test results for starting tracks (Fig. \ref{fig:test_results_starting}). It also contains a summary of the hyperparameter settings used for the different models in track training (table \ref{tab:track_options}) and shower training (table \ref{tab:cascade_options}). The specific settings for each model are indexed by column, where the setting description to each column is given in table \ref{tab:numeric_defs}. Other abbreviations used in describing the parameter settings are given in table \ref{tab:opt_abbreviations}.

\begin{table*}[ht]
  \centering
  \resizebox{\textwidth}{!}{%
  \begin{tabular}{cp{1.5em}p{1.5em}p{1.5em}p{1.5em}p{1.5em}p{1.5em}p{1.5em}p{1.5em}p{1.5em}p{1.5em}p{1.5em}p{1.5em}p{1.5em}p{1.5em}p{1.5em}p{1.5em}}
    \hline
\multicolumn{17}{c}{\textbf{Track training}} \\
    \hline
& \multicolumn{16}{c}{Numeric options (see tables \ref{tab:numeric_defs} and \ref{tab:opt_abbreviations} for definitions)} \\
\cline{2-17}
    Model ID & 1 & 2 & 3 & 4 & 5 & 6 & 7 & 8 & 9 & 10 & 11 & 12 & 13 & 14 & 15 & 16 \\
    \hline
    1 & both & all & yes & yes & md & no & sin & none & no & 96 & 1 & no & hh & no & no & no \\
    2 & both & all & yes & yes & md & yes & sin & none & no & 96 & 1 & no & hh & no & no & no \\
    3 & both & all & no & yes & c3 & no & sin & none & no & 96 & 1 & no & hh & no & no & no \\
    4 & both & all & no & yes & md & no & sin & none & no & 96 & 1 & no & hh & no & no & no \\
    5 & both & all & no & yes & c1 & no & sin & none & no & 96 & 1 & no & hh & no & no & no \\
    6 & both & all & no & yes & md & no & none & none & no & 96 & 1 & no & hh & no & no & no \\
    7 & both & all & no & yes & c2 & no & sin & none & no & 96 & 1 & no & hh & no & no & no \\
    8 & both & all & no & yes & md & no & sin & none & no & 96 & 1 & no & hh & no & no & no \\
    9 & si & all & no & yes & md & no & sin & none & no & 96 & 1 & no & hh & no & no & no \\
    10 & flat & all & no & yes & md & no & sin & none & yes & 96 & 1 & no & hh & no & no & no \\
    11 & si & all & no & no & md & no & sin & none & yes & 96 & 1 & no & hh & no & no & no \\
    12 & flat & all & no & yes & md & no & sin & none & yes & 96 & 1 & yes & hh & no & no & no \\
    13 & flat & all & no & no & md & no & sin & none & yes & 192 & 4 & no & hh & no & no & no \\
    14 & flat & all & yes & no & md & no & sin & none & yes & 96 & 1 & no & hh & no & no & no \\
    15 & flat & all & no & no & md & no & sin & none & yes & 96 & 1 & no & hh & no & no & no \\
    16 & flat & all & no & no & md & no & sin & none & no & 96 & 1 & no & hh & no & no & no \\
    17 & flat & all & no & no & md & no & sin & none & yes & 192 & 1 & no & hh & no & no & no \\
    18 & flat & all & no & no & md & no & sin & none & yes & 96 & 1 & no & hh & no & no & no \\
    19 & flat & all & no & no & md & no & sin & none & yes & 192 & 2 & no & hh & no & no & no \\
    20 & flat & all & no & no & md & no & none & none & yes & 96 & 1 & no & hh & no & no & no \\
    21 & flat & 200 & no & no & md & no & none & $\mathrm{rel}_1$ & yes & 96 & 1 & no & hh & no & no & no \\
    22 & flat & all & no & no & md & no & sin & none & yes & 96 & 1 & no & hh & no & no & no \\
    23 & flat & 200 & no & no & md & no & none & $\mathrm{rel}_4$ & yes & 96 & 1 & no & hh & no & no & no \\
    24 & flat & 200 & no & no & md & no & none & none & yes & 96 & 1 & no & hh & no & no & no \\
    25 & flat & all & no & no & md & no & sin & none & yes & 96 & 1 & no & hh & yes & no & no \\
    26 & flat & 200 & no & no & md & no & none & $\mathrm{rel}_6$ & yes & 96 & 1 & no & hh & no & no & no \\
    27 & flat & all & no & no & md & yes & sin & none & yes & 96 & 1 & no & hh & no & no & no \\
    28 & flat & all & no & no & md & no & sin & none & yes & 96 & 1 & no & hh & no & no & no \\
    29 & flat & all & no & no & md & no & sin & none & yes & 96 & 1 & no & hh & no & yes & no \\
    30 & flat & all & no & no & md & no & sin & none & yes & 96 & 1 & no & hh & no & no & no \\
    31 & flat & 200 & no & no & md & no & none & $\mathrm{rel}_2$ & yes & 96 & 1 & no & hh & no & no & no \\
    32 & flat & all & no & no & md & no & sin & none & yes & 96 & 1 & no & hh & no & no & no \\
    33 & flat & all & no & no & md & no & sin & none & no & 96 & 1 & no & hh & no & no & no \\
    34 & flat & all & no & no & md & no & sin & none & yes & 96 & 1 & no & xyz & no & no & yes \\
    35 & flat & 200 & no & no & md & no & none & $\mathrm{rel}_7$ & yes & 96 & 1 & no & hh & no & no & no \\
    36 & flat & 200 & no & no & md & no & none & $\mathrm{rel}_5$ & yes & 96 & 1 & no & hh & no & no & no \\
    37 & flat & 200 & no & no & md & no & none & $\mathrm{rel}_3$ & yes & 96 & 1 & no & hh & no & no & no \\
    38 & flat & all & no & no & md & no & sin & none & yes & 96 & 1 & no & hh & no & no & yes \\
    39 & flat & all & no & no & md & no & sin & none & yes & 96 & 1 & yes & hh & no & no & no \\
    40 & both & all & - & - & - & - & - & - & no & - & - & - & hh & no & no & no \\
    \hline
  \end{tabular}
  }
   \caption{Options used for the different track training runs. The models are ordered according to their ID which is determined by average total test loss (see Fig. \ref{subfig:testresults_throughgoing}). Explanations for the numeric option mapping are given in table \ref{tab:numeric_defs} and for some abbreviations are given in table \ref{tab:opt_abbreviations}. Model 40 is a GNN and the related transformer-related options are marked with \enquote{-}.}
  \label{tab:track_options}
\end{table*}

\begin{table*}[ht]
  \centering
  \resizebox{\textwidth}{!}{%
  \begin{tabular}{cp{1.5em}p{1.5em}p{1.5em}p{1.5em}p{1.5em}p{1.5em}p{1.5em}p{1.5em}p{1.5em}p{1.5em}p{1.5em}p{1.5em}p{1.5em}p{1.5em}p{1.5em}p{1.5em}}
    \hline
\multicolumn{17}{c}{\textbf{Shower training}} \\
    \hline
& \multicolumn{16}{c}{Numeric options (see tables \ref{tab:numeric_defs} and \ref{tab:opt_abbreviations} for definitions)} \\
\cline{2-17}
    Model ID & 1 & 2 & 3 & 4 & 5 & 6 & 7 & 8 & 9 & 10 & 11 & 12 & 13 & 14 & 15 & 16 \\
    \hline
    1 & flat & all & yes & yes & md & no & none & none & no & 96 & 1 & no & hh & no & no & no \\
    2 & flat & all & yes & yes & md & yes & none & none & no & 96 & 1 & no & hh & no & no & no \\
    3 & flat & 400 & no & yes & md & no & none & none & no & 96 & 1 & no & hh & no & no & no \\
    4 & flat & 200 & yes & yes & md & no & none & $\mathrm{rel}_1$ & no & 96 & 1 & no & hh & no & no & no \\
    5 & flat & all & no & yes & md & no & none & none & no & 96 & 1 & no & hh & no & no & no \\
    6 & flat & 200 & no & no & md & no & none & $\mathrm{rel}_1$ & yes & 96 & 1 & no & hh & no & no & no \\
    7 & flat & 200 & no & no & md & no & none & $\mathrm{rel}_5$ & yes & 96 & 1 & no & hh & no & no & no \\
    8 & flat & 200 & no & yes & c3 & no & none & $\mathrm{rel}_1$ & no & 96 & 1 & no & hh & no & no & no \\
    9 & flat & 200 & no & no & md & no & none & $\mathrm{rel}_4$ & yes & 96 & 1 & no & hh & no & no & no \\
    10 & flat & 200 & no & yes & c1 & no & none & $\mathrm{rel}_1$ & no & 96 & 1 & no & hh & no & no & no \\
    11 & flat & 200 & no & no & md & no & none & $\mathrm{rel}_6$ & yes & 96 & 1 & no & hh & no & no & no \\
    12 & flat & 200 & no & yes & md & no & none & none & no & 96 & 1 & no & hh & no & no & no \\
    13 & flat & 200 & no & yes & c2 & no & none & $\mathrm{rel}_1$ & no & 96 & 1 & no & hh & no & no & no \\
    14 & flat & 200 & no & no & md & no & none & $\mathrm{rel}_2$ & yes & 96 & 1 & no & hh & no & no & no \\
    15 & flat & 200 & no & yes & md & no & none & $\mathrm{rel}_1$ & no & 96 & 1 & no & hh & no & no & no \\
    16 & flat & 200 & no & no & md & no & none & $\mathrm{rel}_7$ & yes & 96 & 1 & no & hh & no & no & no \\
    17 & flat & 200 & no & no & md & no & none & none & yes & 96 & 1 & no & hh & no & no & no \\
    18 & flat & 200 & no & no & md & no & none & $\mathrm{rel}_3$ & yes & 96 & 1 & no & hh & no & no & no \\
    19 & flat & all & no & no & md & no & none & none & yes & 96 & 1 & no & hh & no & no & no \\
    20 & flat & all & no & no & md & no & sin & none & yes & 96 & 1 & no & hh & no & no & no \\
    21 & flat & 200 & - & - & - & - & - & - & no & - & - & - & hh & no & no & no \\
    \hline
  \end{tabular}
  }
   \caption{Options used for the different shower training runs. The models are ordered according to their ID which is determined by average total test loss (see Fig. \ref{subfig:testresults_cascades}). Explanations for the numeric option mapping are given in table \ref{tab:numeric_defs} and for some abbreviations are given in table \ref{tab:opt_abbreviations}. Model 21 is a GNN and the related transformer-related options are marked with \enquote{-}.}
  \label{tab:cascade_options}
\end{table*}

\begin{table*}[ht]
  \centering
  \begin{tabular}{cp{1.5em}p{1.5em}p{1.5em}p{1.5em}p{1.5em}p{1.5em}p{1.5em}p{1.5em}p{1.5em}p{1.5em}p{1.5em}p{1.5em}p{1.5em}p{1.5em}p{1.5em}p{1.5em}}
    \hline
\multicolumn{17}{c}{\textbf{Numeric option abbreviations:}} \\
    \hline
    \multicolumn{17}{l}{\begin{tabular}[t]{@{}l@{}} 1: train data weighting \\ 2: max DOMs \\ 3: nonlinear in-projection \\ 4: ReSi Dual \\ 5: aggregation mode \\ 6: add mean diff. for in-projection \\ 7: abs. pos. encoding \\ 8: rel. pos. encoding \\ 9: bottleneck \\ 10: transformer computing dim \\ 11: numheads \\ 12: extra layer norm \\ 13: rotation type \\ 14: randomize xyz of DOMs during training \\ 15: ignore saturation for COG calc. \\ 16: simple vMF flow \end{tabular}} \\
    \hline
  \end{tabular}
  \caption{A list of numerical identifiers for different model hyperparameters.}
  \label{tab:numeric_defs}
\end{table*}

\begin{table*}[ht]
  \centering
  \begin{tabular}{lp{35em}}
    \hline
    \multicolumn{2}{c}{\textbf{train data weighting}} \\
    \hline
    flat & Flat distribution in deposited energy. \\
    si & Weighting data to a spectrum with spectral index -1.8. \\
    both & Arithmetic mean of spectral index (-1.8) + equal weighting in deposited energy. \\
    \hline
    \multicolumn{2}{c}{\textbf{rel. positional encoding}} \\
    \hline
    $\mathrm{rel}_1$ & Relative value position encoding in first layer, using standard relative value positional encoding. \\
    $\mathrm{rel}_2$ & Relative value position encoding using just the first layer, also adding the absolute input token before projection to the positional encoding input. \\
    $\mathrm{rel}_3$ & Relative value position encoding using just the first layer, but in parallel to another normal encoding. \\
    $\mathrm{rel}_4$ & Relative value position encoding in first layer with both overall input and value token, and in layer 2) and 3) with just the previous value token. \\
    $\mathrm{rel}_5$ & Relative value position encoding in layers 1)-5) using both the absolute input token before the transformer concatenated with the respective value token in each layer. \\
    $\mathrm{rel}_6$ & Relative value position encoding in interleaved layers 1), 3), 6), 9) and 12) using both the absolute input token before the transformer concatenated with the respective value token in each layer. \\
    $\mathrm{rel}_7$ & first layer with both overall input and value token, and in layer 5) and 10) with just the previous value token. \\
    \hline
    \multicolumn{2}{c}{\textbf{aggregation mode}} \\
    \hline
    md & Mean + diagonal variance of all tokens (2 X vector size of just mean). \\
    c1 & Learnable class token, added as a learnable bias in the first layer after QKV projection. \\
    c2 & Learnable class token, added as an extra learnable token in the first layer, treated just as normal token. Like originally introduced in \cite{vision_transformers}.  \\
    c3 & Improved class token that is treated separately from all other tokens, interacts via cross attention and has its own separate MLP layers. In combination with \enquote{ReSi Dual}, it also uses two residual streams.  \\
    \hline
    \multicolumn{2}{c}{\textbf{rotation type}} \\
    \hline
    hh & Three stacked householder reflectors for rotations. \\
    xyz & Standard mean parametrization of rotation matrix. \\
    \hline
  \end{tabular}
  \caption{Descriptions of some option abbreviations.}
  \label{tab:opt_abbreviations}
\end{table*}

\clearpage

\bibliographystyle{unsrt}
\bibliography{main}

\end{document}